\documentclass[11pt]{article}
\usepackage{graphicx} % standard LaTeX graphics for including eps-figure files
\usepackage{cite}     %this package collapses a list of three or more consecutive reference numbers into a range!
\usepackage{amsthm}
\usepackage{amsmath}
\usepackage{amssymb}
\usepackage{amsfonts}

\def \BE {\begin{equation}}
\def \EE {\end{equation}}
\def \BEA { \begin{eqnarray}}
\def \EEA {\end{eqnarray}}
\def \bea { \begin{eqnarray}}
\def \eea {\end{eqnarray}}
\def \be {\begin{equation}}
\def \ee {\end{equation}}

\def \pul {\textstyle{\frac{1}{2}}}
\def \ctvrt {\textstyle{\frac{1}{4}}}

\def \H {\mathcal{H}}
\def \T {\mathcal{T}}
\def \B {\mathcal{B}}

\newcommand{\Dif}{{\rm D}}                                 % upright D

\newcommand{\Kdt}{{\hbox{\tiny Kundt}}}

\newcommand{\dd}{{\rm{d}}} % diferencial tj. stojate d s argumentem
\newcommand{\rovno}{\!\!\!& = &\!\!\!} % rovnitko se zarovnanim pro pouztiti v eqnarray
\newcommand{\eqdef}{\!\!\!& \equiv &\!\!\!} % rovnitko se zarovnanim pro pouztiti v eqnarray
\newcommand{\ssqrt}{{\textstyle\frac1{\sqrt{2}}}} % small inverse sqrt of 2
\newcommand{\HH}{{H}}

\newcommand{\boldu}{\mbox{\boldmath$u$}} % bold u
\newcommand{\bolde}{\mbox{\boldmath$e$}} % bold e
\newcommand{\boldk}{\mbox{\boldmath$k$}} % bold k
\newcommand{\boldl}{\mbox{\boldmath$l$}} % bold l
\newcommand{\boldm}{\mbox{\boldmath$m$}} % bold m
\usepackage[usenames]{color}

\begin{document}

\title{
Black holes and other exact spherical solutions in Quadratic Gravity}

\author{J.~Podolsk\'y$^\star$, R.~\v{S}varc$^\star$,
V.~Pravda$^\diamond$, A.~Pravdov\' a$^\diamond$\\
\vspace{0.05cm} \\
\vspace{0.05cm} \\
{\small $^\star$ Institute of Theoretical Physics, Faculty of Mathematics and Physics,} \\
{\small Charles University, V~Hole\v{s}ovi\v{c}k\'ach~2, 180~00 Prague 8, Czech Republic.} \\
{\small $^\diamond$ Institute of Mathematics, Academy of Sciences of the Czech Republic}, \\
{\small \v Zitn\' a 25, 115 67 Prague 1, Czech Republic.} \\[3mm]
{\small E-mail: \texttt{podolsky@mbox.troja.mff.cuni.cz, robert.svarc@mff.cuni.cz, }}\\
{\small \texttt{pravda@math.cas.cz, pravdova@math.cas.cz}}}

\maketitle

\begin{abstract}
We study static, spherically symmetric vacuum solutions to Quadratic Gravity, extending considerably our previous Rapid Communication [Phys. Rev. D 98, 021502(R) (2018)] on this topic. Using a conformal-to-Kundt metric ansatz, we arrive at a much simpler form of the field equations in comparison with their expression in the standard spherically symmetric coordinates. We present details of the derivation of this compact form of two ordinary differential field equations for two metric functions. Next, we apply analytical methods and express their solutions as infinite power series expansions. We systematically derive all possible cases admitted by such an ansatz, arriving at six main classes of solutions, and provide recurrent formulas for all the series coefficients. These results allow us to identify the classes containing the Schwarzschild black hole as a special case. It turns out that one class contains only the Schwarzschild black hole, three classes admit the Schwarzschild solution as a special subcase, and two classes are not compatible with the Schwarzschild solution at all since they have strictly nonzero Bach tensor. In our analysis, we naturally focus on the classes containing the Schwarzschild spacetime, in particular on a new family of the Schwarzschild--Bach black holes which possesses one additional non-Schwarzschild parameter corresponding to the value of the Bach tensor invariant on the horizon. We study its geometrical and physical properties, such as basic thermodynamical quantities and tidal effects on free test particles induced by the presence of the Bach tensor.  We also compare our results with previous findings in the literature obtained using the standard spherically symmetric coordinates.

\end{abstract}

\vfil\noindent PACS numbers:  04.20.Jb, 04.50.--h, 04.70.Bw,
04.70.Dy, 11.25.--w

% 04.20.Jb Exact solutions
% 04.50.-h Higher-dimensional gravity and other theories of gravity
% 04.50.Kd Modified theories of gravity
% 04.70.Bw Classical black holes
% 04.70.Dy Quantum aspects of black holes, evaporation, thermodynamics
% 04.50.Gh Higher-dimensional black holes, black strings, and related objects
% 11.25.-w Strings and branes


\newpage
\tableofcontents
\newpage

\section{Introduction}
\label{intro}

Soon after Albert Einstein   formulated his General Relativity in November 1915 and David Hilbert found an elegant procedure how to derive Einstein's field equations from the variational principle, various attempts started to extend and generalize this gravity theory. One possible road, suggested by Theodor Kaluza exactly a century ago in 1919, was to consider higher dimensions in an attempt to unify the field theories of gravitation and electromagnetism. In the same year, another road was proposed by Hermann Weyl. In this case, the idea was to derive alternative field equations of a metric theory of gravity by starting with a different action. Instead of using the Einstein--Hilbert Lagrangian of General Relativity, which is simply the Ricci curvature scalar $R$ (a double contraction of a single Riemann tensor), Weyl proposed a Lagrangian containing \emph{contractions of a product of two curvature  tensors}. Such a Lagrangian is thus not linear in curvature --- it is quadratic so that this theory can be naturally called ``quadratic gravity''. Einstein was well aware of these attempts to formulate such alternative theories of gravity, and for some time he also worked on them. Interestingly, expressions for the quadratic gravity theory can be found even in his last writing pad (at the bottom of its last but one page) which he used in spring 1955.

Although it turned out rather quickly that these original classical theories extending General Relativity led to specific conceptual, mathematical and physical problems, the nice ideas have been so appealing that --- the whole century after their conception --- they are still very actively investigated. Both the higher dimensions of the Kaluza--Klein theory and Weyl's higher-order curvature terms in an effective action are now incorporated into the foundations of string theory. Quadratic Gravity (QG) also plays an important role in contemporary studies of relativistic quantum field theories.

Quadratic Gravity is a very natural and quite ``conservative'' extension of the Einstein theory, the most precise gravity theory today. Quadratic terms in the QG Lagrangian can be understood as corrections to General Relativity, which may play a crucial role at extremely high energies. In the search for a consistent quantum gravity theory, which could be applicable near the Big Bang or near spacetime singularities inside black holes, it is important to understand the role of these higher-order curvature corrections.

Interestingly, it was suggested by Weinberg and Deser, and then proved by Stelle \cite{Stelle:77} already in the 1970s that adding the terms quadratic in the curvature to the Einstein--Hilbert action renders gravity renormalizable, see the very recent review \cite{Salvio}. This property is also preserved in the general coupling with a generic quantum field theory. However, due to the presence of higher derivatives, ``massive ghosts'' also appear (the corresponding classical Hamiltonian is unbounded from below). Nevertheless, there is a possibility that these ghosts could be benign \cite{Smilga}. For all these reasons, this QG theory has attracted considerable attention in recent years.

In our work, we are interested in \emph{classical solutions to QG in four dimensions}. It can be easily shown that all Einstein spacetimes obey the vacuum field equations of this theory. However, QG also admits additional vacuum solutions with nontrivial Ricci tensor. In this paper, we focus on such \emph{static, spherically symmetric vacuum solutions} without a cosmological constant. They were first studied in the seminal work \cite{Stelle:1978}, in which three families of such spacetimes were identified by using a power expansion of the metric functions around the origin. The failure of the Birkhoff theorem in {Quadratic} Gravity has also been pointed out therein.  Spherically symmetric solutions were further studied in \cite{Holdom:2002}, where also numbers of free parameters for some of the above-mentioned classes were determined. Recently it has been pointed out in \cite{LuPerkinsPopeStelle:2015,LuPerkinsPopeStelle:2015b,PerkinsPhD} that, apart from the Schwarzschild black hole and other spherical solutions, QG admits a \emph{non-Schwarzschild} spherically symmetric and static black holes.

The field equations of a generic Quadratic Gravity theory form a highly complicated system of fourth-order nonlinear PDEs. Only a few nontrivial exact solutions are thus known so far, and various approximative and numerical methods have had to be used in their studies. Specifically, in the new class of black holes presented in \cite{LuPerkinsPopeStelle:2015}, the two unknown metric
functions of the standard form of spherically symmetric metric were given in terms of two complicated coupled ODEs
which were (apart from the first few orders in the power expansion) solved and analyzed numerically.
Interestingly, all QG corrections to the four-dimensional vacuum Einstein equations for constant Ricci scalar are nicely combined into a conformally well-behaved Bach tensor. Together with a conformal-to-Kundt metric ansatz~\cite{PravdaPravdovaPodolskySvarc:2017}, this leads to a considerably simpler autonomous system of the field equations.  We employed this approach in our recent letters \cite{PodolskySvarcPravdaPravdova:2018} and \cite{SvarcPodolskyPravdaPravdova:2018} for vanishing and nonvanishing cosmological constant, respectively. In \cite{PodolskySvarcPravdaPravdova:2018} we were thus able to present an explicit form of the corresponding nontrivial black-hole spacetimes --- the so-called \emph{Schwarzschild--Bach black holes} with two parameters, a position of the horizon and an additional Bach parameter.  By setting this additional Bach parameter to zero, the Schwarzschild metric of General Relativity is directly recovered.  In the present considerably longer paper, we are now giving the details of the derivation summarized in \cite{PodolskySvarcPravdaPravdova:2018}, and also survey and analysis of other classes of spherically symmetric solutions to Quadratic Gravity.

Our paper is organized as follows. In Sec.~\ref{QGandEWtheory} we recall the Quadratic Gravity and the Einstein--Weyl theory, and we put the corresponding field equations into a convenient form in which the Ricci tensor is proportional to the Bach tensor. In Sec.~\ref{BHmetricsec} we introduce a suitable spherically symmetric metric ansatz
in the conformal-to-Kundt form, and we give relations to the standard metric form. In Sec.~\ref{derivingFE} we overview the derivation of the field equations with various technical details and thorough discussion being postponed to Appendices A--C.
In Sec.~\ref{invariants} expressions for curvature invariants are derived. In Sec.~\ref{integration}  expansions in powers of ${\Delta \equiv r-r_0}$ around a fixed point $r_0$, and for $r \rightarrow \infty$ are introduced. In  Sec.~\ref{expansiont_0} the leading orders in ${\Delta }$ of the field equations are solved and four main classes of solutions are obtained. For these solutions, in Sec.~\ref{description} all coefficients  of the metric functions in the power expansions in $\Delta$ are given in the form of  recurrent formulas, convenient gauge choices are found, and various aspects of the solutions are discussed. Sections \ref{expansiont_INF} and  \ref{description_INF} focus on the same topics as Secs.~\ref{expansiont_0} and \ref{description}, respectively, but this time for expansions $r \rightarrow \infty$. In Sec.~\ref{summary} the relation of the solutions obtained in Secs. \ref{expansiont_0}--\ref{description_INF} (including their special subcases) to the solutions given in the literature is discussed, and summarized in Table \ref{tab:3}. Mathematical  and physical aspects (specific tidal effects and  thermodynamical quantities) of the Schwarzschild--Bach solutions are discussed in Sections~\ref{discussion-and-figures} and \ref{physics}, respectively. Finally, concluding remarks are given in Sec.~\ref{conclusions}.

\newpage

\section{Quadratic Gravity and the Einstein--Weyl theory}
\label{QGandEWtheory}

Quadratic Gravity (QG) is a natural generalization of Einstein's
theory that includes higher derivatives of the metric. Its
action in four dimensions contains additional quadratic terms, namely square of
the Ricci scalar $R$ and a contraction of the Weyl tensor
$C_{abcd}$ with itself \cite{Weyl1919,Bach1921}. In the
absence of matter, the most general QG action generalizing the Einstein--Hilbert action reads\cite{PravdaPravdovaPodolskySvarc:2017}\footnote{In four
dimensions, the Gauss--Bonnet term
${R_{abcd}R^{abcd}-4R_{ab}R^{ab}+R^2}$ does not contribute to the
field equations.}
\be
S = \int \dd^4 x\, \sqrt{-g}\, \Big(
\gamma \,(R-2\Lambda) +\beta\,R^2  - \alpha\, C_{abcd}\, C^{abcd}
\Big), \label{actionQG}
\ee
where ${\gamma=1/G}$ ($G$ is the Newtonian constant), $\Lambda$ is the cosmological constant, and
$\alpha$, $\beta$ are additional QG theory parameters. The
Einstein--Weyl theory is contained as a special case by setting
${\beta=0}$.

\emph{Vacuum field equations} corresponding to the action
(\ref{actionQG}) are
\begin{align}
&\gamma \left(R_{ab} - {\pul} R\, g_{ab}+\Lambda\,g_{ab}\right)-4 \alpha\,B_{ab} \nonumber \\
&\quad +2\beta\left(R_{ab}-\tfrac{1}{4}R\, g_{ab}+ g_{ab}\, \Box - \nabla_b \nabla_a\right) R = 0 \,, \label{GenQGFieldEq}
\end{align}
where $B_{ab}$ is the \emph{Bach tensor} defined as
\be
B_{ab} \equiv \big( \nabla^c \nabla^d + {\pul} R^{cd} \big)C_{acbd} \,. \label{defBach}
\ee
It is traceless, symmetric, and conserved:
\begin{equation}
g^{ab}B_{ab}=0 \,, \qquad B_{ab}=B_{ba} \,, \qquad \nabla^b B_{ab}=0 \,,
\label{Bachproperties}
\end{equation}
and also conformally well-behaved (see expression (\ref{OmBach}) below).

Now, \emph{assuming} ${R=\hbox{const.}}$, the last two terms in (\ref{GenQGFieldEq}) containing covariant derivatives of $R$ vanish. Using (\ref{Bachproperties}), the trace of the field equations thus immediately implies
\be
R=4\Lambda\,.
\label{R=4Lambda}
\ee
By substituting this relation into the field equations (\ref{GenQGFieldEq}), they simplify considerably to
\be
R_{ab}-\Lambda\,g_{ab}=4k\, B_{ab}\,, \qquad \hbox{where}\qquad
k \equiv \frac{\alpha}{\gamma+8\beta\Lambda} \,.
\label{fieldeqsgen}
\ee

In this paper, \emph{we restrict ourselves to} investigation of solutions with \emph{vanishing cosmological constant}~$\Lambda$  (see \cite{SvarcPodolskyPravdaPravdova:2018} for the study of a more general case ${\Lambda\ne0}$). In view of (\ref{R=4Lambda}), this implies vanishing Ricci scalar,
\be
R=0\,,
\label{R=0}
\ee
and the field equations (\ref{fieldeqsgen}) further reduce to a simpler form
\be
 R_{ab}=4k\, B_{ab}\,,
 \label{fieldeqsEWmod}
\ee
where the constant $k$ is now a shorthand for the combination of the theory parameters
${ k \equiv \alpha/\gamma= G\alpha}$.
For ${k=0}$ we recover vacuum Einstein's equations of General Relativity.
Interestingly, all solutions of (\ref{fieldeqsEWmod}) in \emph{Einstein--Weyl gravity} (${\beta=0}$) with ${R=0}$ \emph{are also solutions to general Quadratic Gravity} (${\beta\ne0}$) since for ${\Lambda=0}$ the QG parameter $\beta$ does not contribute to the constant $k$ defined by (\ref{fieldeqsgen}).

\section{Black hole metrics}
\label{BHmetricsec}

For studying static, nonrotating black holes, it is a common approach to employ the canonical form of a general spherically symmetric metric
\begin{equation}
\dd s^2 = -h(\bar r)\,\dd t^2+\frac{\dd \bar r^2}{f(\bar r)}+\bar r^2(\dd \theta^2+\sin^2\theta\,\dd \phi^2) \,.
\label{Einstein-WeylBH}
\end{equation}
In particular, for the famous \emph{Schwarzschild solution} of Einstein's General Relativity \cite{Schwarzschild:1916} (and also of QG), the two metric functions \emph{are the same} and take the well-known form
\begin{equation}
f(\bar{r}) = h(\bar{r})=1-\frac{2m}{\bar{r}} \,.
\label{SchwarzschildBH}
\end{equation}
The metric (\ref{Einstein-WeylBH}) was also used in the seminal papers \cite{LuPerkinsPopeStelle:2015,LuPerkinsPopeStelle:2015b} to investigate generic spherical black holes in Quadratic Gravity, in which it was surprisingly shown, mostly by numerical methods, that such a class contains further black-hole solutions \emph{distinct} from the Schwarzschild solution (\ref{SchwarzschildBH}). It turned out that while the Schwarzschild black hole has ${f=h}$, this non-Schwarzschild black hole is characterized by ${f\not=h}$. However, due to the complexity of the QG field equations (\ref{GenQGFieldEq}) for the classical metric form (\ref{Einstein-WeylBH}), it has not been possible to find an explicit analytic form of the metric functions ${f(\bar{r}), h(\bar{r})}$.

\subsection{A new convenient metric form of the black hole geometry}
\label{BH metric}

As demonstrated in our previous works \cite{PodolskySvarcPravdaPravdova:2018, SvarcPodolskyPravdaPravdova:2018},
it is much more convenient to employ an \emph{alternative metric form} of the spacetimes represented by
(\ref{Einstein-WeylBH}). This is obtained by performing the transformation
\begin{equation}
\bar{r} = \Omega(r)\,, \qquad t = u - \int\! \frac{\dd r}{\H(r)} \,, \label{to static}
\end{equation}
resulting in
\be
\dd s^2 = \Omega^2(r)
\Big[\,\dd \theta^2+\sin^2\theta\,\dd \phi^2 -2\,\dd u\,\dd r+{\cal H}(r)\,\dd u^2 \,\Big]\,.
\label{BHmetric}
\ee
The two new metric functions $\Omega(r)$ and $\H(r)$ are related to
 $f(\bar r)$ and $h(\bar r)$ via simple relations
\be
h = -\Omega^2\, \H \,, \qquad
f = -\left(\frac{\Omega'}{\Omega}\right)^2 \H \,, \label{rcehf}
\ee
where prime denotes the derivative with respect to $r$. Of course, the argument $r$ of both functions $\Omega$
and $\H$ must be expressed in terms of $\bar{r}$ using the inverse of the relation ${\bar{r} = \Omega(r)}$.

The metric \eqref{BHmetric} admits a \emph{gauge freedom} given by a constant rescaling and a shift of~$r$,
\be
r \to \lambda\,r+\nu\,, \qquad u \to \lambda^{-1}\,u \,.
\label{scalingfreedom}
\ee

More importantly, this new black hole metric is \emph{conformal} to a much simpler Kundt-type metric,
\be
\dd	s^2 =\Omega^2(r)\,\dd s^2_{\hbox{\tiny Kundt}}\,.
\label{confrelation}
\ee
Indeed,  ${\dd s^2_{\hbox{\tiny Kundt}}}$ belongs to the
famous class of \emph{Kundt geometries}, which are
nonexpanding, shear-free and twist-free, see \cite{Stephanietal:2003, GriffithsPodolsky:2009}.
In fact, it is a subclass of Kundt spacetimes which is the \emph{direct-product of two 2-spaces},
and is of Weyl algebraic type~D and Ricci type~II \cite{GriffithsPodolsky:2009, PravdaPravdovaPodolskySvarc:2017}. The first part of
\be
\dd s^2_{\hbox{\tiny Kundt}}=\dd \theta^2+\sin^2\theta\,\dd \phi^2 -2\,\dd u\,\dd r+{\cal H}(r)\,\dd u^2
\label{Kundt seed}
\ee
spanned by ${\theta, \phi}$ is a round 2-sphere of Gaussian curvature ${K=1}$,  while the second part spanned by ${u, r}$ is a 2-dim Lorentzian spacetime. With the usual stereographic representation of a 2-sphere given by  ${x+\hbox{i}\, y = 2\tan(\theta/2)\exp(\hbox{i}\phi)}$, this \emph{Kundt seed} metric can be rewritten as
\be
\dd s^2_{\hbox{\tiny Kundt}}=\frac{\dd x^2+\dd y^2}{\big(1+\frac{1}{4}(x^2+y^2)\big)^2}
 -2\,\dd u\,\dd r+{\cal H}(r)\,\dd u^2 \,.
\label{Kundt seed xy}
\ee

\subsection{The black hole horizon}
\label{BH horizon}

In the usual metric form (\ref{Einstein-WeylBH}), the Schwarzschild horizon is defined by the zeros of the same two metric functions $h{({\bar r})=f({\bar r})}$. Due to (\ref{SchwarzschildBH}), it is located at ${{\bar r}_h=2m}$, where $m$ denotes the total mass of the black hole.

In a general case, such a horizon can be defined as the \emph{Killing horizon} associated with the vector field ${\partial_t}$. Its norm is determined by the metric function $-h({\bar r})$. In the regions where ${h({\bar r})>0}$, the spacetime is static and $t$ is the corresponding temporal coordinate. The Killing horizon is generated by the \emph{null vector field} ${\partial_t}$, and it is thus located at a specific radius ${\bar r}_h$ satisfying
\begin{equation}
h \big|_{{\bar r}={\bar r}_h}=0\,. \label{standardhorizon}
\end{equation}

In terms of the new metric form \eqref{BHmetric}, we may similarly employ the vector field
${\partial_u}$ which coincides with ${\partial_t}$ everywhere. Its norm is given by  $\Omega^2\, \H$. Since the conformal factor $\Omega$ is nonvanishing throughout the spacetime, the Killing horizon is uniquely located at a
specific radius $r_h$ satisfying the condition
\begin{equation}
\H \big|_{r=r_h}=0\,. \label{horizon}
\end{equation}
Interestingly, via the relations \eqref{rcehf} this automatically implies ${h({\bar r_h})=0=f({\bar r_h})}$.

It is also important to recall that there is a \emph{time-scaling freedom} of the metric \eqref{Einstein-WeylBH}
\be
t\to  t/\sigma \,,
\label{scaling-t}
\ee
where ${\sigma \ne 0}$ is any constant, which implies ${h\to h\,\sigma^2}$. This freedom can be used to adjust an appropriate value of $h$ at a chosen radius ${\bar r}$. Or, in an asymptotically flat spacetime such as (\ref{SchwarzschildBH}) it could be used to achieve ${h \to 1}$ as ${{\bar r}\to \infty}$, thus enabling us to determine the mass of a black hole.

\subsection{The Kundt seed of the Schwarzschild solution}
\label{Kundt seed of the Schwarzschild}

It is also important to explicitly identify the Kundt seed geometry (\ref{Kundt seed}) which, via the conformal relation (\ref{confrelation}), generates the well-known vacuum \emph{Schwarzschild solution}.
This is simply given by
\begin{equation}
\bar{r}=\Omega(r)=-\frac{1}{r}\,,\qquad
\H(r) = -r^2-2m\, r^3 \,.
\label{Schw}
\end{equation}
Indeed, the first relation implies ${r=-1/\bar{r}}$, so that
${\H(\bar{r}) = - (1-2m/\bar{r})/\bar{r}^2}$. Using (\ref{rcehf}), we easily obtain (\ref{SchwarzschildBH}).
It should be emphasized that the standard physical range ${\bar{r}>0}$ corresponds to ${r<0}$. Also, the auxiliary Kundt coordinate $r$ \emph{increases from negative values to}~$0$, as $\bar{r}$ increases to~$\infty$.

Notice that ${\cal H}$ given by (\ref{Schw}) is simply a \emph{cubic} in the coordinate $r$ of the Kundt  geometry. For ${m=0}$, the Kundt seed with ${{\cal H} = -\,r^2}$ is the Bertotti--Robinson spacetime with the geometry ${S^2\times AdS_2}$ (see chapter~7 of~\cite{GriffithsPodolsky:2009}), and the corresponding conformally related metric (\ref{confrelation}) is just the flat space. It should also be emphasized that, while the Schwarzschild and Minkowski spacetimes are (the simplest) vacuum solutions in Einstein's theory, their Kundt seeds (\ref{Schw}) \emph{are not vacuum solutions} in Einstein's theory since their Ricci tensor is nonvanishing.
In fact, the Bertotti--Robinson geometry is an electrovacuum space of Einstein's theory.

Since conformal transformations preserve the Weyl tensor, both $\dd s^2$ and $\dd s^2_{\hbox{\tiny Kundt}} $ are of the \emph{same algebraic type}. Indeed, in the null frame
${\boldk = \mathbf{\partial}_r}$,
${\boldl = {\textstyle\frac{1}{2}}{\cal H}\,\mathbf{\partial}_r+\mathbf{\partial}_u}$,
${\boldm_i = \big(1+\ctvrt(x^2+y^2)\big)\mathbf{\partial}_i}$,
the only Newman--Penrose Weyl scalar for (\ref{Kundt seed xy}) is ${\Psi_2=-\frac{1}{12}({\cal H}''+2)}$, and both $\boldk$ and $\boldl$ are double principal null directions.
For the specific function (\ref{Schw}), ${\Psi_2=m\,r}$.
The Kundt seed geometry for the Schwarzschild solution is thus of algebraic type~D.
It is conformally flat if, and only if, ${m=0}$, in which case it is the Bertotti--Robinson spacetime.

\subsection{The Robinson--Trautman form of the black hole metrics}
\label{RT}

Recently, we have  proven in~\cite{PravdaPravdovaPodolskySvarc:2017} that \emph{any metric conformal to a Kundt geometry must belong to the class of expanding Robinson--Trautman geometries} (or it remains in the Kundt class). Indeed, performing  a simple transformation
${r(\tilde r)}$ of (\ref{confrelation}), (\ref{Kundt seed xy}), such that
\begin{equation}
r = \int\!\!\frac{\dd \tilde r}{\Omega^2(\tilde r)}\, , \qquad
\HH \equiv \Omega^{2}\, \H \,,
\label{guu_RT}
\end{equation}
we obtain
\begin{equation}
\dd s^2_{\hbox{\tiny RT}} = \Omega^2(\tilde r)\,\frac{\dd x^2 + \dd y^2}{\big(1+\ctvrt(x^2+y^2)\big)^2}
-2\,\dd u\,\dd \tilde r+\HH(\tilde r)\,\dd u^2 \,. \label{confRT}
\end{equation}
This has the canonical form of the Robinson--Trautman class
\cite{Stephanietal:2003, GriffithsPodolsky:2009} with the identification
\be
\Omega_{,\tilde r} = \sqrt{f/h}\,,\qquad  \HH  = - h\,.
\ee
The Schwarzschild black hole is recovered for ${\Omega(\tilde r)=\tilde r}$ that is ${\Omega_{,\tilde r}=1}$,
equivalent to ${f(\bar{r}) = h(\bar{r})}$. Other distinct non-Schwarzschild black hole solutions are identified by
${f(\bar{r}) \ne h(\bar{r})}$. The Killing horizon is obviously given by ${\HH(\tilde r_h)=0}$, corresponding to ${\H(r_h)=0=h(\bar r_h)}$ and ${f(\bar r_h)=0}$.

\section{The field equations}
\label{derivingFE}

The conformal approach to describing and studying black holes and other spherical solutions in Einstein--Weyl gravity and fully general Quadratic Gravity, based on the new form of the metric \eqref{BHmetric}, is very convenient. Due to (\ref{confrelation}), it enables to evaluate easily the Ricci and Bach tensors, entering the field equations (\ref{fieldeqsEWmod}), from the Ricci and Bach tensors of the much simpler Kundt seed metric ${\dd s^2_{\hbox{\tiny Kundt}}}$.
In particular, to derive the explicit form of the field equations, it is possible to proceed as follows:

\vspace{2mm}

\begin{enumerate}

\item Calculate all components of the Ricci and Bach tensors $R_{ab}^\Kdt$ and $B_{ab}^\Kdt$ for the Kundt seed metric $g_{ab}^\Kdt$. Since such a metric (\ref{Kundt seed xy}) is simple, containing only one general metric function of one variable $\H(r)$, its key curvature tensors are also simple. Their explicit form is presented in Appendix~A.

\item Use the well-known geometric relations for the Ricci and Bach tensors of conformally related metrics $g_{ab}^\Kdt$ and ${g_{ab}=\Omega^2 \,g_{ab}^\Kdt}$. Thus it is straightforward to evaluate the curvature tensors $R_{ab}$ and $B_{ab}$ for spherically symmetric geometries, starting from their forms of the Kundt seed calculated in the first step. In particular, since the Bach tensor trivially rescales under the conformal transformation as ${B_{ab} = \Omega^{-2}\,B_{ab}^\Kdt}$, it remains simple. These calculations are performed in Appendix~B.

\item These explicit components of the Ricci and Bach tensors are substituted into the field equations of Quadratic Gravity, which we already reduced to the expression ${R_{ab}=4k\, B_{ab}}$, see \eqref{fieldeqsEWmod}. This immediately leads to a very simple and compact form of these field equations. Moreover,  using the Bianchi identities, it can be shown that the whole system reduces just to two equations \eqref{Eq1C}, \eqref{Eq2C} for the metric functions $\Omega(r)$ and $\H(r)$, see Appendix~C.

\end{enumerate}

\vspace{2mm}

By this procedure, we thus arrive at a remarkably simple form of the field equations (\ref{fieldeqsEWmod}) for spherically symmetric vacuum spacetimes in Einstein--Weyl gravity and  general Quadratic Gravity with ${R=0}$, namely
\emph{two ordinary differential equations} for the \emph{two metric functions} $\Omega(r)$ and ${\cal H}(r)$:
\begin{align}
\Omega\Omega''-2{\Omega'}^2 = &\ \tfrac{1}{3}k\, \B_1 \H^{-1} \,, \label{Eq1}\\
\Omega\Omega'{\cal H}'+3\Omega'^2{\cal H}+\Omega^2
 = &\ \tfrac{1}{3}k \,\B_2  \,. \label{Eq2}
\end{align}
The functions $\B_1(r)$ and $\B_2(r)$ denote \emph{two independent components of the Bach tensor},
\bea
&& \B_1 \equiv {\cal H}{\cal H}''''\,, \label{B1}\\
&& \B_2 \equiv {\cal H}'{\cal H}'''-\tfrac{1}{2}{{\cal H}''}^2 +2\,. \label{B2}
\eea

Recall also the relation (\ref{R=0}), that is ${R=0}$, which is a trace of the field equations  \eqref{fieldeqsEWmod}. This relation takes the explicit form
\begin{equation}
{\cal H}\Omega''+{\cal H}'\Omega'+{\textstyle \frac{1}{6}} ({\cal H}''+2)\Omega = 0 \,,
 \label{trace}
\end{equation}
see (\ref{barR}). Indeed, it immediately follows from  \eqref{Eq1}, \eqref{Eq2}: just subtract from the derivative of the second equation the first equation multiplied by $\H'$ (and divide the result by $6\Omega'$).

It is a great advantage of our conformal approach with the convenient form of the new metric (\ref{BHmetric}) that the field equations (\ref{Eq1}), (\ref{Eq2}) are \emph{considerably simpler} than the previously used field equations for the standard metric \eqref{Einstein-WeylBH}. Moreover, they form an \emph{autonomous system}, which means that the differential equations \emph{do not explicitly depend on the radial variable $r$}. This will be essential for solving such a system, finding their analytic solution in the generic form \eqref{rozvojomeg0}, \eqref{rozvojcalH0} or \eqref{rozvojomegINF}, \eqref{rozvojcalHINF} in subsequent Section~\ref{integration}.

\section{Fundamental scalar invariants and geometric classification}
\label{invariants}
For a geometrical and physical interpretation of spacetimes that are solutions to the field equations (\ref{Eq1}), (\ref{Eq2}), it will be crucial to investigate the behaviour of scalar curvature invariants constructed from the Ricci, Bach, and Weyl tensors themselves. A direct calculation yields
\begin{align}
R_{ab}\, R^{ab} &=  16k^2\, B_{ab} B^{ab} \,, \label{invR}\\
B_{ab}\, B^{ab} &=  \tfrac{1}{72}\,\Omega^{-8}\,\big[(\B_1)^2 + 2(\B_1+\B_2)^2\big] \,,\label{invB}\\
C_{abcd}\, C^{abcd} &=  \tfrac{1}{3}\,\Omega^{-4}\,\big({\cal H}'' +2\big)^2 \,. \label{invC}
\end{align}
To derive these expressions, we have used the field equations, the quantities (\ref{RT_R rr})--(\ref{RT_R xx}), (\ref{Bach rr})--(\ref{Bach xx}), (\ref{WeyliK})--(\ref{WeylfK}), and relations (\ref{contraEinstein-WeylBHC}), (\ref{confrel}), (\ref{OmBach}) together with ${C_{abcd}\,C^{abcd}=\Omega^{-4}\, C_{abcd}^\Kdt\, C^{abcd}_\Kdt}$ which follows from the invariance of the Weyl tensor under conformal transformations.

It is interesting to observe from (\ref{invB}) and  (\ref{Bach rr})--(\ref{Bach xx}) with (\ref{OmBach}) that
\be
B_{ab}=0\quad \hbox{if, and only if,}\quad  B_{ab}\,B^{ab} =0\,.
\label{Bach=0iffINV=0}
\ee
Moreover,
\be
C_{abcd}\,C^{abcd}=0\quad \hbox{implies}\quad  B_{ab} =0\,,
\label{Weylinv=0thenBach=0}
\ee
because the relation ${{\cal H}'' +2=0}$ substituted into \eqref{invB} gives ${B_{ab}\,B^{ab} =0}$,
i.e., ${B_{ab} =0}$ due to \eqref{Bach=0iffINV=0}.

Notice also that the \emph{first Bach component} ${\B_1=\H \H''''}$
\emph{always vanishes on the horizon} where ${\H=0}$, see the condition \eqref{horizon}.

In view of the key invariant \eqref{invB}, there are
\emph{two geometrically distinct classes of solutions} to
(\ref{Eq1}), (\ref{Eq2}), depending on the Bach tensor
${B_{ab}}$. The first simple case corresponds to ${B_{ab}=0}$,
while the much more involved second case, not allowed in General Relativity, arises when
${B_{ab}\ne0}$. This invariant classification has geometrical and physical consequences. In particular, the distinction of spacetimes with ${B_{ab}=0}$ and with ${B_{ab}\ne0}$ can be detected by measuring geodesic deviation of test particles, see Section~\ref{geodeviation} below.

\subsection{${B_{ab}=0}$: Uniqueness of Schwarzschild} \label{integration:Schw}

First, let us assume the metrics (\ref{BHmetric}) such that  ${B_{ab}=0}$  everywhere.
In view of \eqref{Bach=0iffINV=0} and \eqref{invB},
this condition requires  ${\B_1=0=\B_2}$, that is
\be
{\cal H}''''=0\,,\qquad
{\cal H}'{\cal H}'''-{\textstyle\frac{1}{2}}{{\cal H}''}^2 +2 =0\,.
\label{Bab=0-RHS}
\ee
Therefore, all left-hand sides and right-hand sides of equations
(\ref{Eq1})  and (\ref{Eq2}) %(\ref{Neq_rr})--(\ref{Neq_xx})
\emph{vanish separately}, i.e.,
\be
\Omega\Omega''=2{\Omega'}^2\,,\qquad
\Omega\Omega'{\cal H}'+3\Omega'^2{\cal H}+\Omega^2 =0\,.
\label{Bab=0-LHS}
\ee
The first equations of (\ref{Bab=0-RHS}) and (\ref{Bab=0-LHS}) imply that ${\cal H}$ must be
\emph{at most cubic}, and $\Omega^{-1}$ must be \emph{at most linear} in $r$. Using a coordinate freedom \eqref{scalingfreedom} of the metric (\ref{BHmetric}), without loss of generality we obtain ${\Omega=-1/r}$. The remaining equations (\ref{Bab=0-RHS}), (\ref{Bab=0-LHS}) then admit a unique solution
\begin{equation}
\Omega(r)=-\frac{1}{r}\,,\qquad
{\cal H}(r) = -r^2-2m\, r^3   \,.
\label{IntegrSchwAdS}
\end{equation}
Not surprisingly, this is exactly  the Schwarzschild solution of General Relativity, see equation (\ref{Schw}). Thus we have verified that the \emph{Schwarzschild black hole spacetime is the only possible solution with vanishing Bach tensor}.
Its corresponding scalar invariants (\ref{invR})--(\ref{invC}) are
\be
R_{ab}\, R^{ab} = 0 = B_{ab}\, B^{ab}\,,\qquad
C_{abcd}\, C^{abcd} = 48\,m^2\,r^6 \,.
\label{SchwarzInvariants}
\ee
Clearly, for ${m\not=0}$ there is a curvature singularity at ${r\to\infty}$ corresponding to ${\bar{r}=\Omega(r)=0}$.\footnote{For brevity, in this paper the symbol ${r\to\infty}$ means ${|r|\to\infty}$, unless the sign of $r$ is explicitly specified.}

\subsection{${B_{ab}\ne0}$: New types of solutions to QG}
\label{integration:nonSchw}

Many other spherically symmetric vacuum solutions to Quadratic Gravity and Einstein--Weyl gravity exist when the Bach tensor is nontrivial. They are \emph{much more involved, and do not exist in General Relativity}. Indeed, the field equations (\ref{fieldeqsEWmod}) imply  ${R_{ab}=4k\, B_{ab}\ne0}$, which is in contradiction with vacuum Einstein's equations ${R_{ab}=0}$.

In the rest of this paper, we now concentrate on these new spherical spacetimes in QG, in particular on black holes generalizing the Schwarzschild solution. First, we integrate the field equations (\ref{Eq1}),
(\ref{Eq2}) for the metric functions $\Omega(r)$ and ${\cal H}(r)$. Actually, we
demonstrate that there are several classes of such solutions with ${B_{ab}\ne0}$. After their explicit identification and description, we will analyze their geometrical and physical properties.

\section{Solving the field equations}
\label{integration}

For nontrivial Bach tensor (${\B_1, \B_2 \ne0}$), the right-hand sides of the field equations (\ref{Eq1}), (\ref{Eq2}) are nonzero so that the nonlinear system of two ordinary differential equations for $\Omega(r)$, ${\cal H}(r)$ is coupled in a complicated way. Finding explicitly its general solution seems to be hopeless. However, \emph{it is possible to write the admitted solutions analytically, in terms of (infinite) mathematical series expressed in powers of the radial coordinate~$r$}.

In fact, there are \emph{two natural possibilities}. The first is the expansion in powers of the parameter ${\Delta \equiv r-r_0}$ which expresses the solution around any finite value $r_0$ (including ${r_0=0}$). The second possibility is the expansion in powers of $r^{-1}$ which is applicable for large values of $r$. Let us now investigate both these cases.

\subsection{Expansion in powers of~${\Delta \equiv r-r_0}$}
\label{expansio_DElta}

It is a great advantage that \eqref{Eq1}, \eqref{Eq2}
is an \emph{autonomous system}. Thus we can find the metric functions in the form of an \emph{expansion in powers of $r$  around any fixed value} ${r_0}$,
\begin{eqnarray}
\Omega(r) \rovno \Delta^n   \sum_{i=0}^\infty a_i \,\Delta^{i}\,, \label{rozvojomeg0}\\
\H(r)     \rovno \Delta^p \,\sum_{i=0}^\infty c_i \,\Delta^{i}\,, \label{rozvojcalH0}
\end{eqnarray}
where
\be
\Delta\equiv r-r_0\,, \label{DElta}
\ee
and $r_0$ is
\emph{any real constant}.\footnote{There may also exist other solutions such that their expansion contains logarithmic or exponential terms in $r$.}
In particular, in some cases this allows us to find solutions close to any black hole horizon~$r_h$ by choosing ${r_0=r_h}$.

It is assumed that ${i=0, 1, 2, \ldots}$ are integers, so that the metric functions are
expanded in integer steps of ${\Delta=r-r_0}$. On the other hand, the \emph{dominant real powers}
$n$ and $p$ in the expansions (\ref{rozvojomeg0}) and (\ref{rozvojcalH0})
\emph{need not be} positive integers. We only assume that ${a_0\not=0}$ and ${c_0\not=0}$, so that
the coefficients $n$ and $p$ are uniquely defined as the leading powers.

By inserting  \eqref{rozvojomeg0}--\eqref{DElta} into the field equations \eqref{Eq1}, \eqref{Eq2}, we prove in Section~\ref{expansiont_0} that \emph{only 4 classes of solutions of this form are allowed}, namely
\be
[n,p]=[-1,2]\,,\qquad
[n,p]=[0,1]\,,\qquad
[n,p]=[0,0]\,,\qquad
[n,p]=[1,0]\,.
\label{4classes}
\ee
In subsequent Section~\ref{description}, it will  turn out that the only possible solution in the class  ${[n,p]=[-1,2]}$ is the Schwarzschild black hole \eqref{Schw} for which the Bach tensor vanishes. Explicit Schwarzschild--Bach black holes with ${B_{ab}\ne0}$ are contained in the classes ${[0,1]}$ and ${[0,0]}$. The fourth class ${[n,p]=[1,0]}$ represents singular solutions without horizon, and it is equivalent to the class ${(s,t)=(2,2)}$ identified previously in \cite{Stelle:1978,LuPerkinsPopeStelle:2015b,PerkinsPhD}.

\subsection{Expansion in powers of~$r^{-1}$}
\label{expansion_INF}

Analogously, we may study and classify all possible solutions to the QG field equations for an asymptotic expansion as  ${r\rightarrow \infty}$. Instead of \eqref{rozvojomeg0}, \eqref{rozvojcalH0} with \eqref{DElta}, for very large $r$ we can assume that the metric functions $\Omega(r)$,  $\mathcal{H}(r)$  are expanded in \emph{negative powers} of $r$ as
\begin{eqnarray}
\Omega(r)      \rovno r^N   \sum_{i=0}^\infty A_i \,r^{-i}\,, \label{rozvojomegINF}\\
\mathcal{H}(r) \rovno r^P \,\sum_{i=0}^\infty C_i \,r^{-i}\,. \label{rozvojcalHINF}
\end{eqnarray}

Inserting the series (\ref{rozvojomegINF}), (\ref{rozvojcalHINF}) into the field equations \eqref{Eq1}, \eqref{Eq2}, it can be shown that \emph{only 2 classes of such solutions are allowed}, namely
\be
[N,P]=[-1,3]^\infty\,,\qquad
[N,P]=[-1,2]^\infty\,,
\label{2classes}
\ee
see Section~\ref{expansiont_INF}. In subsequent Section~\ref{description_INF}, it will be shown that the class ${[N,P]=[-1,3]^\infty}$ represents the Schwarzschild--Bach black holes, whereas the class ${[N,P]=[-1,2]^\infty}$ is a specific Bachian generalization of a flat space which does not correspond to a black hole.

\section{Discussion of solutions using the expansion in powers of~$\Delta$}
\label{expansiont_0}

By inserting the series (\ref{rozvojomeg0}), (\ref{rozvojcalH0}) into the first field equation (\ref{Eq1}), the following key relation is obtained
\begin{align}
&\sum_{l=2n-2}^{\infty}\Delta^{l}\sum^{l-2n+2}_{i=0}a_i\, a_{l-i-2n+2}\,(l-i-n+2)(l-3i-3n+1) \nonumber \\
& \hspace{35.0mm}=\tfrac{1}{3}k \sum^{\infty}_{l=p-4}\Delta^{l}\,c_{l-p+4}\,(l+4)(l+3)(l+2)(l+1) \,.
\label{KeyEq1}
\end{align}
The second field equation (\ref{Eq2}) puts further constraints on the admitted solutions, namely
\begin{align}
&\sum_{l=2n+p-2}^{\infty}\Delta^{l}\sum^{l-2n-p+2}_{j=0}\sum^{j}_{i=0}a_i\,a_{j-i}\,c_{l-j-2n-p+2}\,(j-i+n)(l-j+3i+n+2)
%\nonumber \\
%& \hspace{10.0mm}
+\sum_{l=2n}^{\infty}\Delta^{l}\sum^{l-2n}_{i=0}a_i\,a_{l-i-2n}
%-\Lambda \sum_{l=4n}^{\infty}\Delta^{l}\sum^{l-4n}_{m=0}\bigg(\sum^{m}_{i=0}a_i\,a_{m-i}\bigg)\bigg(\sum^{l-m-4n}_{j=0}a_j\,a_{l-m-j-4n}\bigg)
\nonumber \\
& = \tfrac{1}{3}k \bigg[2+\sum^{\infty}_{l=2p-4}\Delta^{l}\sum^{l-2p+4}_{i=0}c_{i}\,c_{l-i-2p+4}\,(i+p)(l-i-p+4)(l-i-p+3)(l-\tfrac{3}{2}i-\tfrac{3}{2}p+\tfrac{5}{2})\bigg]\,.
\label{KeyEq2}
\end{align}
A considerably simpler is the additional (necessary but not
sufficient) condition following from the trace equation
(\ref{trace}) which reads
\begin{align}
&\sum_{l=n+p-2}^{\infty}\Delta^{l}\sum^{l-n-p+2}_{i=0}c_i\,a_{l-i-n-p+2}\,\big[(l-i-p+2)(l+1)+\tfrac{1}{6}(i+p)(i+p-1)\big] %\nonumber \\
%& \hspace{50mm}
=-\tfrac{1}{3}\sum^{\infty}_{l=n}\Delta^{l}\,a_{l-n}
%\tfrac{2}{3}\Lambda \sum^{\infty}_{l=3n}\Delta^{l}\sum^{l-3n}_{j=0}\sum^{j}_{i=0}a_i\,a_{j-i}\,a_{l-j-3n}
\,.
\label{KeyEq3}
\end{align}

Now we analyze the consequences of the equations (\ref{KeyEq1})--%, (\ref{KeyEq2}),
(\ref{KeyEq3}).

First, by comparing the corresponding coefficients of the same powers
of $\Delta^l$ on both sides of the key relation (\ref{KeyEq1}), we can express
the coefficients $c_j$ in terms of (products of) $a_j$.
Moreover, the \emph{terms with the lowest order} put further restrictions. In particular, comparing the lowest
orders on both sides (that is ${l=2n-2}$ and ${l=p-4}$) it is obvious that \emph{we have to discuss
three distinct cases}, namely:
\begin{itemize}
\item \textbf{Case I}: ${\ \ 2n-2<p-4}$\,, \ i.e.,  ${\ p>2n+2}$\,,
\item \textbf{Case II}: ${\ 2n-2>p-4}$\,,  \ i.e.,  ${\ p<2n+2}$\,,
\item \textbf{Case III}: ${2n-2=p-4}$\,,   \ i.e.,  ${\ p=2n+2}$\,.
\end{itemize}
Now let us systematically derive all possible solutions in these three distinct cases.

\subsection{\textbf{Case I}}

In this case, ${2n-2<p-4}$, so that the \emph{lowest} order in the key equation (\ref{KeyEq1}) is on the \emph{left hand} side, namely $\Delta^l$ with ${l=2n-2}$, and this yields the condition
\begin{equation}
n(n+1)=0 \,.
\label{KeyEq1CaseI}
\end{equation}
There are thus only two possible cases, namely ${n=0}$ and ${n=-1}$. Next, it is convenient to apply
the equation (\ref{KeyEq3}) whose lowest orders on its both sides are
\begin{equation}
\big[6n(n+p-1)+p(p-1)\big]c_0\,\Delta^{n+p-2}+\cdots=-2\,\Delta^{n}+\cdots \,.
\label{KeyEq3CaseI}
\end{equation}
For ${n=0}$, these powers are ${\Delta^{p-2}}$
and ${\Delta^{0}}$, respectively, but ${p-2>2n=0}$ by the definition of Case~I.
The lowest order ${0=-2\Delta^{0}}$ thus leads to a contradiction.
Only the possibility ${n=-1}$ remains, for which \eqref{KeyEq3CaseI} reduces to
\begin{equation}
(p-3)(p-4)c_0\,\Delta^{p-3}+\cdots =-2\,\Delta^{-1}+\cdots
\,.
\label{KeyEq3CaseIn=-1}
\end{equation}
Since ${c_0\ne0}$, the only possibility is ${p=2}$, in which case ${c_0=-1}$.
\vspace{5mm}

\noindent
\textbf{To summarize}: The only possible class of solutions in Case~I is given by
\begin{equation}
[n,p]=[-1,2]\qquad \hbox{with}\quad c_0=-1\,.
\label{CaseI_summary}
\end{equation}

\subsection{\textbf{Case II}}

In this case, ${2n-2>p-4}$, so that the \emph{lowest} order in the key equation (\ref{KeyEq1}) is on the \emph{right hand} side, namely $\Delta^l$ with ${l=p-4}$, and this gives the condition
\begin{equation}
p(p-1)(p-2)(p-3)=0 \,.
\label{KeyEq1CaseII}
\end{equation}
Thus there are four possible cases, namely ${p=0}$, ${p=1}$, ${p=2}$, and ${p=3}$. Equation \eqref{KeyEq3} has the lowest orders on both sides the same as given by equation \eqref{KeyEq3CaseI},
that is
\begin{align}
\hbox{for}\quad p=0:\qquad &
\big[6n(n-1)\big]c_0\,\Delta^{n-2}+\cdots=-2\,\Delta^{n}+\cdots&&
\hbox{necessarily}\quad n=0, 1\,,\\
\hbox{for}\quad p=1:\qquad &
\big[6n^2\big]c_0\,\Delta^{n-1}+\cdots=-2\,\Delta^{n}+\cdots&&
\hbox{necessarily}\quad n=0\,,\\
\hbox{for}\quad p=2:\qquad &
\big[6n(n+1)+2\big]c_0\,\Delta^{n}+\cdots=-2\,\Delta^{n}+\cdots&&
(3n^2+3n+1)c_0=-1\,,\label{contrp=2c0}
\\
\hbox{for}\quad p=3:\qquad &
\big[6n(n+2)+6\big]c_0\,\Delta^{n+1}+\cdots=-2\,\Delta^{n}+\cdots&&
\hbox{not compatible}\,.
\end{align}
Moreover, the lowest orders of all the terms in the field equation \eqref{KeyEq2} for the case ${p=2}$, implying ${n>0}$, are
\be
3a_0^2\,[n(3n+2)c_0+1]\,\Delta^{2n} +2k(c_0^2 -1)
+ \cdots =0 \,, \label{eq2rozvoj0omeg}
\ee
which requires ${c_0=\pm 1}$, but the constraint \eqref{contrp=2c0} ${3n^2+3n+1=\pm 1}$ cannot be satisfied for ${n>0}$.
\vspace{5mm}

\noindent
\textbf{To summarize}: The only possible three classes of solutions in Case~II are given by
\be
 [n,p]=[0,1]\,,\qquad
 [n,p]=[0,0]\,,\qquad
 [n,p]=[1,0]\,.
\label{CaseII_summary}
\ee

\subsection{\textbf{Case III}}

Now ${2n-2=p-4}$, that is  ${n=-1+p/2}$ equivalent to ${p=2n+2}$. In such a case, the \emph{lowest} order in the key equation (\ref{KeyEq1})
is \emph{on both sides}, namely $\Delta^l$ with ${l=p-4}$. This implies the condition
\be
  p(p-2)\big[3a_0^ 2+4kc_0(p-1)(p-3)\big]=0\,.
 \label{KeyEq1CaseIII}
\ee
There are three subcases to be considered, namely
${p=0}$,  ${p=2}$, and ${3a_0^2=-4kc_0 (p-1)(p-3)}$ with ${p\not= 0,1,2,3}$.
This corresponds to
${n=-1}$,  ${n=0}$, and ${3a_0^2=-4kc_0(4n^2-1)}$ with ${n\not= -1,-1/2,0,1/2}$, respectively.
The leading orders of the trace equation \eqref{KeyEq3} on both sides are
\bea
2(11n^2+6n+1) c_0\,\Delta^{3n}  +\cdots \rovno -2\, \Delta^n+\cdots  \,. \label{eqtr00omegIII}
\eea
Consequently, we obtain
\begin{align}
&\hbox{for}\quad n=-1\Leftrightarrow p=0:\quad &
12c_0\,\Delta^{-3}+\cdots=-2\,\Delta^{-1}+\cdots& \qquad\hbox{not compatible}\,,\label{contrp=2c0IIIa}\\
&\hbox{for}\quad n=0\Leftrightarrow p=2:\quad & 2c_0+\cdots=-2+\cdots&\qquad c_0=-1\,,\label{contrp=2c0IIIb}\\
&\hbox{for}\quad 3a_0^2=4kc_0(1-4n^2): & (11n^2+6n+1)c_0+\cdots=0&\qquad
\hbox{not compatible}\,. \label{contrp=2c0IIIc}
\end{align}
The incompatibility  in the cases \eqref{contrp=2c0IIIa} and \eqref{contrp=2c0IIIc} are due to
the fact that ${c_0\ne 0}$ and ${11n^2+6n+1}$ is always positive. In the case \eqref{contrp=2c0IIIb},
we employ the  field equation \eqref{KeyEq2} which for ${n=0, p=2}$ gives the condition
${3a_0^2+2k(c_0^2-1) = 0}$. Since ${c_0=-1}$ implies ${a_0=0}$, we again end up in a contradiction.

\vspace{5mm}

\noindent
\textbf{To summarize}: There are no possible solutions in Case~III.

\section{Description and study of all possible solutions in powers of~$\Delta$}
\label{description}

Let us analyze all spherically symmetric solutions contained in the possible four classes \eqref{CaseI_summary} and \eqref{CaseII_summary} contained in Case~I and Case~II, respectively.

\subsection{Uniqueness of the Schwarzschild black hole in the class ${[n,p]=[-1,2]}$}
\label{Schw_[n,p]=[-1,2]}

Starting with the only admitted class ${[n,p]=[-1,2]}$ in the Case~I, see \eqref{CaseI_summary}, now we prove that \emph{the only solution in this class is the Schwarzschild solution}  with vanishing Bach tensor. Such a solution can be easily identified within the complete form (\ref{rozvojomeg0})--(\ref{DElta}), with ${r_0=0}$,  using the expression (\ref{IntegrSchwAdS}) as
\bea
&& a_0=-1\,,\hspace{26mm}a_i=0\quad \forall\ i\ge 1\,,\\
&& c_0=-1\,,\quad c_1=-2m\,,\quad c_i=0\quad \forall\ i\ge 2\,,
\eea
where $m$ is a free parameter.

Let us prove the uniqueness. The full key equation \eqref{KeyEq1} for ${n=-1}$ ${p=2}$ reads
\begin{align}
2a_1a_0\,\Delta^{-3}
+ 6a_2a_0\,\Delta^{-2}
+12a_3a_0\,\Delta^{-1}
%+(20a_4a_0+2a_1a_3-2a_2^2)\,\Delta^{0}
&
+\sum_{l=0}^{\infty}\Delta^{l}\sum^{l+4}_{i=0}a_i\, a_{l+4-i}\,(l+3-i)(l+4-3i)
\nonumber \\
&
=%8k\,c_{2}\, \Delta^{0}+
\tfrac{1}{3}k \sum^{\infty}_{l=0}\Delta^{l}\,c_{l+2}\,(l+4)(l+3)(l+2)(l+1) \,,
\label{KeyEq1[n,p]=[-12]}
\end{align}
which necessarily implies
\be
a_1=0\,,\qquad a_2=0\,,\qquad a_3=0\,,
\label{Schwinitcond1a}
\ee
and
\be
\sum^{l+4}_{i=0}a_i\, a_{l+4-i}\,(l+3-i)(l+4-3i)
=\tfrac{1}{3}k \,c_{l+2}\,(l+4)(l+3)(l+2)(l+1)\qquad \forall\ l\ge 0\,,
\label{Schwinitcond1b}
\ee
that is
\be
(l+4)(l+5)a_0a_{l+4}
=\tfrac{1}{3}k \,c_{l+2}\,(l+4)(l+3)(l+2)(l+1)
-\sum^{l+3}_{i=1}a_i\, a_{l+4-i}\,(l+3-i)(l+4-3i)\qquad \forall\ l\ge 0\,.
\label{Schwinitcond1c}
\ee
The second field equation \eqref{KeyEq2}, using (\ref{Schwinitcond1a}), takes the explicit form
\begin{align}
-c_2a_0^2\,\Delta^{0}&+ \sum_{l=1}^{\infty}\Delta^{l}\sum^{l+2}_{j=0}\sum^{j}_{i=0}a_i\,a_{j-i}\,c_{l-j+2}\,(j-i-1)(l-j+3i+1)
+\sum_{l=1}^{\infty}\Delta^{l}\sum^{l+2}_{i=0}a_i\,a_{l-i+2}\nonumber \\
&
= \tfrac{1}{3}k \,\sum^{\infty}_{l=1}\Delta^{l}\sum^{l}_{i=0}c_{i}\,c_{l-i}\,(i+2)(l-i+2)(l-i+1)
(l-\tfrac{3}{2}i-\tfrac{1}{2})\,,
\label{KeyEq2[n,p]=[-12]}
\end{align}
which implies
\be
c_2=0\,.
\label{Schwinitcond2}
\ee
However, instead of solving (\ref{KeyEq2[n,p]=[-12]}) for a general $l$, it is convenient to
employ the ``trace equation'' \eqref{KeyEq3}
\be
\sum^{l+1}_{i=0}c_i\,a_{l+1-i}\,\big[(l+1)(l-i)+\tfrac{1}{6}(i+1)(i+2)\big]
=-\tfrac{1}{3}\,a_{l+1}  \qquad \forall\ l\ge 2\,.
\label{KeyEq3[n,p]=[-12]}
\ee
This can be rewritten as
\be
(l-1)l\,a_0\,c_{l+1}=6l(l+1)\,a_{l+1}-\sum^{l}_{i=1}c_i\,a_{l+1-i}\,\big[6(l+1)(l-i)+(i+1)(i+2)\big]
\qquad \forall\ l\ge 2\,,
\label{KeyEq3[n,p]=[-12]b}
\ee
i.e., by relabeling the index ${l \to l+2}$, as
\be
(l+1)(l+2)\,a_0\,c_{l+3}=6(l+2)(l+3)\,a_{l+3}-\sum^{l+2}_{i=1}c_i\,a_{l+3-i}\,\big[6(l+3)(l+2-i)+(i+1)(i+2)\big]
\qquad \forall\ l\ge 0\,.
\label{KeyEq3[n,p]=[-12]c}
\ee

Now, we employ the mathematical induction. Let us assume that for some ${l\ge 0}$
\bea
&& a_i=0 \qquad \forall\ i=1,\ldots,l+3\,,\\
&& c_i=0 \qquad \forall\ i=2,\ldots,l+2\,.
\eea
For ${l=0}$ this is true due to \eqref{Schwinitcond1a}, \eqref{Schwinitcond2}.
Then the field equation \eqref{Schwinitcond1c} reduces to
\be
(l+4)(l+5)\,a_0\,a_{l+4}=0\,,\label{caseIbeq1}
\ee
while equation \eqref{KeyEq3[n,p]=[-12]c} gives
\be
(l+1)(l+2)\,a_0\,c_{l+3}=0\,.
\ee
This obviously implies ${a_{l+4}=0}$ and  ${c_{l+3}=0}$, completing the induction step.

Therefore, \emph{all} coefficients $a_i$ for $i\geq 1$ and
\emph{all} $c_i$ for $i\geq 2$ vanish, which means that the
only possible solution in Case I is
\be
\Omega=\frac{a_0}{\Delta}\,,\quad \mathcal{H}=-\Delta^2 +c_{1}\Delta^{3}\,.
\label{Schwarzschild[-1,2]}
\ee
With the coordinate freedom \eqref{scalingfreedom}, enabling us to set ${a_0=-1}$ and ${\Delta=r}$, this is exactly the explicit
Schwarzschild solution \eqref{IntegrSchwAdS}.
\vspace{5mm}

\noindent
\textbf{To conclude}: The class of solutions ${[n,p]=[-1,2]}$ represents spherically symmetric Schwarzschild solution \eqref{IntegrSchwAdS}, and it is the only solution in this class.

\subsection{Schwarzschild--Bach black holes in the class ${[n,p]=[0,1]}$: near the horizon}
\label{SchwaBach_[n,p]=[0,1]}

Now we will prove that this second class represents spherically symmetric \emph{non-Schwarzschild solutions to QG} that describe
\emph{black holes with nonvanishing Bach tensor}. Thus it is natural to call this family \emph{Schwarzschild--Bach black holes}. The first three terms in the expansion of the full solution take the explicit form
\begin{eqnarray}
\Omega(r)      \rovno  -\frac{1}{r} + \frac{b}{r_h^2}(r-r_h)
-\frac{b}{r_h^3}\Big(2 +\frac{1}{8k r_h^2}+b\Big)(r-r_h)^2+\ldots\,, \label{IIbOmegaFULL}\\
\mathcal{H}(r) \rovno
(r-r_h)\bigg[ \frac{r^2}{r_h}
+ 3b\,(r-r_h)+\frac{b}{r_h}\Big(4-\frac{1}{2k r_h^2} + 3b \Big)(r-r_h)^2 + \ldots \bigg]\,,
\label{IIbH0FULL}
\end{eqnarray}
where $r_h$ localizes the \emph{black hole horizon} since ${\H(r_h)=0}$. In fact, for the \emph{whole} class ${[n,p]=[0,1]}$, the metric function $\H$ given by \eqref{rozvojcalH0}, \eqref{DElta} takes the generic form ${\H(r) = (r-r_0)\,\big(c_0+c_1(r-r_0)+\ldots\big)}$, which means that ${r=r_0}$ is the root of $\H$, and thus the horizon. Therefore, we can identify the constant $r_0$ (around which the solution is expanded) with the location of geometrical/physical horizon,
\be
r_0\equiv r_h \,. \label{r0=rh}
\ee

When the additional new \emph{``Bach parameter''} $b$ in \eqref{IIbOmegaFULL}, \eqref{IIbH0FULL} is set to zero, the Bach tensor vanishes, and this solution reduces to the Schwarzschild spacetime (\ref{IntegrSchwAdS}) with ${r_h=-1/(2m)}$.

Let us systematically derive the complete analytic form of these Schwarzschild--Bach black holes, leading to \eqref{IIbOmegaFULL}, \eqref{IIbH0FULL}. The equation \eqref{KeyEq1} for ${[n,p]=[0,1]}$ gives
 \be
\sum^{l+1}_{i=0}a_i\, a_{l+2-i}\,(l+2-i)(l+1-3i) =\tfrac{1}{3}k
\,c_{l+3}\,(l+4)(l+3)(l+2)(l+1) \,, \label{KeyEq1[n,p]=[01]}
\ee
where ${l\ge 0}$. Relabeling  ${l \to l-1 }$, we thus obtain
\be
c_{l+2}=\frac{3}{k\,(l+3)(l+2)(l+1)l}\,\sum^{l}_{i=0}a_i \,
a_{l+1-i}(l+1-i)(l-3i)  \qquad \forall\ l\ge 1\,,
\label{nonSchwinitcondc}
\ee
which enables us to express all coefficients $c_{l+2}$ in terms of ${a_0,\ldots, a_{l+1}}$, starting from $c_3$.
In the lowest nontrivial order ${l=0}$,
the ``trace equation'' \eqref{KeyEq3}  implies
\be
a_1=-\frac{a_0}{3c_0}(1+c_1)\,,
\label{nonSchwinitcond3}
\ee
while for higher orders ${l=1, 2, \ldots}$, yields
\be
a_{l+1}=\frac{-1}{(l+1)^2\,c_0}\,\Big[\tfrac{1}{3}\,a_{l}+\sum^{l+1}_{i=1}
     c_i\,a_{l+1-i}\,\big[(l+1)(l+1-i)+\tfrac{1}{6}i(i+1)\big]\Big]
 \qquad \forall\ l\ge 1\,,
\label{nonSchwinitconda}
\ee
which expresses all $a_{l+1}$ in terms of ${a_0,\ldots,
a_l}$ and ${c_1,\ldots, c_{l+1}}$.
Finally,  in the lowest
nontrivial order ${l=0}$, the field equation \eqref{KeyEq2} gives the constraint
${6kc_0c_2=3a_0(a_0+a_1c_0)+2k(c_1^2-1)}$. Using \eqref{nonSchwinitcond3},
this becomes
\be
c_2=\frac{1}{6kc_0}\big[a_0^2(2-c_1)+2k(c_1^2-1)\big]\,.
\label{nonSchwinitcond2}
\ee

There are thus \emph{three free initial parameters}, namely
$a_0$, $c_0$, and $c_1$ (apart from ${r_0=r_h}$). Using \eqref{nonSchwinitcond3},
\eqref{nonSchwinitcond2}, we obtain ${a_1, c_2}$, and then
$a_{l+1}$, $c_{l+2}$  for all ${l = 1, 2, \ldots }$ by the
alternate application of the \emph{recurrent relations}
\eqref{nonSchwinitconda}, \eqref{nonSchwinitcondc}. This gives
the complete analytic solution.

Now, the scalar invariants (\ref{invB}), (\ref{invC}) evaluated at ${r=r_h\equiv r_0}$ take the form
\be
B_{ab}\,B^{ab}(r_h) = \Big(  \frac{1-c_1^2+3c_0c_2}{3 a_0^4} \Big)^2 \,,\qquad
C_{abcd}\, C^{abcd}(r_h) = \frac{4}{3 a_0^4}(1 + c_1)^2 \,.\label{BInv2}
\ee
\emph{The Bach tensor is in general nonvanishing}. In fact, for a
physical interpretation of this family of solutions, it is
convenient to introduce a new parameter $b$ proportional to
${1-c_1^2+3c_0c_2}$. Setting ${b=0}$ then gives the necessary
condition for the Bach tensor to vanish. In view of \eqref{nonSchwinitcond2},
such \emph{Bach parameter}~$b$ can be defined simply as
\be
 b \equiv \tfrac{1}{3}(c_1-2)\,, \label{b_definice}
\ee
so that the Bach scalar invariant \eqref{BInv2} at the black hole horizon $r_h$ becomes
\be
B_{ab}\,B^{ab}(r_h) = \frac{b^2}{4 k^2 a_0^4 } \,.
\label{BachInvariant}
\ee
Using  $b$ as the dimensionless key parameter in the expansion \eqref{rozvojomeg0}, \eqref{rozvojcalH0}, the
recurrent relations \eqref{nonSchwinitconda}, \eqref{nonSchwinitcondc} readily yield an explicit solution of the field equations in the form
\begin{align}
& a_1 = -\frac{a_0}{c_0}  \Big( 1 + b \Big) \,,\label{IIb_expansiona}\\
& a_2 = +\frac{a_0}{c_0^2}\Big( 1 + \big(2+\tfrac{a_0^2}{8k}\big)b+b^2 \Big) \,,\nonumber\\
& a_3 = -\frac{a_0}{c_0^3}\Big( 1 + \tfrac{1}{9}\big(25+\tfrac{29a_0^2}{8k}+\tfrac{a_0^4}{16k^2}\big)b
+\tfrac{1}{9}\big(23+\tfrac{35a_0^2}{8k}\big)b^2+\tfrac{7}{9}b^3 \Big) \,, \ldots\,,
\nonumber
\end{align}
and
\begin{align}
& c_1 = 2 + 3 b \,,\label{IIb_expansionc}\\
& c_2 = \frac{1}{c_0}\Big( 1 + \big(4-\tfrac{a_0^2}{2k}\big)b + 3 b^2 \Big) \,,\nonumber\\
& c_3 = \frac{a_0^4}{32k^2c_0^2}\, b\,,\nonumber\\
& c_4 = \frac{a_0^2}{30kc_0^3}\, b\,\Big(\big(1-\tfrac{5a_0^2}{4k}-\tfrac{a_0^4}{32k^2}\big)
+\big(2-\tfrac{13a_0^2}{8k}\big)b+ b^2 \Big)\,, \ldots\,,
\nonumber
\end{align}
and so on, where $a_0$, $c_0$, and $b$ are three free parameters.

\subsubsection{Identification of the Schwarzschild black hole}

Now, it is possible to \emph{identify the Schwarzschild black hole}. This is defined geometrically by
the property that its Bach tensor vanishes. In view of \eqref{BachInvariant}, it requires to set the key parameter $b$ to zero. Interestingly, with ${b=0}$, the expansion coefficients \eqref{IIb_expansiona}, \eqref{IIb_expansionc} simplify enormously to
\bea
&& a_i=a_0\,\Big(\!\!-\frac{1}{c_0}\Big)^i\quad \hbox{for all}\ i \ge 0 \,,\\
&& c_1=2\,, \quad c_2=\frac{1}{c_0}\,, \quad c_i=0\quad \hbox{for all}\ i \ge 3\,.
\eea
The first sequence clearly corresponds to a \emph{geometrical series}, while the second series is \emph{truncated to
a polynomial of the 3rd order}. The metric functions thus take the explicit closed form
\begin{eqnarray}
\Omega(r)      \rovno  a_0 \,\sum_{i=0}^\infty  \,\Big(\!\!-\frac{\Delta}{c_0}\Big)^i
 =\frac{a_0\,c_0}{c_0+\Delta}=\frac{a_0\,c_0}{r-r_h+c_0}\,, \label{IIbOmega}\\
\mathcal{H}(r) \rovno  c_0(r-r_h)+2(r-r_h)^2+c_0^{-1}(r-r_h)^3 \,. \label{IIbH0}
\end{eqnarray}
Using the gauge freedom \eqref{scalingfreedom} (a constant rescaling and shift of the coordinate $r$), \emph{we are free to chose}
\be
a_0=-\frac{1}{c_0}\,,\qquad c_0=r_h\,,\qquad
\label{IIb_a0}
\ee
so that the metric functions become
\be
{\bar r}=\Omega(r) = -\frac{1}{r}\,, \qquad
\mathcal{H}(r) = -r^2+\frac{r^3}{r_h} = \big(r-r_h\big)\frac{r^2}{r_h}  \,.
\label{IIbH0Schw}
\ee
Clearly, there is a \emph{black hole horizon located at}  ${r_h}$. This is the \emph{Schwarzschild horizon} given by the usual condition ${\,h=1-2m/\bar{r}=0\,}$. In terms of ${r=-1/{\bar r}}$, it is equivalent to ${r_h=-1/(2m)}$. Thus for the case ${b=0}$, we have fully recovered the standard form of the Schwarzschild solution, since the metric functions \eqref{IIbH0Schw} are exactly the same as \eqref{IntegrSchwAdS}.

\subsubsection{More general Schwarzschild--Bach black holes}

When ${b\not=0}$, the corresponding solution given by \eqref{nonSchwinitconda}, \eqref{nonSchwinitcondc},
that is \eqref{IIb_expansiona}, \eqref{IIb_expansionc}, can be naturally interpreted
as \emph{generalized black holes with a nontrivial Bach tensor} whose invariant value
${B_{ab}\,B^{ab}}$ at the horizon is proportional to $b^2$, according to \eqref{BachInvariant}.
Moreover, as ${b \to 0}$ we explicitly obtain the Schwarzschild black hole \eqref{IIbH0Schw}.
Using the summation of the ``background'' terms independent of $b$ as in \eqref{IIbOmega}, and the same gauge fixing \eqref{IIb_a0}, it is possible to write this solution explicitly as \eqref{IIbOmegaFULL}, \eqref{IIbH0FULL}. Recall that $r_h$ \emph{still gives the exact value of the horizon} even if  $b$ is now nonzero, see the text below equation \eqref{IIbH0FULL}.

To express a \emph{general solution in this class completely}, it is convenient to introduce coefficients $\alpha_i, \gamma_i$ as \emph{those parts of} $a_i, c_i$, respectively, which \emph{do not involve} the ${b=0}$ Schwarzschild ``background'', i.e., using the following definitions:
\bea
a_i \eqdef a_i(b=0)-\frac{b}{r_h}\,\frac{\alpha_i}{(-r_h)^i}\,,\qquad\hbox{where}\quad
a_i(b=0)\equiv\frac{1}{(-r_h)^{1+i}}   \,,\label{def_alphai}\\
%c_i \eqdef c_i(b=0)\,+ b\,r_h\,\frac{\gamma_i}{(r_h)^i}\\
c_1 \eqdef 2 + 3b\,\gamma_1 \,,\qquad
c_2 \equiv \frac{1}{r_h}+3b\,\frac{\gamma_2}{r_h} \,,\qquad
c_i \equiv 3b\,\frac{\gamma_i}{(r_h)^{i-1}}\quad \hbox{for all}\ i \ge 3\,.
 \label{def_gammai}
\eea
With the natural gauge choice \eqref{IIb_a0}, the complete solution then takes the explicit form
\bea
\Omega(r) \rovno -\frac{1}{r}
-\frac{b}{r_h}\,\sum_{i=1}^\infty\alpha_i\,\Big(1-\frac{r}{r_h}\Big)^i \,,
\label{Omega_[0,1]}\\
\mathcal{H}(r) \rovno (r-r_h)\bigg[  \,\frac{r^2}{r_h}
+3b\,r_h\,\sum_{i=1}^\infty \gamma_i\,\Big(\frac{r}{r_h}-1\Big)^i\,\bigg] \,,
\label{H_[0,1]}
\eea
with the initial coefficients
\begin{equation}
\alpha_1=1 \,, \qquad \gamma_1=1\,, \qquad \gamma_2 = \frac{1}{3}\Big(4-\frac{1}{2kr_h^2}+3b\Big) \,,
\label{alphasgammaIIbinitial}
\end{equation}
and all other coefficients $\alpha_l, \gamma_l$ for any ${l \ge 1}$ given by the recurrent relations
(defining ${\alpha_0=0}$)
\begin{align}
\alpha_{l+1}= &\, \frac{1}{(l+1)^2}\Big[\alpha_l\big(2l^2+2l+1\big)-\alpha_{l-1}l^2
-3\sum_{i=1}^{l+1}(-1)^i\,\gamma_i\,(1+b\,\alpha_{l+1-i})\big[(l+1)(l+1-i)+\tfrac{1}{6}i(i+1)\big]\Big],
   \nonumber\\
\gamma_{l+2}= &\, \frac{(-1)^{l+1}}{kr_h^2\,(l+3)(l+2)(l+1)l}\,\sum_{i=0}^{l}
   \big(\alpha_i+\alpha_{l+1-i}(1+b\,\alpha_i) \big)(l+1-i)(l-3i) \,,
   \label{alphasIIbgeneral}
\end{align}
which follow from \eqref{nonSchwinitconda} and \eqref{nonSchwinitcondc} for
$a_{l+1}$ and $c_{l+2}$, respectively.
The first terms generated by these relations are
\begin{align}
& \alpha_2 = 2+\frac{1}{8kr_h^2}+b \,, \nonumber\\
& \alpha_3 = \frac{1}{9}\Big(25+\frac{29}{8kr_h^2}+\frac{1}{16k^2r_h^4}\Big)
    +\frac{1}{9}\Big(23+\frac{35}{8kr_h^2}\Big)\,b+\frac{7}{9}\,b^2 \,, \ldots\,,
\label{alphasIIb0}\\
& \gamma_3 = \frac{1}{96k^2r_h^4} \,, \nonumber\\
& \gamma_4 = \frac{1}{18kr_h^2}\Big(\frac{1}{5}-\frac{1}{4kr_h^2}-\frac{1}{160k^2r_h^4}\Big)
    +\frac{3}{720kr_h^2}\Big(16-\frac{13}{kr_h^2}\Big)\,b+\frac{1}{90kr_h^2}\,b^2 \,, \ldots\,,
\label{gammasIIb0}
\end{align}
yielding \eqref{IIbOmegaFULL}, \eqref{IIbH0FULL}.

This family of spherically symmetric black-hole spacetimes \eqref{Omega_[0,1]}, \eqref{H_[0,1]} in  Einstein--Weyl/Quadratic Gravity depends on \emph{two parameters} with a \emph{clear geometrical
and physical interpretation}, namely:
\begin{itemize}
\item The parameter $r_h$ identifies the
    \emph{horizon position}. Indeed, ${r=r_h}$ is the root of the metric function $\H(r)$ given by \eqref{H_[0,1]}.
\item The dimensionless \emph{Bach parameter} $b$ \emph{distinguishes} the Schwarzschild solution (${b=0}$) from the more general Schwarzschild--Bach black hole spacetime with nonzero Bach tensor (${b\ne0}$).
\end{itemize}

In fact, we have chosen the parameter $b$ in such a way that it
\emph{determines the value of the Bach tensor \eqref{B1}, \eqref{B2} on the
horizon} $r_h$, namely
\be
\B_1(r_h) = 0\,, \qquad
\B_2(r_h) = -\frac{3}{kr_h^2}\,b \,. \label{bonhorizon}
\ee
Thus on the horizon, the invariants \eqref{invB} and \eqref{invC} of the Bach and Weyl tensors  take the values
\be
B_{ab}\,B^{ab}(r_h) = \frac{r_h^4}{4 k^2}\,b^2 \,,\qquad
C_{abcd}\, C^{abcd}(r_h) = 12\,r_h^4\,(1+b)^2  \,, \label{BCInvariants_[0,1]}
\ee
respectively.
\vspace{5mm}

\noindent
\textbf{To conclude}: The class of solutions ${[n,p]=[0,1]}$ represents spherically symmetric Schwarzschild--Bach  black holes (abbreviated as Schwa--Bach), expressed in terms of the series \eqref{Omega_[0,1]}, \eqref{H_[0,1]} \emph{around the horizon} $r_h$, i.e., for the special choice ${r_0=r_h}$. These Schwa--Bach black holes include and generalize the well-known Schwarzschild black hole.

\vspace{5mm}

Restricting to Einstein's theory, corresponding to ${k=0}$, requires ${a_0+a_1c_0=0}$, see the constraint above equation \eqref{nonSchwinitcond2}. Substituting this into \eqref{nonSchwinitcond3}, we obtain ${c_1=2}$, and thus ${b=0}$. This again confirms that the only possible spherical vacuum solution in General Relativity is the Schwarzschild solution.

Let us finally remark that the explicit recurrent relations  \eqref{alphasIIbgeneral} can be rewritten in a slightly more compact form if we relabel the index $l$ to ${j \equiv l+1}$, so that the relations for any ${j \ge 2}$  become
\begin{align}
\alpha_{j}= &\, \frac{1}{j^2}\Big[\alpha_{j-1}\big(2j^2-2j+1\big)-\alpha_{j-2}(j-1)^2
-3\sum_{i=1}^{j}(-1)^i\,\gamma_i\,(1+b\,\alpha_{j-i})\big[j(j-i)+\tfrac{1}{6}i(i+1)\big]\Big]\,,
   \nonumber\\
\gamma_{j+1}= &\, \frac{(-1)^{j}}{kr_h^2\,(j+2)(j+1)j(j-1)}\,\sum_{i=0}^{j-1}
   \big(\alpha_i+\alpha_{j-i}(1+b\,\alpha_i) \big)(j-i)(j-1-3i) \,.
   \label{alphasgammasgeneral_[0,1]}
\end{align}

\subsection{Schwarzschild--Bach black holes in the class ${[n,p]=[0,0]}$: near a generic point}
\label{SchwaBach_[n,p]=[0,0]}

This more general class of possible spherically symmetric vacuum solutions to QG (see \eqref{CaseII_summary}) \emph{may, as a special case, also represent the family of Schwarzschild--Bach black holes} with nonvanishing Bach tensor. In contrast to the previous case ${[n,p]=[0,1]}$, the expansion is now considered around \emph{an arbitrary fixed value} $r_0$ which is \emph{distinct from the position of the black hole horizon} $r_h$,
\be
r_0 \ne r_h\,. \label{r0_NOT=_rh}
\ee
Indeed, for ${[n,p]=[0,0]}$ the metric function $\H$ given by \eqref{rozvojcalH0}, \eqref{DElta} is ${\H(r) = c_0+c_1(r-r_0)+\ldots}$, where ${c_0\ne0}$, so that the value ${r=r_0}$ is \emph{not} the root of $\H$ and thus cannot be the horizon.

In such a case, the first few terms in the expansion of the full solution take the explicit form
\begin{eqnarray}
\Omega(r)      \rovno  -\frac{1}{r} + b_1\,\frac{r_h}{2r_0^3}\,\frac{(r-r_0)^2}{r_h-r_0}  +\ldots\,,
\label{[0,0]_OmegaFULL}\\
\mathcal{H}(r) \rovno
\big(r-r_h\big)\frac{r^2}{r_h} + (b_1-b_2)\,r_0(r-r_0) -3b_2\,(r-r_0)^2
  \nonumber\\
 && +\frac{(b_2-b_1)\big(1+\gamma+\frac{1}{2kr_0^2}\big)-2(2+3\gamma)b_2+3b_2^2}{(1+3\gamma+b_1-b_2)\,r_0}  \,(r-r_0)^3  \ldots \,,
\label{[0,0]_HFULL}
\end{eqnarray}
where $b_1$ and $b_2$ are \emph{two independent Bach parameters} proportional to values of the two components of the Bach tensor at $r_0$. By setting ${b_1=0=b_2}$, the Schwarzschild solution (which has vanishing Bach tensor) is immediately obtained.

Let us derive this analytic form of the Schwa--Bach black holes. For ${[n,p]=[0,0]}$ the complete solution to \eqref{Eq1}, \eqref{Eq2} of the form \eqref{rozvojomeg0}--\eqref{DElta} is given by the Taylor expansions
\be
\Omega(r)  = a_0 + \sum_{i=1}^\infty a_i \,(r-r_0)^{i}\,,\qquad
\H(r)      = c_0 + \sum_{i=1}^\infty c_i \,(r-r_0)^{i}\,.
\label{rozvoj[0,0]}
\ee
The key equation \eqref{KeyEq1} for ${n=0=p}$, after relabeling ${l \to l-1}$, gives
\be
c_{l+3}=\frac{3}{k\,(l+3)(l+2)(l+1)l}\,\sum^{l}_{i=0}
a_i \,a_{l+1-i}(l+1-i)(l-3i)  \qquad \forall\ l\ge 1\,.
\label{[0,0]initcondc}
\ee
Equation \eqref{KeyEq3}, relabeling  ${l \to l-1}$, implies
\be
a_{l+1}=\frac{-1}{l(l+1)\,c_0}\,\Big[\tfrac{1}{3}\,a_{l-1}+\sum^{l+1}_{i=1}
     c_i\,a_{l+1-i}\,\big[l(l+1-i)+\tfrac{1}{6}i(i-1)\big]\Big]
 \qquad \forall\ l\ge 1\,.
\label{[0,0]initconda}
\ee
Finally, the field equation \eqref{KeyEq2} in the lowest
nontrivial order ${l=0}$ gives one additional constraint
\be
c_3=\frac{1}{6kc_1}\big[3a_0(a_0+a_1c_1)+9a_1^2c_0+2k(c_2^2-1)\big]\,.
\label{[0,0]initcond2}
\ee

Thus there are \emph{five free initial parameters}, namely ${a_0,\, a_1,\, c_0,\, c_1,\, c_2}$ (in addition to $r_0$). All
the remaining coefficients $a_{l+1}$, $c_{l+3}$ in \eqref{rozvoj[0,0]} are then obtained by
applying the recurrent relations \eqref{[0,0]initconda}, \eqref{[0,0]initcondc}, respectively, starting as
\bea
&& a_2 = -\frac{1}{6 c_0}\,\big[a_0 + 3 a_1 c_1 + a_0 c_2\big]\,, \ldots \\
&& c_4 = -\frac{1}{24 kc_0}\,\big[6 a_1^2 c_0 + a_0 (a_0 + 3 a_1 c_1 + a_0 c_2)\big]\,,  \ldots\,.
\label{[0,0]initcond3}
\eea

Now we show that \emph{three of the five initial parameters} (namely ${a_0, a_1, c_0}$) \emph{can be conveniently fixed using the gauge freedom} in such a way that the Schwarzschild solution and flat Minkowski background are uniquely identified and directly seen.

\subsubsection{Identification of the Schwarzschild black hole}

Specific geometry can be identified by the scalar invariants \eqref{invB}, \eqref{invC} with \eqref{B1}, \eqref{B2}. In particular, the Bach invariant evaluated at ${r=r_0}$ is
\begin{align}
& B_{ab}\, B^{ab} (r_0) =  \frac{1}{72\,a_0^8}\,\big[ (\B_1)^2 + 2(\B_1+\B_2)^2\big]\,,\nonumber\\
& \hbox{where}\quad \B_1 (r_0)= 24 c_0c_4  \,,\qquad  \B_2 (r_0)= 2(3 c_1 c_3  - c_2^2 +  1)\,.
\label{[0,0]_BachInvariant}
\end{align}
\emph{\!Vanishing of the Bach tensor} (${B_{ab}=0 \Leftrightarrow \B_1=0=\B_2}$), which uniquely identifies the Schwarzschild solution, thus requires ${c_4=0}$ and ${3 c_1 c_3  - c_2^2 +  1 =0}$. In combination with \eqref{[0,0]initcond3}, \eqref{[0,0]initcond2}, this implies two necessary conditions
\be
c_1=-\frac{a_0}{a_1}\Big(1+3\frac{a_1^2}{a_0^2}c_0\Big)\,,\qquad
c_2=2+3\frac{a_1^2}{a_0^2}c_0\,
\label{[0,0]BachzeroA}
\ee
that only depend on the fraction ${a_1/a_0}$ and $c_0$. Interestingly, for such a choice of parameters the recurrent relations \eqref{[0,0]initconda}, \eqref{[0,0]initcondc} give a very simple complete solution
\be
a_i=a_0\,\Big(\frac{a_1}{a_0}\Big)^i\quad \hbox{for all}\ i \ge 0 \,,\qquad
c_3=-\frac{a_1}{a_0}\Big(1+\frac{a_1^2}{a_0^2}c_0\Big)\,, \quad c_i=0\quad \hbox{for all}\ i \ge 4\,.
\ee
The first sequence clearly yields a geometrical series, while the second series is truncated to
the 3rd-order polynomial. Thus the metric functions  take the closed form
\begin{eqnarray}
\Omega(r)      \rovno  a_0 \,\sum_{i=0}^\infty  \,\Big(\frac{a_1}{a_0}\Delta\Big)^i
 =\frac{a_0^2}{a_0-a_1\Delta}=\frac{a_0^2}{(a_0+a_1r_0)-a_1r}\,, \label{[0,0]Omega}\\
\mathcal{H}(r) \rovno  c_0+c_1(r-r_0)+c_2(r-r_0)^2+c_3(r-r_0)^3 \,. \label{[0,0]H0}
\end{eqnarray}
Using the gauge freedom  \eqref{scalingfreedom}, the most convenient choice
\be
a_0=-\frac{1}{r_0}\,,\qquad a_1=\frac{1}{r_0^2}
\label{[0,0]_a0a1}
\ee
can always be made, so that the metric functions reduce to
\be
{\bar r}=\Omega(r) = -\frac{1}{r}\,, \qquad
\mathcal{H}(r) = \big(r-r_0\big)\frac{r^2}{r_0}+\frac{c_0}{r_0^3}\,r^3  \,.
\label{[0,0]_Schw}
\ee
Notice that this function $\mathcal{H}$ can be rewritten as
\be
\mathcal{H}(r) = -r^2+\frac{r^3}{r_h} = \big(r-r_h\big)\frac{r^2}{r_h} \,,
\qquad\hbox{where}\qquad
r_h \equiv \frac{r_0^3}{r_0^2+c_0}  \,.
\label{[0,0]_SchwSHIFT}
\ee
This is exactly the standard form \eqref{IIbH0Schw} of the Schwarzschild solution, with the \emph{black hole horizon located at} $r_h$ (clearly the root of $\H$). Thus the constant $c_0$ is uniquely determined in terms of the physical/geometrical parameter $r_h$ (the horizon) and an arbitrary parameter $r_0$ (entering the expansion variable ${\Delta=r-r_0}$) as
\be
c_0 \equiv \gamma\,r_0^2\,,\qquad\hbox{where}\qquad\gamma\equiv\frac{r_0}{r_h}-1\,,\qquad r_0 \ne r_h  \,.
\label{[0,0]_DEF c0}
\ee
Thus we have proven that all solutions in the class ${[n,p]=[0,0]}$ \emph{with vanishing Bach tensor are equivalent
to the Schwarzschild black hole solution}, as also identified in the classes ${[n,p]=[0,1]}$ and ${[n,p]=[-1,2]}$, see
expressions \eqref{IIbH0Schw} and \eqref{Schwarzschild[-1,2]}, respectively. The \emph{main difference} is that in the class ${[n,p]=[0,1]}$,
it is possible (and, in fact, necessary) to choose the expansion parameter $r_0$ equal to the horizon $r_h$, see \eqref{r0=rh}, naturally allowing to expand the solution around the black hole horizon, while in the present case of the class ${[n,p]=[0,0]}$, \emph{such a choice is forbidden} ($r_0$ is \emph{not} the root of $\mathcal{H}$).
Indeed, for the choice ${r_0=r_h}$, the expression \eqref{[0,0]_DEF c0}
would lead to ${c_0=0}$ which is not allowed. Otherwise the constant $r_0$, determining the initial position
around which the solution is expanded, can be chosen \emph{arbitrarily}.

These conclusions are consistent with the behavior of the Weyl curvature invariant \eqref{invC} at $r_0$,
\be
 C_{abcd}\, C^{abcd}(r_0)  =   12\,\frac{r_0^6}{r_h^2}  = 48 m^2r_0^6 \,,
\ee
where we have used the conditions \eqref{[0,0]_SchwSHIFT} and the Schwarzschild horizon position ${r_h=-1/(2m)}$. This invariant value at the horizon agrees with \eqref{SchwarzInvariants}. For ${m=0}$, flat Minkowski background
is obtained, corresponding to  ${c_2=-1}$, that is ${c_0=-r_0^2}$, in which case ${\mathcal{H}(r) = -r^2}$, and there is no horizon.

\subsubsection{More general black hole solutions with nontrivial Bach tensor}

Returning to the generic case \eqref{[0,0]initcondc}--\eqref{[0,0]initcond3}
in the class ${[n,p]=[0,0]}$ with nonvanishing Bach tensor, it is now necessary to
\emph{introduce two distinct Bach parameters $b_1$ and $b_2$}, corresponding to the \emph{two components
$\B_1(r_0)$ and $\B_2(r_0)$} of the Bach tensor \eqref{B1} and \eqref{B2}, respectively, evaluated at $r_0$. They enter \eqref{[0,0]_BachInvariant} via the coefficients $c_4$ and $c_3$, which are expressed in terms of the two remaining initial parameters $c_1$ and $c_2$ using \eqref{[0,0]initcond3} and \eqref{[0,0]initcond2}. For ${B_{ab}=0}$,
they take the form \eqref{[0,0]BachzeroA}, i.e., with the gauge \eqref{[0,0]_a0a1} and fixing \eqref{[0,0]_DEF c0}, ${c_1=(1+3\gamma)\,r_0}$ and ${c_2=2+3\gamma}$. It turns out to be useful to define two dimensionless
Bach parameters $b_1$ and $b_2$ via the relations
\be
c_1 = (1+3\gamma+b_1-b_2)\,r_0 \,,\qquad
c_2 =  2+3\gamma-3b_2 \,,
\label{Bach_c1c2}
\ee
that is
\be
 b_1 \equiv  \tfrac{1}{3}\big(-1-6\gamma-c_2+3c_1/r_0\big) \,,\qquad
 b_2 \equiv  \tfrac{1}{3}\big(2+3\gamma-c_2\big) \,.
\label{Bach_b1b2}
\ee
Then $b_1$ and $b_2$ are \emph{directly proportional} to the two Bach tensor components $\B_1(r_0)$ and $\B_2(r_0)$,
\be
b_1=\tfrac{1}{3}kr_0^2\,\B_1(r_0) \,,\qquad b_2=\tfrac{1}{3}kr_0^2\,\big(\B_1(r_0)+\B_2(r_0)\big)\,,
\label{[0,0]_BachInvariantb1b2B1B2}
\ee
and the Bach invariant at $r_0$ is simply expressed as
\be
B_{ab}\, B^{ab} (r_0)= \frac{r_0^4}{8k^2}\,\big( b_1^2 + 2\, b_2^2\,\big) \,.
\label{[0,0]_BachInvariantb1b2}
\ee

With the parametrization by $b_1$, $b_2$ introduced in \eqref{Bach_c1c2}, assuming again the natural gauge \eqref{[0,0]_a0a1} and fixing \eqref{[0,0]_DEF c0}, the coefficients $a_i$, $c_i$ of the explicit solution
\eqref{[0,0]initcondc}--\eqref{[0,0]initcond2} are then given as
\begin{align}
&
a_0 = -\frac{1}{r_0}\,,\qquad
a_1 =\frac{1}{r_0^2}\,,\qquad
a_2 = -\frac{1}{r_0^3}-\frac{b_1}{2\gamma\, r_0^3}\,, \ldots\,,\label{[0,0]_expansionaINFa} \\
& c_0 = \gamma\,r_0^2 \,,\qquad
c_1 = (1+3\gamma)\,r_0+(b_1-b_2)\,r_0 \,,\qquad
c_2 = 2+3\gamma-3b_2 \,,\nonumber\\
&
c_3 = \frac{(1+\gamma)(1+3\gamma)-2(2+3\gamma)b_2+3b_2^2+(b_2-b_1)/(2kr_0^2)}{(1+3\gamma+b_1-b_2)\,r_0}  \,,\qquad
c_4 = \frac{b_1}{8k\gamma r_0^4}\,, \ldots\,.
\label{[0,0]_expansioncINFc}
\end{align}
For ${b_1=0=b_2}$, we immediately recover the Schwarzschild solution \eqref{[0,0]_Schw}, that is \eqref{[0,0]_SchwSHIFT}. In a generic case, the complete solution can be understood as the Schwarzschild black hole ``background'' modified by a nonzero Bach tensor, encoded in the terms that are proportional to  (powers of) the dimensionless  Bach parameters $b_1$ and~$b_2$.
The expansion of this full solution takes the explicit form \eqref{[0,0]_OmegaFULL}, \eqref{[0,0]_HFULL}.

\subsubsection{Identification of the Schwa--Bach black hole solutions ${[0,1]}$ in the class ${[0,0]}$ }

Now a natural question arises about the explicit relation between the form \eqref{IIbOmegaFULL}, \eqref{IIbH0FULL} and the form  \eqref{[0,0]_OmegaFULL}, \eqref{[0,0]_HFULL} of the family of Schwarzschild--Bach black holes. The problem is that we cannot simply express the single Bach parameter $b$ in terms of the two parameters ${b_1, b_2}$. The reason is that $b$ determines the value of the Bach tensor \emph{at the horizon} $r_h$, namely
\be
\B_1(r_h) = 0\,, \qquad
\B_2(r_h) = -\frac{3}{kr_h^2}\,b \,, \label{bonhorizonN}
\ee
while $b_1$ and $b_2$ determine its two independent values \emph{at any given} $r_0$
\be
\B_1(r_0) = \frac{3}{kr_0^2}\,b_1\,,\qquad
\B_2(r_0) = \frac{3}{kr_0^2}\,(b_2 - b_1)\,,
\label{[0,0]_BachInvariantb1b2B1B2N}
\ee
see \eqref{bonhorizon} and  \eqref{[0,0]_BachInvariantb1b2B1B2}, respectively. Since the functions ${\B_1(r),\, \B_2(r)}$ are complicated, the relations between the constants $b$ and ${b_1,\, b_2}$ are obscured.

However, this problem can be circumvent by the following procedure. In order to explicitly identify the Schwa--Bach black hole solution \eqref{Omega_[0,1]}, \eqref{H_[0,1]}, expressed around the horizon $r_h$ in the class ${[0,1]}$, within the generic class ${[0,0]}$ given by \eqref{rozvoj[0,0]}, we just have to determine its five free parameters ${a_0,\, a_1,\, c_0,\, c_1,\, c_2}$ properly. Instead of considering \eqref{[0,0]_expansionaINFa}, \eqref{[0,0]_expansioncINFc}, we can simply \emph{evaluate the functions} \eqref{Omega_[0,1]}, \eqref{H_[0,1]} (and their derivatives) at ${r=r_0}$, and then compare them with the Taylor expansions \eqref{rozvoj[0,0]} (and their derivatives) evaluated at ${r=r_0}$, obtaining\footnote{Of course, provided $r_0$ is within the convergence radius od \eqref{Omega_[0,1]}, \eqref{H_[0,1]}.}
\begin{align}
a_0 & = -\frac{1}{r_0}-\frac{b}{r_h}\sum_{i=1}^\infty \alpha_i\,\Big(1-\frac{r_0}{r_h}\Big)^i \,, \label{[0,0]a0}\\
a_1 & = \frac{1}{r_0^2}+\frac{b}{r_h^2}\sum_{i=1}^\infty i\,\alpha_i\,\Big(1-\frac{r_0}{r_h}\Big)^{i-1} \,, \\
c_0 & = (r_0-r_h)\bigg[\frac{r_0^2}{r_h}+3b\,r_h\sum_{i=1}^\infty \gamma_i\,\Big(\frac{r_0}{r_h}-1\Big)^i\, \bigg] \,, \\
c_1 & = (3r_0-2r_h)\frac{r_0}{r_h}+3b\,r_h\sum_{i=1}^\infty (i+1)\, \gamma_i\,\Big(\frac{r_0}{r_h}-1\Big)^i \,, \\
c_2 & = (3r_0-r_h)\frac{1}{r_h}+\frac{3}{2}b\,\sum_{i=1}^\infty i(i+1)\,\gamma_i\,\Big(\frac{r_0}{r_h}-1\Big)^{i-1} \label{[0,0]c2}\,.
\end{align}
Then using  the recurrent relations \eqref{[0,0]initcondc}--\eqref{[0,0]initcond2}, we are able to express the Schwarzschild--Bach black holes using the complete expansion around \emph{any value} $r_0$ and just a \emph{single Bach parameter} $b$ which determines the value of the Bach tensor at the horizon $r_h$.

When ${b=0}$, the coefficients $a_i$ form a geometrical series, and the metric functions simplify to
\eqref{[0,0]_Schw}, \eqref{[0,0]_SchwSHIFT} which is again the Schwarzschild solution \eqref{Schw}.
Both the classes ${[0,0]}$ and ${[0,1]}$ with ${B_{ab}=0}$ thus reduce
to the Schwarzschild black hole. The difference is that in the class ${[0,1]}$
the radial distance parameter $r_0$ is \emph{equal} to $r_h$, while
${r_0 \ne r_h}$ can be chosen \emph{arbitrarily} in the class ${[0,0]}$.

\subsubsection{Formal limit  ${r_0 \to r_h}$ }

Let us consider a ``consistency check'' between the two series expressing the Schwa--Bach black hole solution, namely \eqref{IIbOmegaFULL}, \eqref{IIbH0FULL} in the class ${[0,1]}$ and \eqref{[0,0]_OmegaFULL}, \eqref{[0,0]_HFULL} in the class ${[0,0]}$.

To this end, let us denote temporarily the coefficients in the class ${[0,0]}$ by ${\hat c}_i$ and ${\hat a}_i$.
The limit ${r_0 \to r_h}$ in \eqref{[0,0]a0}--\eqref{[0,0]c2} can be trivially performed, just by setting ${r_0=r_h}$, leading to the relations
\bea
{\hat a}_0 \rovno -\frac{1}{r_h} \equiv a_0\,,\qquad
{\hat a}_1 = \frac{1}{r_h^2}(1+b) \equiv a_1\,,\\
{\hat c}_0 \rovno 0\,,\qquad
{\hat c}_1 = r_h \equiv c_0\,,\qquad
{\hat c}_2 = 2+3b \equiv c_1\,,
\eea
where equation \eqref{IIb_a0} and the first relations in \eqref{IIb_expansiona}, \eqref{IIb_expansionc} have also been employed.
By comparing \eqref{nonSchwinitcondc} and \eqref{[0,0]initcondc} it is also seen that ${\hat c}_{j+1}$ satisfies the same recurrent relation as $c_j$, so that the functions $\H$ agree.
Moreover, from the relation \eqref{[0,0]initconda} it follows that the condition
${\hat c}_0=0$ requires
\be
0=\tfrac{1}{3}\,\hat a_{l-1} +l^2\, {\hat c}_1\,{\hat a}_l
+\sum^{l+1}_{i=2}\hat c_i\,\hat a_{l+1-i}\left[l(l+1-i)+\tfrac{1}{6}i(i-1)\right].
\ee
This implies
\be
{\hat a}_l =-\frac{1}{l^2\, {\hat c}_1} \Big[\tfrac{1}{3}\,\hat a_{l-1}
+\sum^{l}_{i=1}\hat c_{i+1}\, \hat a_{l-i}\left[l(l-i)+\tfrac{1}{6}i(i+1)\right] \Big],
\ee
which (with the identification ${\hat c_{i+1}=c_i}$) is equivalent to the recurrent expression \eqref{nonSchwinitconda} for $a_{l+1}$, so that the functions $\Omega$ also agree.
In other words, in the limit ${r_0\to r_h}$ we obtain
\be
{\hat c}_0 \to 0\,,\ \  \
{\hat c}_{j+1} \to c_j\,,\ \ \
{\hat a}_j \to a_j \quad \hbox{for all}\ j \geq 0\,,
\ee
demonstrating the consistency of the two expressions for the Schwa--Bach black holes in these two classes of solutions.

\subsection{Bachian singularity in the class ${[n,p]=[1,0]}$}
\label{SchwaBach_[n,p]=[1,0]}

This last possible class \eqref{4classes} of spherically symmetric vacuum solutions represents spacetimes which are \emph{not} black holes with horizon localized at $r_0$. Instead, it seems to be a specific family containing a naked singularity with ${B_{ab} \ne 0}$.

The key equation \eqref{KeyEq1} for ${[n,p]=[1,0]}$, relabeling   ${l \to l-3}$, gives
\be
c_{l+1}=\frac{3}{k\,(l+1)l(l-1)(l-2)}\,\sum^{l-3}_{i=0}
a_i \,a_{l-3-i}(l-2-i)(l-5-3i)  \qquad \forall\ l\ge 3\,,
\label{(2,2)initcondc} \ee
expressing $c_{l+1}$, starting from ${c_4}$.
Equation \eqref{KeyEq3} in the lowest order ${l=0}$ implies
\be
a_1=-\frac{a_0c_1}{2c_0}\,,
\label{(2,2)initcond3}
\ee
and in higher orders
\be
a_{l+1}=\frac{-1}{(l+1)(l+2)\,c_0}\,\Big[\tfrac{1}{3}\,a_{l-1}+\sum^{l+1}_{i=1}
     c_i\,a_{l+1-i}\,\big[(l+1)(l+2-i)+\tfrac{1}{6}i(i-1)\big]\Big]
 \qquad \forall\ l\ge 1\,.
\label{(2,2)initconda}
\ee
Finally, the field equation \eqref{KeyEq2} in the lowest
nontrivial order ${l=0}$ gives the condition
\be
c_3=\frac{1}{6kc_1}\big[9a_0^2c_0+2k(c_2^2-1)\big]\,.
\label{(2,2)initcond2}
\ee
All coefficients $a_{l+1}$, $c_{l+1}$  are obtained by
applying the recurrent relations
\eqref{(2,2)initconda}, \eqref{(2,2)initcondc}. This yields
an explicit solution
\be
\Omega(r)  = (r-r_0)\Big[a_0 + \sum_{i=1}^\infty a_i \,(r-r_0)^{i}\Big]\,,\qquad
\H(r)      = c_0 + \sum_{i=1}^\infty c_i \,(r-r_0)^{i}\,,
\label{rozvoj[1,0]}
\ee
where
\bea
% a_1 \rovno -\frac{a_0 c_1}{2 c_0}\,, \nonumber\\
a_2 \rovno -\frac{a_0}{18c_0^2}\,\big[c_0(1+7c_2)-6c_1^2\big]\,, \nonumber\\
a_3 \rovno -\frac{a_0 }{36 k c_0^3 c_1}\,\big[18 a_0^2 c_0^3 + k[4 c_0^2 (c_2^2-1) - 2 c_0 c_1^2 (1 + 10 c_2)
	+9 c_1^4]\big]\,, \ldots \,,\nonumber\\
% c_3 \rovno \frac{1}{6kc_1}\big[9a_0^2c_0+2k(c_2^2-1)\big]\,, \nonumber\\
c_4 \rovno -\frac{a_0^2}{4 k}\,, \qquad
c_5 = \frac{3 a_0^2 c_1}{40 k c_0}\,, \ldots \,,
\eea
and ${a_0,\, c_0,\, c_1,\, c_2}$ are \emph{four initial parameters} (apart from $r_0$), but not all of them are independent. Due to the gauge freedom \eqref{scalingfreedom}, we can set, for example, ${a_0=1}$ and also ${r_0=0}$.

To determine the main geometric properties we employ the scalar invariants \eqref{invB}, \eqref{invC}, which read
\be
B_{ab}\, B^{ab}(r)  = \frac{3 c_0^2}{4 a_0^4 k^2} \frac{1}{(r-r_0)^8}+\ldots \,,\qquad
C_{abcd}\, C^{abcd}(r) = \frac{4}{3 a_0^4}\frac{(1 + c_2)^2}{(r-r_0)^4}+\ldots \,.\label{(2,2)BInv2}
\ee
\emph{The Bach tensor $B_{ab}$ is thus nonvanishing near}~$r_0$.  And since ${R_{ab}=4k\,  B_{ab} \ne 0}$,
this class of solutions \emph{does not contain Ricci-flat subcases}. The Bach invariant always diverges at ${r=r_0}$, and there is also a \emph{Weyl curvature singularity} at ${r=r_0}$ (maybe unless ${c_2=-1}$).

Moreover, for \eqref{rozvoj[1,0]} the expressions \eqref{to static}--\eqref{rcehf} in the limit ${r\rightarrow r_0}$ behave
as
\bea
&& {\bar r}=\Omega(r) \sim   a_0(r-r_0) \to 0 \,, \label{(2,2)horizon}\\
&& h \sim  -c_0\,{\bar r}^2 \to   0\,, \qquad
   f \sim  -a_0^2c_0\, ({\bar r})^{-2} \to   \infty\,. \label{(2,2)horizonC}
\eea
It shows a very specific and unusual behavior of the metric functions $f$ and $h$ close to the curvature singularity at ${{\bar r}=0 }$, in terms of the physical radial coordinate $\bar r$.

This class ${[n,p]=[1,0]}$ of solutions corresponds to the family which has been
identified in \cite{Stelle:1978,LuPerkinsPopeStelle:2015b, HoldomRen:2017} as ${(s,t)=(2,2)}$, and nicknamed \emph{(2,2)-family} in \cite{PerkinsPhD}, see Section~\ref{summary} for more details.

\section{Discussion of solutions using the expansion in powers of~$r^{-1}$}
\label{expansiont_INF}

By inserting the series (\ref{rozvojomegINF}), (\ref{rozvojcalHINF}), that is
\begin{equation}
\Omega(r)    = r^N   \sum_{i=0}^\infty A_i \,r^{-i}\,, \qquad
\mathcal{H}(r) = r^P \,\sum_{i=0}^\infty C_i \,r^{-i}\,,\label{rozvojomagAcalHINF}
\end{equation}
into the key field equation (\ref{Eq1}), we obtain the relation
\begin{align}
&\sum_{l=-2N+2}^{\infty}r^{-l}\sum^{l+2N-2}_{i=0}A_i\,A_{l-i+2N-2}\,(l-i+N-2)(l-3i+3N-1) \nonumber \\
& \hspace{45.0mm}=\tfrac{1}{3}k\sum^{\infty}_{l=-P+4}r^{-l}\,C_{l+P-4}\,(l-4)(l-3)(l-2)(l-1) \,.
\label{KeyEq1INF}
\end{align}
The second field equation (\ref{Eq2}) puts further constraints, namely
\begin{align}
&\sum_{l=-2N-P+2}^{\infty}r^{-l}\sum^{l+2N+P-2}_{j=0}\sum^{j}_{i=0}A_i\,A_{j-i}\,C_{l-j+2N+P-2}
\,(j-i-N)(l-j+3i-N-2) \nonumber \\
& \hspace{10.0mm} +\sum_{l=-2N}^{\infty}r^{-l}\sum^{l+2N}_{i=0}A_i\,A_{l-i+2N}
\nonumber \\
& = \tfrac{1}{3}k \bigg[2+\sum^{\infty}_{l=-2P+4}r^{-l}\sum^{l+2P-4}_{i=0}C_{i}\,C_{l-i+2P-4}\,(i-P)(l-i+P-4)(l-i+P-3)
(l-\tfrac{3}{2}i+\tfrac{3}{2}P-\tfrac{5}{2})\bigg]\,.
\label{KeyEq2INF}
\end{align}
The supplementry condition following from the ``trace equation'' (\ref{trace}) reads
\begin{align}
&\sum_{l=-N-P+2}^{\infty}r^{-l}\sum^{l+N+P-2}_{i=0}C_i\,A_{l-i+N+P-2}\,\big[(l-i+P-2)(l-1)+\tfrac{1}{6}(i-P)(i-P+1)\big] \nonumber \\
& \hspace{80mm}=-\tfrac{1}{3}\sum^{\infty}_{l=-N}r^{-l}\,A_{l+N}
\,.
\label{KeyEq3INF}
\end{align}

By comparing the corresponding coefficients of the same powers of $r^{-l}$ on both sides of the relation (\ref{KeyEq1INF}),  we can express the coefficients $C_j$ in terms of $A_j$s.
Moreover, the \emph{terms with the lowest order} imply that we have to discuss \emph{three distinct cases}, namely:
\begin{itemize}
\item \textbf{Case I}$^\infty$: ${\ \ -2N+2<-P+4}$\,, \ i.e.,  ${\ P<2N+2}$\,,
\item \textbf{Case II}$^\infty$: ${\ -2N+2>-P+4}$\,,  \ i.e.,  ${\ P>2N+2}$\,,
\item \textbf{Case III}$^\infty$: ${-2N+2=-P+4}$\,,   \ i.e.,  ${\ P=2N+2}$\,.
\end{itemize}
Let us derive all possible solutions in these cases.

\subsection{\textbf{Case I}$^\infty$}

In the case, ${-2N+2<-P+4}$, the \emph{highest} order in the key equation (\ref{KeyEq1INF}) is on the \emph{left hand} side, namely $r^{-l}$ with ${-l=2N-2}$, which yields the condition
\begin{equation}
N(N+1)=0 \,.
\label{KeyEq1CaseIINF}
\end{equation}
The only two admitted cases are ${N=0}$ and ${N=-1}$.
The highest orders on both sides of equation (\ref{KeyEq3INF}) are
\begin{equation}
\big[6N(N+P-1)+P(P-1)\big]C_0\,r^{N+P-2}+\cdots=-2\,r^N+\cdots \,.
\label{KeyEq3CaseIINF}
\end{equation}
For ${N=0}$, these powers are ${r^{P-2}}$
and ${r^0}$, respectively, but ${P-2<2N=0}$ by the definition of Case~I$^\infty$.
The highest order ${0=-2r^0}$ thus leads to a contradiction. Similarly,
for the second possibility ${N=-1}$, the powers are ${r^{P-3}}$
and ${r^{-1}}$, respectively, but ${P-3<2N-1=-3<-1}$. The highest order is thus ${0=-2r^{-1}}$, which is again a contradiction.
\vspace{5mm}

\noindent
\textbf{To summarize}: There are no possible solutions in Case~I$^\infty$.

\subsection{\textbf{Case II}$^\infty$}

In this case, ${-2N+2>-P+4}$, so that the \emph{highest} order in the key equation (\ref{KeyEq1INF}) is on the \emph{right hand} side, namely $r^{-l}$ with ${l=-P+4}$, which gives the condition
\begin{equation}
P(P-1)(P-2)(P-3)=0 \,.
\label{KeyEq1CaseIIINF}
\end{equation}
Thus there are four possible cases, namely ${P=0}$, ${P=1}$, ${P=2}$, and ${P=3}$. Equation \eqref{KeyEq3INF} has the highest orders on both sides as given by equation \eqref{KeyEq3CaseIINF},
that is
\begin{align}
\hbox{for}\quad P=0:\qquad &
\big[6N(N-1)\big]C_0\,r^{N-2}+\cdots=-2\,r^N+\cdots&&
\hbox{not compatible}\,,\\
\hbox{for}\quad P=1:\qquad &
\big[6N^2\big]C_0\,r^{N-1}+\cdots=-2\,r^N+\cdots&&
\hbox{not compatible}\,,\\
\hbox{for}\quad P=2:\qquad &
\big[6N(N+1)+2\big]C_0\,r^N+\cdots=-2\,r^N+\cdots&&
(3N^2+3N+1)C_0=-1\,,\label{contrp=2c0INF}
\\
\hbox{for}\quad P=3:\qquad &
\big[6N(N+2)+6\big]C_0\,r^{N+1}+\cdots=-2\,r^N+\cdots&&
\hbox{necessarily}\quad N=-1\,.
\end{align}
The highest orders of all terms in equation \eqref{KeyEq2INF} for the case ${P=2}$, implying ${N<0}$, are
\be
3A_0^2\,[N(3N+2)C_0+1]\,r^{2N} +2k(C_0^2 -1)
+ \cdots =0 \,, \label{eq2rozvoj0omegINF}
\ee
which requires ${(3N^2+2N)C_0=-1}$. Together with constraint \eqref{contrp=2c0INF} this implies ${N=-1}$, ${C_0=-1}$.
\vspace{5mm}

\noindent
\textbf{To summarize}: The only possible two classes of solutions in Case~II$^\infty$ are given by
\be
 [N,P]=[-1,3]^\infty \,, \qquad
 [N,P]=[-1,2]^\infty \,. \label{CaseII_summaryINF}
\ee

\subsection{\textbf{Case III}$^\infty$}

Now, ${-2N+2=-P+4}$, that is  ${N=-1+P/2}$ and ${P=2N+2}$. In such a case, the \emph{highest} order in the key equation (\ref{KeyEq1INF}) is \emph{on both sides}, namely $r^{-l}$ with ${l=2-2N}$. This implies the condition
\be
  P(P-2)\big[3A_0^ 2+4kC_0(P-1)(P-3)\big]=0\,.
 \label{KeyEq1CaseIIIINF}
\ee
There are three subcases to be considered, namely
${P=0}$,  ${P=2}$, and ${3A_0^2=-4kC_0 (P-1)(P-3)}$ with ${P\not= 0,1,2,3}$.
This corresponds to
${N=-1}$,  ${N=0}$, and also ${3A_0^2=-4kC_0(4N^2-1)}$ with $N\not= -1, -1/2, 0, 1/2$, respectively.
The leading orders of the trace equation \eqref{KeyEq3INF} on both sides are
\bea
2(11N^2+6N+1) C_0\,r^{3N}  +\cdots \rovno -2\, r^N+\cdots  \,. \label{eqtr00omegIIIINF}
\eea
Consequently, we obtain
\begin{align}
&\hbox{for}\quad N=-1\,,\ P=0:\quad &
12C_0\,r^{-3}+\cdots=-2\,r^{-1}+\cdots& \qquad\hbox{not compatible}\,,\label{contrp=2c0IIIaINF}\\
&\hbox{for}\quad N=0\,,\ P=2:\quad & 2C_0+\cdots=-2+\cdots&\qquad C_0=-1\,,\label{contrp=2c0IIIbINF}\\
&\hbox{for}\quad 3A_0^2=4kC_0(1-4N^2): & (11N^2+6N+1)C_0+\cdots=0&\qquad
\hbox{not compatible}\,. \label{contrp=2c0IIIcINF}
\end{align}
The incompatibility  in the case \eqref{contrp=2c0IIIcINF} is due to the fact that ${11N^2+6N+1}$ is always positive. In the case \eqref{contrp=2c0IIIbINF}, we employ the  field equation \eqref{KeyEq2INF}, which for ${N=0, P=2}$ requires
${3A_0^2+2k(C_0^2-1) = 0}$. Since ${C_0=-1}$ would imply ${A_0=0}$, we also end up in a contradiction.

\vspace{5mm}

\noindent
\textbf{To summarize}: There are no possible solutions in Case~III$^\infty$.

\section{Description and study of all possible solutions in powers of~$r^{-1}$}
\label{description_INF}

Now we derive and investigate spherically symmetric solutions in the domain as ${r \to \infty}$ by completely solving the equations (\ref{KeyEq1INF}), (\ref{KeyEq2INF}), and their consequence (\ref{KeyEq3INF}). As it has been  proven in previous Section~\ref{expansiont_INF}, there are only two distinct cases \eqref{CaseII_summaryINF} to be discussed.\\

\subsection{Schwarzschild--Bach black holes in the class ${[-1,3]^\infty}$: near the singularity}
\label{Schw_[N,P]=[-1,3]}

In the class given by ${N=-1}$, ${P=3}$ in the expansion \eqref{rozvojomegINF}, \eqref{rozvojcalHINF} in \emph{negative} powers of~$r$, the only possible black hole solutions are
\begin{eqnarray}
\Omega(r)     \rovno  -\frac{1}{r}
+ \frac{B}{r}\,\bigg(\frac{2}{9}\,\frac{r_h^3}{r^3}
+\frac{1}{6}\,\frac{r_h^4}{r^4}
+\frac{2}{15}\,\frac{r_h^5}{r^5} +\ldots\bigg)\,, \label{IIdOmegaFULL}\\
\mathcal{H}(r) \rovno (r-r_h)\frac{r^2}{r_h}
+ B\,\bigg( r_h^2-\frac{1}{90k}\,\frac{r_h^3 }{r^3}
-\frac{1}{140k}\,\frac{r_h^4}{r^4}-\frac{1}{210k}\,\frac{r_h^5}{r^5}+ \ldots \bigg) \,. \label{IIdH0FULL}
\end{eqnarray}
These solutions represent the \emph{class of Schwarzschild--Bach black holes} in Quadratic Gravity/the Einstein--Weyl theory. By setting ${B=0}$, the \emph{Schwarzschild solution} \eqref{IIbH0Schw} is again obtained, with the \emph{horizon located at} $r_h$.

In the limit ${r\to\infty}$, the relation \eqref{to static} implies ${{\bar r}=\Omega(r)\sim -1/r \to 0 }$. In such a limit, \emph{the curvature singularity at ${\bar r = 0}$ is approached}, where ${\H \to \infty}$. Moreover, from the relations \eqref{rcehf} it follows that ${h({\bar r})\sim 1/( r_h \, \bar r) \to \infty}$ and ${f({\bar r})\sim h({\bar r})}$. Thus both metric functions of \eqref{Einstein-WeylBH}  \emph{diverge exactly in the same way as for the Schwarzschild solution, independently of the Bach parameter} $B$.

Let us derive this class of solutions.
The key equation \eqref{KeyEq1INF}, relabeling  ${l \to l+2 }$, implies
\be
C_{l+1}=\frac{3}{k\,(l-2)(l-1)l(l+1)}\,\sum^{l-2}_{i=0}
A_i\,A_{l-2-i}(l-1-i)(l-2-3i)  \qquad \forall\ l\ge 3\,,
\label{nonSchwinitcondcINF} \ee which gives all $C_{l+1}$ in
terms of ${A_0,\ldots, A_{l-2}}$, starting form ${C_4=0}$. The
trace equation \eqref{KeyEq3INF} yields \be
A_{l}=\frac{-1}{l^2\,C_0}\,\Big[\tfrac{1}{3}\,A_{l-1}
  +\sum^{l}_{i=1} C_i\,A_{l-i}\,\big[\,l(l-i)+\tfrac{1}{6}i(i+1)\big]\Big]
 \qquad \forall\ l\ge 1\,,
\label{nonSchwinitcondaINF}
\ee
which expresses all $A_{l}$
in terms of ${A_0,\ldots, A_{l-1}}$ and ${C_1,\ldots, C_{l}}$.
Finally, the second field equation \eqref{KeyEq2INF} in the
lowest nontrivial order ${l=0}$ gives the additional constraint
\be
C_2=\frac{C_1^2-1}{3C_0}\,.
\label{nonSchwinitcond2INF}
\ee
Therefore, in this case there are \emph{four free parameters}, namely
${A_0, C_0, C_1, C_3}$. Using \eqref{nonSchwinitcond2INF} we
obtain~${C_2}$, and then $A_l$, $C_{l+3}$  for all ${l\ge 1}$  by the application of the recurrent relations
\eqref{nonSchwinitcondaINF}, \eqref{nonSchwinitcondcINF}.

\subsubsection{Identification of the Schwarzschild black hole}

The scalar invariants \eqref{invB}, \eqref{invC} for \eqref{rozvojomegINF}, \eqref{rozvojcalHINF}
now take the form
\be
B_{ab}\, B^{ab}(r \to \infty) = \Big(45 \frac{C_0}{A_0^4}\, C_6\Big)^2  \,,
\qquad
C_{abcd}\, C^{abcd}(r \to \infty) \sim 12 \frac{C_0^2}{A_0^4}\,r^6  \,.
\label{BInv2INF}
\ee
Since ${A_0, C_0}$ are nonzero by definition, the necessary condition for
the \emph{Bach tensor to vanish} (which geometrically identifies the classical Schwarzschild solution) is
\be
C_6 = 0\,.
\label{C_6=0}
\ee
Interestingly, for such a setting, the expansion coefficients simplify enormously to
\bea
&& A_i=A_0\,\Big(\!\!-\frac{C_1+1}{3 C_0}\,\Big)^i\quad \hbox{for all}\ i \ge 0 \,,\\
&& C_2=\frac{C_1^2-1}{3C_0}\,, \quad C_3=\frac{(C_1+1)^2(C_1-2)}{27 C_0^2 }\,, \quad C_i=0\quad
\hbox{for all}\ i \ge 4\,.
\eea
The first sequence is a geometrical series, while the second series is
truncated to the 3rd-order polynomial. Thus the metric functions can be written in the
closed form
\begin{eqnarray}
\Omega(r)      \rovno  \frac{A_0}{r}   \,\sum_{i=0}^\infty \,\Big(\!\!-\frac{C_1+1}{3C_0\,r}\Big)^i
 =\frac{A_0}{r+(C_1+1)/(3C_0)}\,, \label{IIdOmega}\\
\mathcal{H}(r) \rovno  C_0\,r^3+C_1\,r^2+\frac{C_1^2-1}{3C_0}\, r+\frac{(C_1+1)^2(C_1-2)}{27 C_0^2 } \,. \label{IIdH0}
\end{eqnarray}
In view of \eqref{scalingfreedom}, we are free to chose the gauge
\be
A_0=-1\,,\qquad C_1=-1\,,
\label{IId_a0}
\ee
so that the metric functions become
\be {\bar r}=\Omega(r) = -\frac{1}{r}\,,
\qquad \mathcal{H}(r) = -r^2+C_0\,r^3 \,.
\label{IIdH0Schw}
\ee
This is exactly the \emph{Schwarzschild black hole metric} in the form \eqref{IntegrSchwAdS} and \eqref{IIbH0Schw}.
It also identifies the physical meaning of the coefficient $C_0$ as
\begin{equation}
C_0 = \frac{1}{r_h}  \,,
\label{IIdH0SchwC0}
\end{equation}
where $r_h$ determines the \emph{horizon position}, the root of $\mathcal{H}$ given by \eqref{IIdH0Schw}. Of course, the Schwarzschild horizon is given by ${r_h=-1/(2m)}$, i.e., ${C_0=-2m}$. All free parameters of such solution are thus fixed and fully determined.

\subsubsection{More general Schwarzschild--Bach black holes}

For the physical interpretation of the more general solutions in this family, it is convenient to introduce the \emph{Bach parameter $B$ proportional to ${C_6}$} entering \eqref{BInv2INF}, which for the gauge choice \eqref{IId_a0} reads
${ C_6 = -C_3 / (90k C_0)}$. We also naturally require $B$ to be a \emph{dimensionless} parameter, so that the
best choice seems to be
\be
B \equiv C_0^2C_3 = \frac{C_3}{r_h^2} \,.
\label{b_definiceINF}
\ee
With such $B$ as the key parameter in the expansions
\eqref{rozvojomegINF}, \eqref{rozvojcalHINF} and the
same natural gauge \eqref{IId_a0}, the recurrent
relations \eqref{nonSchwinitcondaINF}, \eqref{nonSchwinitcondcINF}
yield an explicit solution of the field equations in a simple form
\begin{align}
& A_0 = -1\,,\qquad
A_1 = 0\,,\qquad
A_2 = 0\,,\nonumber\\
& A_3 = \frac{2}{9}\,r_h^3\,B \,, \qquad
A_4 = \frac{1}{6}\,r_h^4\,B \,, \qquad
A_5 = \frac{2}{15}\,r_h^5\,B \,, \nonumber\\
& A_6 = \frac{1}{9}\,r_h^6\,\Big( 1-\frac{7}{360kr_h^2}-\frac{10}{9}\,B \Big)\,B\,, \ldots\,,\label{IId_expansionaINFa} \\
& C_0 = r_h^{-1} \,,\qquad
C_1 = -1 \,,\qquad
C_2 = 0 \,,\nonumber\\
& C_3 = r_h^2\,B \,,\qquad
C_4 = 0 \,,\qquad
C_5 = 0 \,,\nonumber\\
& C_6 = -\frac{1}{90k}\,r_h^3\,B  \,,\qquad
C_7 = -\frac{1}{140k}\,r_h^4\,B  \,,\qquad
C_8 = -\frac{1}{210k}\,r_h^5\,B \,, \ldots\,.
\label{IId_expansioncINFc}
\end{align}
This gives the explicit expansion \eqref{IIdOmegaFULL}, \eqref{IIdH0FULL}.

The corresponding scalar invariants \eqref{BInv2INF} at ${\bar r = 0}$ are
\be
B_{ab}\, B^{ab}(r \to \infty) = \frac{r_h^4}{4k^2}\,B^2  \,,\qquad
C_{abcd}\, C^{abcd} (r \to \infty) \sim \frac{12}{r_h^2}\,r^6 \to \infty \,,
\label{BInv2INFfinal}
\ee
which can be compared with the invariants \eqref{BCInvariants_[0,1]} evaluated at the horizon $\bar r_h$
\be
B_{ab}\,B^{ab}(r_h) = \frac{r_h^4}{4 k^2}\,b^2 \,,\qquad
C_{abcd}\, C^{abcd}(r_h) = 12\,r_h^4\,(1+b)^2  \,, \label{BInv2final}
\ee
obtained previously for the class ${[n,p]=[0,1]}$ of the Schwarzschild--Bach black holes. There is a striking similarity between the two expressions for ${B_{ab}\,B^{ab}}$, and thus we could be inclined to directly identify the Bach parameter~$B$ with the parameter $b$. However, is should again be emphasized that $B$ determines the value of the Bach invariant \emph{at the Weyl curvature singularity} ${\bar r=0}$, while $b$ determines its value \emph{at the horizon}~$\bar r_h$. And these values are, in general, distinct.

\subsection{Bachian vacuum in the class ${[N,P]=[-1,2]^\infty}$}
\label{Schw_[N,P]=[-1,2]}

Finally, it remains to analyze the second possibility \eqref{CaseII_summaryINF} in the Case~II$^\infty$.
For ${N=-1}$, ${P=2}$ the key equation \eqref{KeyEq1INF}, relabeling ${l \to l+2}$,
gives
\be
C_l=\frac{3}{k\,(l-2)(l-1)l(l+1)}\,\sum^{l-2}_{i=0}
A_i \,A_{l-2-i}(l-1-i)(l-2-3i)  \qquad \forall\ l\ge 3\,.
\label{[-1,2]initcondc}
\ee
Equation \eqref{KeyEq3INF} in its lowest orders ${l=1, 2 }$ puts the constraints
\begin{equation}
 A_1=\tfrac{1}{2}A_0C_1 \,, \qquad  C_0=-1\,,
\end{equation}
and for higher $l$ implies
\be
A_{l-1}=\frac{1}{l(l-1)}
\sum^{l-1}_{i=1} C_i\,A_{l-1-i}\,\big[(l-1)(l-i)+\tfrac{1}{6}(i-2)(i-1)\big]
 \qquad \forall\ l\ge 3\,.
\label{[-1,2]initconda}
\ee
The equation \eqref{KeyEq2INF} gives no additional constraint.
There are thus \emph{three free parameters}, namely ${A_0, C_1, C_2}$, and all other
coefficients are determined by the relations \eqref{[-1,2]initcondc}, \eqref{[-1,2]initconda},
starting as
\begin{align}
& A_2 = \frac{A_0}{3}(C_1^2 + C_2)\,,\qquad
A_3 = \frac{ A_0 }{4}C_1 (C_1^2 + 2C_2)\,,\nonumber\\
& A_4 = \frac{A_0 }{5}\Big(C_1^4 + 3 C_1^2 C_2 + C_2^2 +\frac{A_0^2}{192 k}(C_1^2 + 4 C_2)\Big)\,,  \ldots\,,
\label{[-1,2]_expansionaINFa} \\
& C_3 = 0 \,,\qquad
C_4 = \frac{A_0^2 }{240 k}(C_1^2 + 4C_2) \,,\qquad
C_5 = \frac{A_0^2}{240 k} C_1 (C_1^2 + 4C_2) \,,\nonumber\\
& C_6 = \frac{A_0^2 }{67200 k^2}\Big(3A_0^2+ 4k(59 C_1^2 + 26 C_2)\Big)(C_1^2 + 4C_2)  \,, \ldots\,.
\label{[-1,2]_expansioncINFc}
\end{align}

\subsubsection{Identification of flat Minkowski space}

Now, for very large $r$ the scalar invariants \eqref{invB}, \eqref{invC} behave as
\be
B_{ab}\, B^{ab}(r\to\infty) = \frac{300}{A_0^8}\,C_4^2  \,,\qquad
C_{abcd}\, C^{abcd}(r\to\infty) \sim \frac{12}{A_0^4\,r^4}\,C_4^2  \,.
\label{[-1,2]_Inv_INF}
\ee
Interestingly, they \emph{remain finite}, so that for ${r\to\infty}$ there is \emph{no physical singularity}. Moreover, for ${C_4 \ne0}$ they are nonzero. In fact, the necessary condition for
both the Bach and Weyl tensor invariants to vanish is ${C_4=0}$, that is ${C_1^2 + 4C_2 =0}$. For
such a choice, we obtain the relation ${C_2=-\frac{1}{4} C_1^2}$, and then all the coefficients
\eqref{[-1,2]_expansionaINFa}, \eqref{[-1,2]_expansioncINFc} simplify enormously to
${A_i=A_0\,(\frac{1}{2}C_1)^i}$ for all $i$, and ${C_i=0}$ for all ${i\ge3}$. The metric functions thus reduce to
\BE
\Omega(r)  =  \frac{A_0}{r}   \,\sum_{i=0}^\infty \,\Big(\frac{C_1}{2\,r}\Big)^i
 =\frac{A_0}{r-\tfrac{1}{2}C_1}\,, \qquad
\mathcal{H}(r) =  -(r-\tfrac{1}{2}C_1)^2 \,. \label{[-1,2]_Omega_H0}
\EE
Using the gauge freedom \eqref{scalingfreedom} we can always set
\be
A_0=-1\,,\qquad C_1=0\,,
\label{[-1,2]_a0}
\ee
and the functions take the trivial form
\be {\bar r}=\Omega(r) = -\frac{1}{r}\,,
\qquad \mathcal{H}(r) = -r^2 \,.
\label{[-1,2]_solution}
\ee
In view of \eqref{IIdH0Schw}, \eqref{IIdH0SchwC0}, we conclude that the case ${C_4=0}$
gives the Schwarzschild metric with trivial value ${C_0=-2m=0}$ which is
\emph{just flat Minkowski space without any horizon} (formally ${r_h=\infty}$).
Of course, for flat space, both the Bach and the Weyl tensor vanish everywhere.

\subsubsection{Bachian vacuum}

Now, the complete class of solutions ${[N,P]=[-1,2]^\infty}$ can be naturally analyzed if we introduce the \emph{Bach parameter $B_v$ proportional to ${C_4}$} because, due to \eqref{[-1,2]_Inv_INF}, such solutions admit general Bach and Weyl tensors. With the same gauge \eqref{[-1,2]_a0}, we observe from \eqref{[-1,2]_expansioncINFc} that
${ C_4 = C_2/(60 k)}$, so that it is more convenient to choose the equivalent parameter $C_2$, instead. The simplest choice is
\be
B_v \equiv C_2
\,.
\label{beta_definiceINF}
\ee
With the only remaining parameter $B_v$ (in this case it is not dimensionless), the coefficients
\eqref{[-1,2]_expansionaINFa}, \eqref{[-1,2]_expansioncINFc} simplify to
\begin{align}
& A_0 = -1\,,\quad
A_1 = 0\,,\quad
A_2 = -\frac{1}{3}\,B_v\,,\quad
A_3 = 0\,, \nonumber\\
& A_4 = -\frac{1}{5}\,\Big(\frac{1}{48k}+B_v \Big)\,B_v \,, \quad
A_5 = 0 \,, \ldots   \label{[-1,2]_expansionaINFaa} \\
& C_0 = -1 \,,\quad
C_1 = 0 \,,\quad
C_2 = B_v \,,\quad
C_3 = 0 \,,\nonumber\\
& C_4 = \frac{1}{60k}\,B_v  \,,\quad
C_5 = 0  \,,\quad
C_6 = \frac{1}{700k}\,\Big(\frac{1}{8k}+\frac{13}{3}B_v \Big)\,B_v \,, \ldots
\label{[-1,2]_expansioncINFcc}
\end{align}
yielding an explicit solution
\begin{eqnarray}
\Omega(r)     \rovno  -\frac{1}{r}
- B_v\,\bigg(\frac{1}{3\,r^3}
+\frac{1}{5\,r^5}\,\Big(\frac{1}{48k}+B_v \Big) +\ldots\bigg)\,, \label{[-1,2]_OmegaFULL}\\
\mathcal{H}(r) \rovno -r^2
+ B_v\,\bigg( 1+\frac{1}{60k\,r^2}
+\frac{1}{700k\,r^4}\,\Big(\frac{1}{8k}+\frac{13}{3}B_v \Big)
+ \ldots \bigg) \,. \label{[-1,2]_H0FULL}
\end{eqnarray}
The corresponding scalar invariants \eqref{[-1,2]_Inv_INF} now read
\be
 B_{ab}\, B^{ab} (r\to\infty)= \frac{1}{12k^2}\,B_v^2  \,,\qquad
 C_{abcd}\, C^{abcd} (r\to\infty) \sim \frac{1}{300k^2}\,\frac{B_v^2}{r^4} \to 0  \,.
\label{[-1,2]_Inv_INF_beta}
\ee
Therefore, we may  conclude that this class of  metrics ${[N,P]=[-1,2]^\infty}$ can be understood as a \emph{one-parameter Bachian generalization of flat space} \eqref{[-1,2]_solution} (that is the limit of black hole solutions without mass and horizon) \emph{with a nonzero Bach tensor whose magnitude is determined by the parameter $B_v$}, i.e., the ``massless limit'' of the previous class ${[N,P]=[-1,3]^\infty}$.

Interestingly, in the limit ${r\to\infty}$, the expressions  \eqref{to static}, \eqref{rcehf} now imply
\bea
&& {\bar r}=\Omega(r) \sim   -1/r \to 0 \,, \label{[-1,2]A}\\
&& h \sim 1\,,\qquad
   f \sim 1\,. \label{[-1,2]C}
\eea
Both the metric functions $h$ and $f$ thus remain nonzero and finite, i.e., in this limit we are not approaching a horizon nor a singularity. In fact, for ${\bar r \to 0}$ the metric \eqref{Einstein-WeylBH} \emph{becomes conformally flat}. Interestingly, the Bach invariants \eqref{[-1,2]_Inv_INF_beta} and \eqref{BInv2INFfinal} are very similar.

\subsection{Consistency check of the limit $[-1,3]^\infty\,\rightarrow\, [-1,2]^\infty$}

Let us consider a ``consistency check'' between the class of solutions $[-1,3]^\infty$,
described by \eqref{nonSchwinitcondcINF}--\eqref{nonSchwinitcond2INF},
and the class $[-1,2]^\infty$, described by 	\eqref{[-1,2]initcondc}--\eqref{[-1,2]initconda},
where the coefficients will now be denoted by hats.

The transition from $[-1,3]^\infty$ to $[-1,2]^\infty$ requires
\be
C_0\rightarrow\ 0\,,\qquad
C_i\rightarrow\ \hat C_{i-1},\ \ \ i\geq 1\,,\qquad
A_i\rightarrow\ \hat A_i,\ \ i\geq 0\,.
\ee
The relation \eqref{nonSchwinitcondaINF} for ${l=1}$, that is $3C_0A_1=-A_0(1+C_1)$, in this limit leads to
\be
C_1\rightarrow\ -1,\ \ \ \mbox{i.e.}\ \ \ \hat C_0=-1 \,,
\ee
while the relations \eqref{nonSchwinitcondcINF} for $C_{l+1}$ and \eqref{[-1,2]initcondc} for $\hat C_l$
remain the same. Moreover, the relation \eqref{nonSchwinitcondaINF} for~$A_l$,
\be
-l^2\,C_0A_{l}=\tfrac{1}{3}\,A_{l-1}
+\sum^{l}_{i=1} C_i\,A_{l-i}\,\big[\,l(l-i)+\tfrac{1}{6}i(i+1)\big]
\qquad \forall\ l\ge 1\,,
\ee
for $C_0=0$  leads to
\be
\hat A_{l-1}
=\frac{1}{l(l-1)}
\sum^{l-1}_{i=1} \hat C_i\,\hat A_{l-1-i}\,\big[(l-1)(l-i)+\tfrac{1}{6}(i-2)(i-1)\big]
\qquad \forall\ l\ge 2\,,
\ee
which is exactly \eqref{[-1,2]initconda} and thus concludes the consistency check.

Note that from the free parameters of the family  $[-1,3]^\infty$, two parameters
become determined, namely ${C_0\rightarrow0}$,	${C_1\rightarrow\hat C_0=-1}$, and one parameter ${C_2\rightarrow\hat C_1}$ becomes undetermined	since $3C_0C_2=C_1^2-1\rightarrow 0$. Therefore, four free parameters ${A_0,C_0,C_1,C_3}$ of the $[-1,3]^\infty$ family reduce to three free parameters
${\hat A_0,\hat C_1, \hat C_2}$	of the $[-1,2]^\infty$ family.

\section{Summary and relations to previous results}
\label{summary}

In this section, let us summarize all the distinct and explicit families of spherically symmetric vacuum spacetimes in QG, expressed both in powers of ${\Delta \equiv r-r_0}$ and $r^{-1}$. Moreover, we identify these families with solutions previously discussed in the literature.

In particular, in \cite{Stelle:1978,LuPerkinsPopeStelle:2015,LuPerkinsPopeStelle:2015b},
various classes of static spherically symmetric solutions to higher-derivative gravity equations were identified and denoted by the symbol $(s,t)$, using the \emph{standard spherically symmetric form} \eqref{Einstein-WeylBH}. Such a classification was based on the powers $s$ and $t$ of the \emph{leading terms} of a Laurent expansion of the two metric functions, namely\footnote{To make the identification, we have relabeled the arguments of the metric functions $A(r), B(r)$ of \cite{LuPerkinsPopeStelle:2015b} to $\bar r$.}
\bea
  f^{-1}({\bar r}) \rovno A({\bar r}) \sim {\bar r}^{\,s}\,, \label{rcef2}\\
  h({\bar r})      \rovno B({\bar r}) \sim {\bar r}^{\,t} \,,\label{rceh2}
\eea
in the domain ${\bar r \to 0}$. It was shown in \cite{Stelle:1978,LuPerkinsPopeStelle:2015b} that there are \emph{three main solution families} corresponding to the following choices  of  ${(s,t)}$:
\bea
 (s,t) \rovno (0,0)_0\,, \label{rodina1}\\
 (s,t) \rovno (1,-1)_0\,, \label{rodina2}\\
 (s,t) \rovno (2,2)_0\,, \label{rodina3}
\eea
where the subscript ``$\,_0$'' indicates the expansion around the origin ${\bar r =0}$.

In addition, the following \emph{three families} ${(w,t)}$ were identified in
\cite{LuPerkinsPopeStelle:2015b,PerkinsPhD}
using a series expansion around a \emph{finite} point ${{\bar r}\to {\bar r}_0 \ne0}$:
\bea
	(w,t)=(1,1)_{{\bar r}_0}\,,\\
	(w,t)=(0,0)_{{\bar r}_0}\,,\\
	(w,t)=(1,0)_{{\bar r}_0}\,,
\eea
where
\be
w=-s \,,
\ee
that is  ${f \sim {\bar r}^{\,w}}$ and ${h \sim {\bar r}^{\,t}}$. The subscript ``$\,_{{\bar r}_0}$'' indicates the expansion around ${\bar r_0}$.

In fact, \emph{we have recovered all these families} of solutions  in the present paper, and \emph{we have also identified some additional families}.

To find the specific mutual relations, first let us note that from the relation \eqref{to static} between the spherically symmetric radial coordinate $\bar r$ and the Kundt coordinate $r$, that is ${\bar r=\Omega(r)}$,
it follows using \eqref{rozvojomeg0} and \eqref{rozvojomegINF} that
\begin{itemize}
\item
 ${\bar r \rightarrow 0\,\,\,}$
for ${r\rightarrow r_0}$, ${n>0}$, and also for ${r\rightarrow \infty}$,  ${N<0}$\,,
\item
 ${\bar r \rightarrow \bar r_0\,}$
for ${r\rightarrow r_0}$, ${n=0}$, and also for $r\rightarrow\ \infty$, ${N=0}$\,,
\item
  ${\bar r \rightarrow \infty}$
for ${r\rightarrow r_0}$, ${n<0}$, and also for $r\rightarrow\ \infty$, ${N>0}$\,.
\end{itemize}

Now let us find a relation between the powers $(s,t)$ introduced by \eqref{rcef2} and \eqref{rceh2}, respectively, and the coefficients ${[n, p]}$ employed in this paper. They are the analogous leading powers of the two metric functions $\Omega$ and $\H$, respectively.
For ${n\not=0}$, such a relation is found using the expressions \eqref{rcehf} with ${\bar{r} = \Omega(r)}$  and \eqref{rozvojomeg0}, \eqref{rozvojcalH0} for ${r \to r_0}$.
It turns out that
\be
  s = \frac{2-p}{n}\,,\qquad
  t = 2+\frac{p}{n}\,. \label{st-np}
\ee
Analogously, using \eqref{rozvojomegINF}, \eqref{rozvojcalHINF}, we obtain the relations
\be
  s = \frac{2-P}{N}\,,\qquad
  t = 2+\frac{P}{N}\, \label{st-NP}
\ee
for the asymptotic expansion of the metric functions as  ${r\rightarrow \infty}$.
Thus, for ${n\not=0}$ and ${N\not=0}$, it  immediately follows that
\begin{itemize}
\item the family ${(s,t)=(0,0)_0}$ corresponds to ${[N,P]=[-1,2]^\infty}$\,,
\item the family ${(s,t)=(0,0)^ \infty}$ corresponds to ${[n, p]=[-1,2]}$\,,
\item the family ${(s,t)=(1,-1)_0}$ corresponds to ${[N,P]=[-1,3]^\infty}$\,,
\item the family ${(s,t)=(2,2)_0}$ corresponds to ${[n,p]=[1,0]}$\,,
\end{itemize}
where the superscript ``$\,^\infty$'' in ${(0,0)^ \infty}$ indicates the expansion as ${{\bar r}\to\infty}$.

The two admitted cases \eqref{4classes} with ${n=0}$ have to be analyzed separately (there are no cases \eqref{2classes} with ${N=0}$). In the generic case when ${a_1\not=0}$, using \eqref{rcehf}, \eqref{rozvojomeg0}, \eqref{rozvojcalH0}, we obtain that
\be
 w=p\,, \qquad t=p\,.
\ee
Therefore, for ${n=0}$ and ${a_1\not=0}$ we conclude that
\begin{itemize}
	\item the family ${(w,t)=(0,0)_{\bar r_0}}$ corresponds to ${[n, p]=[0,0]}$\,,
	\item the family ${(w,t)=(1,1)_{\bar r_0}}$ corresponds to ${[n,p]=[0,1]}$\,,
\end{itemize}
\emph{completing the identification of all our main six classes of solutions}. Note that
for ${n=0}$, ${a_1\not=0}$ the relation between $\Delta$ and $\bar \Delta$ is ${\bar\Delta \equiv \bar r -\bar r_0 \sim a_1\Delta}$. Therefore, a series expansion with \emph{integer steps in} $\Delta$ corresponds to a series expansion with \emph{integer steps in} $\bar\Delta$ in the physical radial coordinate~$\bar r$.

All four possible generic families compatible with the field equations as ${r \to r_0}$
and the series expansion \eqref{rozvojomeg0}--\eqref{DElta}
are summarized in Table~\ref{tbl:01}, while
the two cases compatible with the field equations as ${r \to \infty}$
and \eqref{rozvojomegINF}, \eqref{rozvojcalHINF}, are summarized in Table~\ref{tbl:02}. We also indicate their physical interpretation and the corresponding Section, in which these solutions are described and studied.

\begin{table}[h!]
\begin{center}
\begin{tabular}{|c||c||c|c|c|c|c|}
\hline
Class $[n,p]$ &   Family $(s,t)$ & Interpretation & Section\\[0.5mm]
       \hline\hline
$[-1,2]$ &   $(0,0)^\infty$ & Schwarzschild black hole          & \ref{Schw_[n,p]=[-1,2]} \\[1mm]
$[0,1]$  &   $(-1,1)_{\bar r_0}$ & Schwarzschild--Bach black holes (near the horizon)  & \ref{SchwaBach_[n,p]=[0,1]} \\[1mm]
$[0,0]$  &   $(0,0)_{\bar r_0}$ & generic solution, including the Schwa--Bach black holes & \ref{SchwaBach_[n,p]=[0,0]}\\[1mm]
$[1,0]$  &   $(2,2)_0$ & Bachian singularity (near the singularity)       &   \ref{SchwaBach_[n,p]=[1,0]}

\\[1mm]
\hline
\end{tabular} \\[2mm]
\caption{All possible generic types of solutions to Quadratic Gravity and
the Einstein--Weyl theory that can be written as the power series \eqref{rozvojomeg0}--\eqref{rozvojcalH0}
expanded around any constant value~$r_0$.
}
\label{tbl:01}
\end{center}
\end{table}

\vspace{5mm}

%\noindent

\begin{table}[h]
\begin{center}
\begin{tabular}{|c||c||c|c|c|c|c|}
\hline
Class $[N,P]^\infty$ &   Family $(s,t)$ & Interpretation & Section\\[0.5mm]
       \hline\hline
$[-1,3]^\infty$  &   $(1,-1)_0$ & Schwarzschild--Bach black holes (near the singularity) & \ref{Schw_[N,P]=[-1,3]} \\[1mm]
$[-1,2]^\infty$  &   $(0,0)_0$  & Bachian vacuum  (near the origin)   & \ref{Schw_[N,P]=[-1,2]}\\[1mm]
\hline
\end{tabular} \\[2mm]
\caption{All possible generic types of solutions to Quadratic Gravity and
the Einstein--Weyl theory that can be written as the power series
\eqref{rozvojomegINF}, \eqref{rozvojcalHINF}
expanded as ${r\to\infty}$.
}
\label{tbl:02}
\end{center}
\end{table}

\subsection{Special subclasses with $n=0$}
\label{relation_to_previous_results 2}

In addition to the above six main classes of solutions, in the case given by ${n=0}$ we have identified some other special subclasses, including a new one. These are \emph{not} given as integer steps  in $\bar r$
or ${\bar\Delta}$, so that these are additional classes from the point of view of expansions in powers of ${\bar r -\bar r_0}$ in the physical radial coordinate. In our Kundt coordinate $r$, they just naturally appear as special cases of the solutions with ${n=0}$, namely when ${a_1=0\not=a_2}$ and ${a_1=0=a_2}$.

When ${a_1=0\not=a_2}$, the relation is ${\bar\Delta\sim a_2\,\Delta^2}$, and thus a series
expansion with integer steps in $\Delta$ leads to \emph{(half integer) steps} $\bar \Delta^{1/2}$.  Using \eqref{rcehf}, in such a case we obtain
\be
w=\frac{p}{2}+1\,,\qquad t=\frac{p}{2}\,.
\ee
For  ${a_1=0}$ and ${a_2\not=0}$, we thus conclude that
\begin{itemize}
	\item the family $(w,t)=(\tfrac{3}{2},\tfrac{1}{2})_{\bar r_0,1/2}$
 corresponds to ${[n, p]=[0,1]_{a_1=0}}$,
	\item the family ${(w,t)=(1,0)_{\bar r_0,1/2}}$ corresponds to ${[n,p]=[0,0]_{a_1=0}}$.
\end{itemize}

Analogously, when ${a_1=0=a_2}$ and ${a_3\not=0}$, the relation is ${{\bar \Delta}\sim a_3\,\Delta^3}$, and
thus integer steps in $\Delta$ corresponds to steps in $\bar \Delta^{1/3}$.  The relations are now
\be
w=\frac{p+4}{3}\,,\qquad t=\frac{p}{3}\,.
\ee
Thus for ${a_1=0=a_2}$ and ${a_3\not=0}$, we conclude that
\begin{itemize}
	\item the family $(w,t)=(\tfrac{4}{3},0)_{\bar r_0,1/3}$
	%	{\sqrt{(\bar r-\bar r_0)}}$
	corresponds to ${[n, p]=[0,0]_{a_1=0=a_2}}$.
\end{itemize}

Concerning the geometrical and physical interpretation of these special solutions, it can be generally said that the classes with ${n=0}$ contain (among other solutions) black holes and wormholes.
In particular, the class ${[n=0,p=1]}$  represents a \emph{black hole spacetime} since
 it admits a Killing horizon at  ${r_h=r_0}$, see \eqref{horizon}.
As pointed out in \cite{LuPerkinsPopeStelle:2015b}, a \emph{wormhole spacetime} is
characterized by admitting a finite value of ${{\bar r}_0}$ where
${f=0}$ while ${h\not=0}$. Therefore, for a series expansion around this point, necessarily  ${n=0=p}$
(since $\H \not= 0$),  and  ${a_1=0}$ (since $\Omega'=0$).
Thus wormholes may appear only in the class  ${[0,0]_{a_1=0}}$.

The family of solutions ${(\tfrac{3}{2},\tfrac{1}{2})_{\bar r_0,1/2}}$
was identified in \cite{LuPerkinsPopeStelle:2015b} and interpreted in \cite{PerkinsPhD} as
an ``unusual'' type of a horizon. However, it was stated therein that it is a solution to QG only for ${\beta\not =0}$, which implies ${R\not=0}$. Thus it seems that this class does not coincide with our class ${[0,1]_{a_1=0}}$  since, for all our classes, ${R=0}$ by assumption.

Our family ${[0,0]_{a_1=0}}$ corresponds to the family ${(1,0)_{\bar r_0,1/2}}$ of
 \cite{LuPerkinsPopeStelle:2015b,PerkinsPhD}, while our family $[0,0]_{a_1=0=c_1=c_3}$,
where \emph{only even powers} in $\Delta$ are considered (indicated by the subscript ``$\,_E$''),
corresponds to the family $(1,0)_{\bar r_0,E}$ of \cite{LuPerkinsPopeStelle:2015b,PerkinsPhD}.
Both these families describe  wormholes with two different (half-integer wormhole)
and two same patches (integer wormhole), respectively, see \cite{PerkinsPhD}. Note that the
Bach invariant \eqref{invB} for wormholes in the ${[0,0]_{a_1=0}}$ class
 is always nonvanishing.

To our knowledge,  the specific family ${[0,0]_{a_1=0=a_2}}$ \emph{has not yet been considered},
and it corresponds to a new family ${(\tfrac{4}{3},0)_{\bar r_0,1/3}}$ in the notation
of \cite{LuPerkinsPopeStelle:2015b}.

It also seems that the \emph{generic solution} $[0,0]$, with the highest number of
free parameters, can be connected  to all other solutions, and it represents an expansion around a generic point in these spacetimes.

In Table \ref{tab:3}, we summarize all the classes and subclasses found and identified both in the physical and Kundt coordinates, grouped according to the regions in which the expansions are taken in the usual radial coordinate~$\bar r$.

\begin{table}[h!]
	\begin{center}
		\begin{tabular}{|l|l|l|c|l|}
			\hline
			 Family &  $[n,p]$ or $[N,P]^\infty\!\!$ & Parameters  & Free param. & Interpretation
			\\
            \hline %\[-3mm]
            \hline
		$(s,t)$ & \multicolumn{4}{|c|}{ ${\bar r\rightarrow 0}$ }\\
            \hline %\[-3mm]
			$(2,2)_0$& $[1,0]$&
			$a_0,c_0,c_1,c_2,r_0$ & $5\rightarrow 3$& Bachian singularity (nS)\\
			$(2,2)_{0,E}$& $[1,0]_{c_1=0=c_3}$ &
             $a_0,c_0,r_0$ & $3\rightarrow 1$ & Bachian singularity (nS)\\
			$(1,-1)_0$&  $[-1,3]^\infty$&
			$A_0,C_0,C_1,C_3$& $4\rightarrow 2$& Schwa--Bach black holes (S)\\
			$(0,0)_0$	& $[-1,2]^\infty$&
            $A_0,C_1,C_2$& $3\rightarrow 1$& Bachian vacuum (nS)\\
			\hline %\\[-3mm]
		$(w,t)$ & \multicolumn{4}{|c|}{  ${\bar r\rightarrow \bar r_0}$    }\\
			\hline %\\[-3mm]
			$(1,1)_{\bar r_0}$&$[0,1]$ &
			$a_0,c_0,c_1,r_0=r_h$ & $4\rightarrow 2$& Schwa--Bach black holes (S)\\
			$(3/2,1/2)_{\bar r_0,1/2}\!\!$ & $[0,1]_{a_1=0}$&
			$a_0,c_0,r_0$& $3\rightarrow 1$& ``unusual'' horizon (nS)\\
			$(0,0)_{\bar r_0}$&$[0,0]$&
			$a_0,a_1,c_0,c_1,c_2,r_0\!$& $6\rightarrow 4$ & generic solution (S)\\
			$(1,0)_{\bar r_0,1/2}$&$[0,0]_{a_1=0}$&
			$a_0,c_0,c_1,c_2,r_0$& $5\rightarrow 3$& half-integer wormhole (nS) \\
			$(1,0)_{\bar r_0,E}$&$[0,0]_{a_1=0=c_1=c_3}$&
			$a_0,c_0,r_0$& $3\rightarrow 1$& symmetric wormhole (nS) \\
			$(4/3,0)_{\bar r_0,1/3}$&  $[0,0]_{a_1=0=a_2}$&
			$a_0,c_0,c_1,r_0$& $4\rightarrow 2$&  not known (nS) --- new \\ \hline%\\[-3mm]
		$(s,t)$ & \multicolumn{4}{|c|}{  ${\bar r\rightarrow \infty}$    }\\
			\hline %\\[-3mm]			
			$(0,0)^\infty$& $[-1,2]$&
            $a_0,c_1,r_0$ & $3\rightarrow 1$& Schwarzschild black hole (S) \\ \hline
		\end{tabular}  \\[2mm]
\caption{All solutions, sorted according to the physical regions in which the expansions are taken.		
The subscripts ``$\,_0$'', ``$\,_{{\bar r}_0}$'' and the superscript ``$\,^\infty$'' denote solutions $(s,t)$ or $(w,t)$ near ${\bar r=0}$,  ${\bar r={\bar r_0}}$, and  ${\bar r\rightarrow\infty}$, respectively. The subscript ``$\,_E$'' indicates that only even powers are present in the expansion, while ``$\,_{1/2}$'' and ``$\,_{1/3}$''indicate that fractional powers are present. Specific number of free parameters is given before and after removing two parameters by the gauge freedom \eqref{scalingfreedom} in the Kundt coordinates. In physical coordinates, only one parameter can be removed by rescaling \eqref{scaling-t}. The symbols ``(S)'' or ``(nS)'' indicate that a class of solutions contains or does \emph{not} contain the Schwarzschild black hole, respectively. }
	\label{tab:3}	
	\end{center}
\end{table}

\section{Discussion and analysis of the Schwarzschild--Bach black holes}
\label{discussion-and-figures}

In this section, we discuss the behavior of the series expressing the Schwarzschild--Bach black hole solutions \eqref{Omega_[0,1]}, \eqref{H_[0,1]}. For our analysis, we choose the same values of the parameters as in our previous paper \cite{PodolskySvarcPravdaPravdova:2018}, namely
 ${r_h=-1}$,  ${k = 0.5}$, ${b=0.3633018769168}$. Such a very special value of $b$ is ``close'' to the asymptotically flat case.\footnote{We obtained this value from the Mathematica code kindly provided by H. L\"u, cf. also \cite{PerkinsPhD} for a very close value of $b$.}

\begin{figure}[h!]
\begin{center}
	\includegraphics[height=62mm]{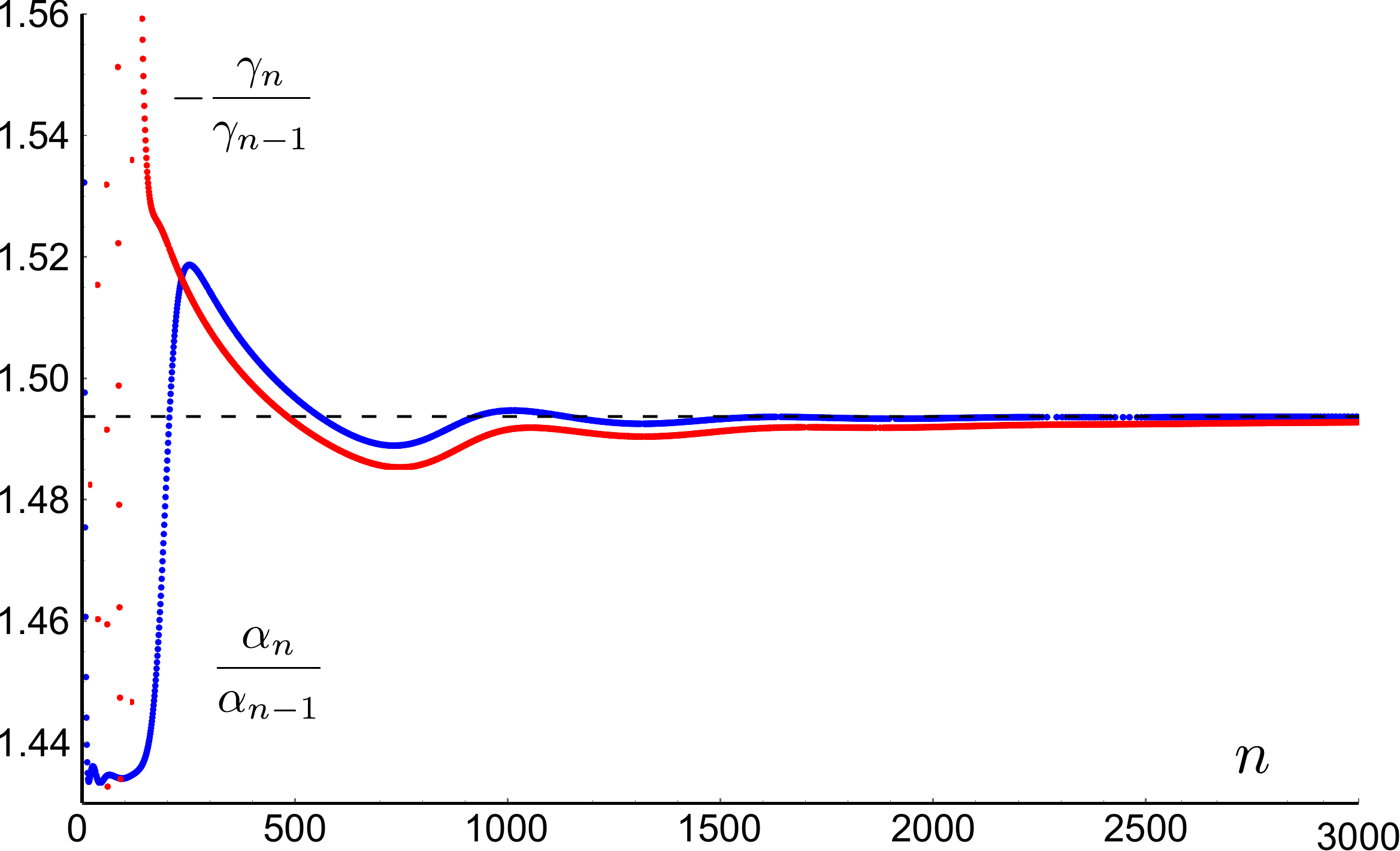}
\end{center}
	\caption{The Schwarzschild--Bach solution ${[0,1]}$ given by \eqref{Omega_[0,1]}, \eqref{H_[0,1]}. The ratios $\frac{\alpha_n}{\alpha_{n-1}}$ (blue) and 		${-\frac{\gamma_n}{\gamma_{n-1}}}$ (red)  for the first 3000 coefficients $\alpha_i$ and  $\gamma_i$ given by the recurrent formula \eqref{alphasIIbgeneral} are ploted.   }
	\label{Figasymp}
\end{figure}

The key observation for estimating the radius of convergence can be made from Figure~\ref{Figasymp}. Interestingly, the ratios of subsequent terms ${\frac{\alpha_n}{\alpha_{n-1}}}$ and ${-\frac{\gamma_n}{\gamma_{n-1}}}$ given by the recurrent relations \eqref{alphasIIbgeneral} are \emph{approaching a constant asymptotically}.
This  suggests that both series given by $\alpha_n$ and $\gamma_n$ behave as \emph{geometric series for large}~$n$, with the ratio $q$ being apparently \emph{equal for both the series}. Therefore,
the series for $\Omega$ and $\cal H$, given by \eqref{Omega_[0,1]}, \eqref{H_[0,1]}, should be convergent for ${-1-\frac{1}{q}<r<-1+\frac{1}{q}}$, where  ${q \approx 1.494}$, that is in the interval ${r\in(-1.67,-0.33)}$.

\begin{figure}[h!]
\begin{center}
	\includegraphics[height=50mm]{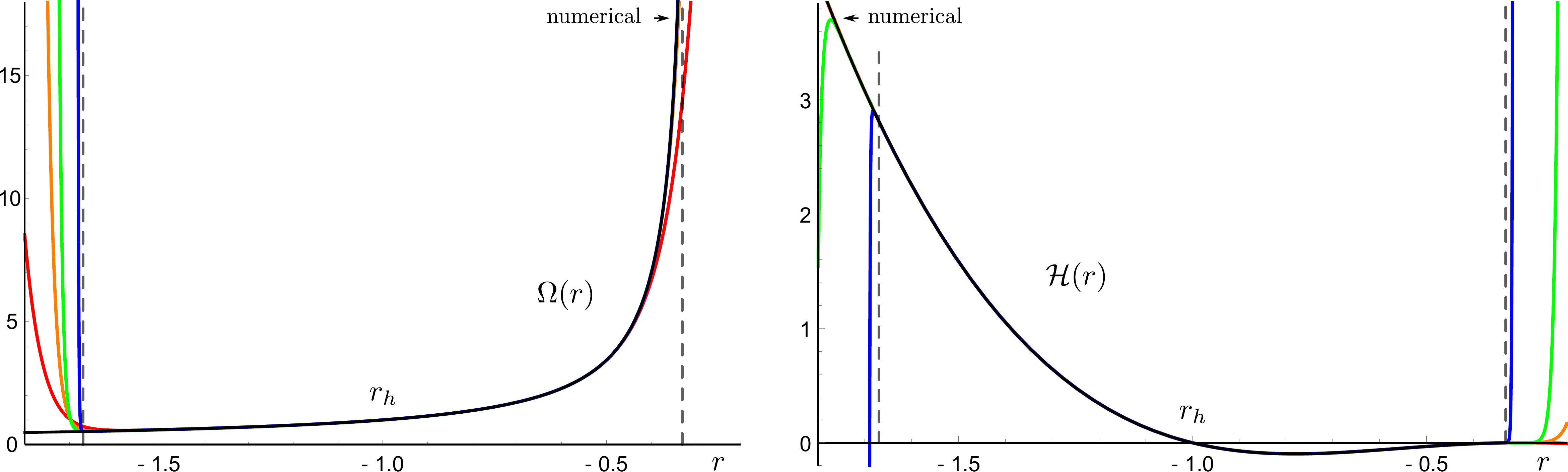}
%	\hspace{2mm}	\includegraphics[height=50mm]{fig2b.pdf}
\end{center}
	\caption{The metric functions $\Omega(r)$ (left) and $\H(r)$ (right) for the Schwarzschild--Bach solution $[0,1]$. The first 20 (red), 50 (orange), 100 (green), and 500 (blue) terms of the series \eqref{Omega_[0,1]}, \eqref{H_[0,1]} for $\Omega$ and $\H$ are also compared with a numerical solution (black). Boundaries of the domain of convergence are denoted by vertical dashed lines. Within this radius of convergence, all these functions overlap with the numerical solution, except the lowest shown 20th order of $\Omega$ near the top right corner on the left graph. }
		\label{fig:OmegaH}
\end{figure}

Figure~\ref{fig:OmegaH} illustrates the convergence of the metric functions $\Omega(r)$ and $\H(r)$  in the Kundt coordinate~$r$. In the domain of convergence, denoted by vertical dashed lines, the solution fully agrees with the numerical solution of the field equations.

\begin{figure}[h!]
\begin{center}
	\includegraphics[height=48mm]{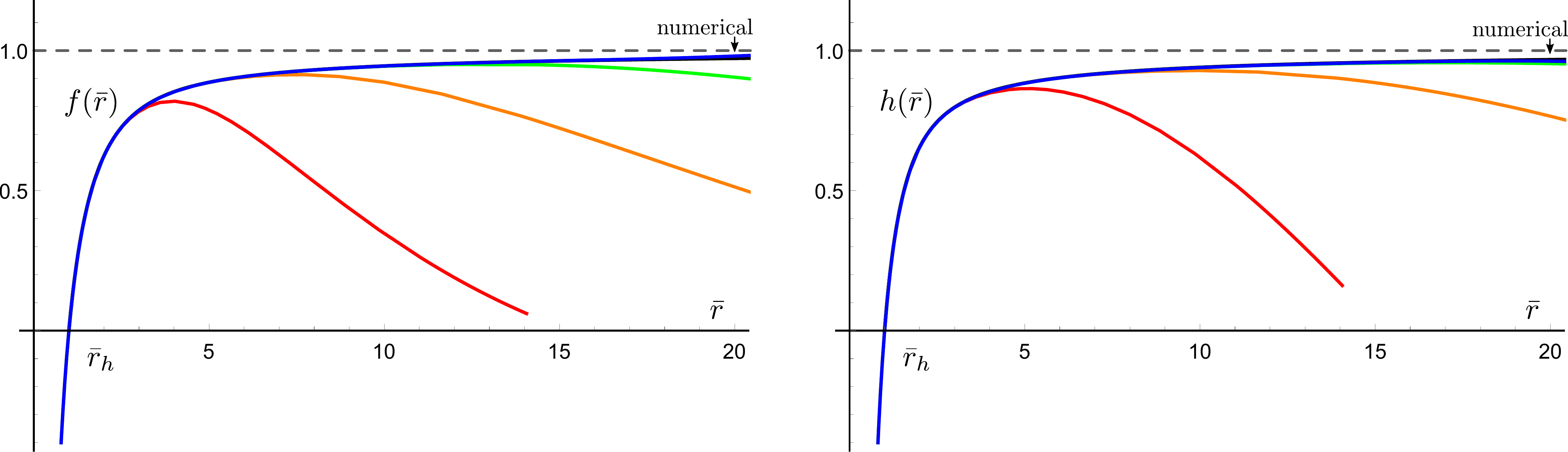}
%	\hspace{2mm}	\includegraphics[height=48mm]{fig3b.pdf}
\end{center}
	\caption{The metric functions $f(\bar r)$ (left) and $h(\bar r)$ (right) for the Schwarzschild--Bach solution [0,1] in the standard coordinates. The first 20 (red), 50 (orange), 100 (green), and 300 (blue) terms of the series are plotted. A numerical solution (black) overlaps with the blue curve, even far above the horizon located at ${{\bar r}_h=1}$ (here up to ${\bar r = 20\, \bar r_h}$).}
	\label{fig:fh}
\end{figure}

For comparison, Figure~\ref{fig:fh} illustrates the convergence of the corresponding metric functions $f(\bar r)$ and $h(\bar r)$ in the standard spherically symmetric coordinates. The solution quickly converges, and approaches a numerical solution even at a large distance from the horizon located at ${{\bar r}_h =1}$.

From the value of ${\Omega(r)\equiv \bar r}$ at the \emph{lower} boundary of the domain of convergence shown in Figure~\ref{fig:OmegaH}, we can easily read off its value ${\bar r \approx  0.53}$ in the usual radial coordinate. In contrast, the value of the coordinate $\bar r$  given by $\Omega(r)$ at the \emph{upper} boundary remains unclear since it depends on the precise value of the series \eqref{Omega_[0,1]} at the upper boundary of the domain of convergence. In fact, we cannot even say with certainty that the radius of convergence in the standard spherical coordinate $\bar r $ is finite --- it may well extend up to ${\bar r \to \infty}$.

\begin{figure}[h!]
\begin{center}
	\includegraphics[height=50mm]{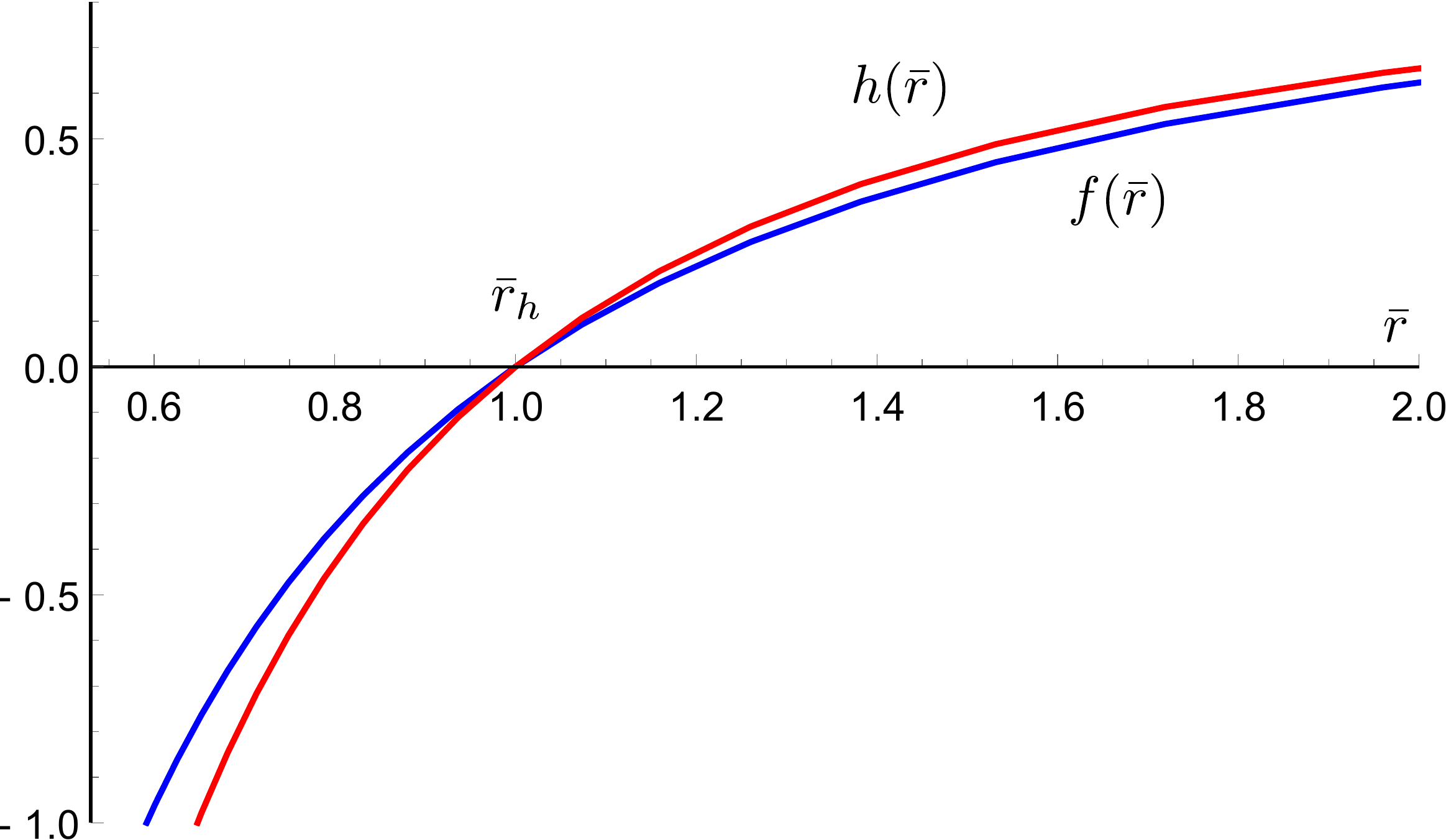}
\end{center}
	\caption{The metric functions $f(\bar r)$ (blue) and $h(\bar r)$ (red) in the near-horizon region for the Schwarzschild--Bach solution [0,1]. These two functions are clearly distinct. They both vanish at the horizon, located here at ${{\bar r}_h=1}$}
	\label{fig:fhnearhorizon}
\end{figure}

Finally, it is illustrative to show explicitly that, in contrast to the Schwarzschild solution,
the metric functions $f(\bar r)$ and $h(\bar r)$ for the Schwarzschild--Bach black holes are \emph{not equal}. This is clearly seen from their plots in Figure~\ref{fig:fhnearhorizon}.

\newpage
\

\newpage

\begin{figure}[h!]
\begin{center}
	\includegraphics[height=55mm]{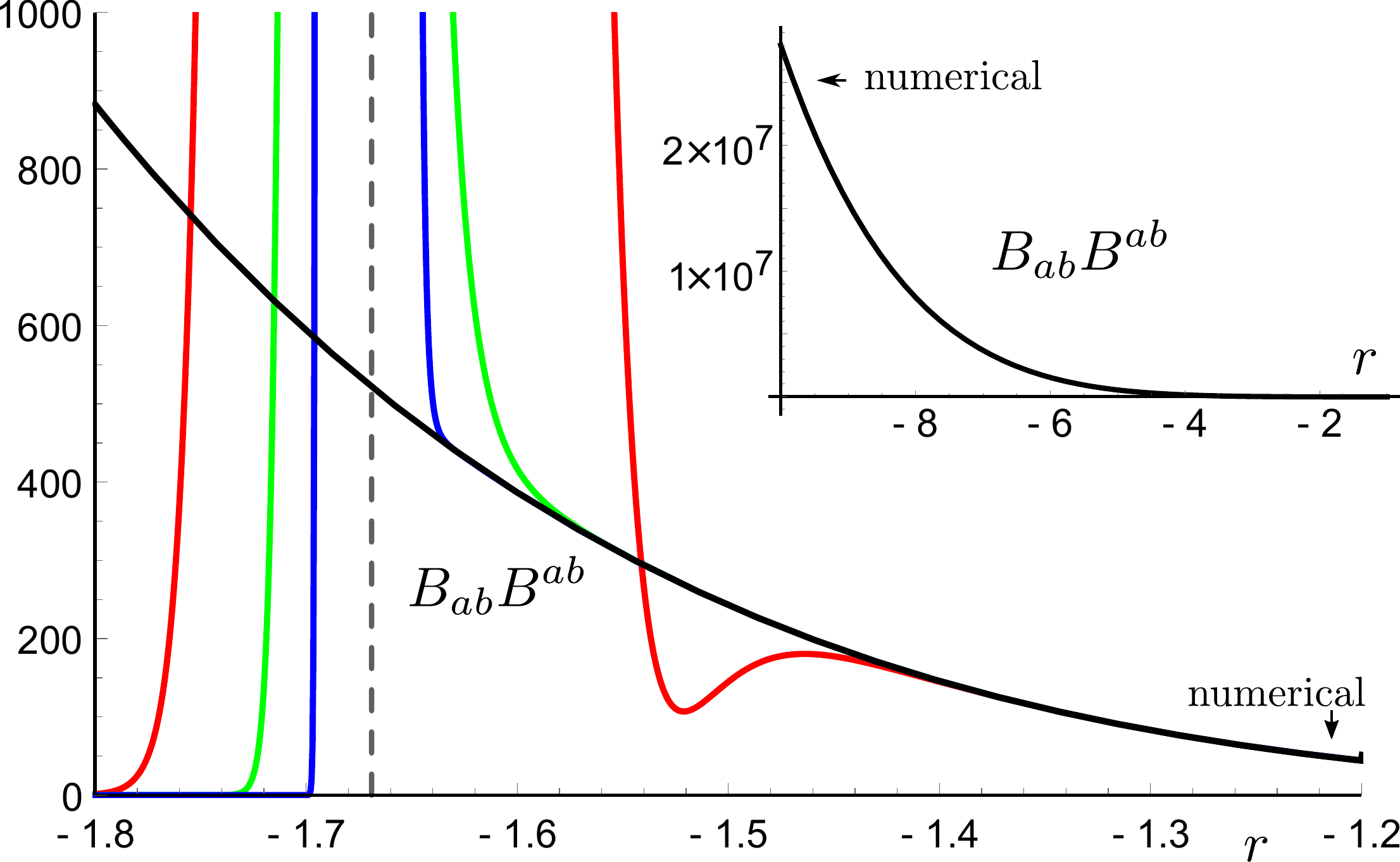}
\end{center}
	\caption{The Bach invariant \eqref{invB} inside the horizon of the Schwarzschild--Bach black holes ${[0,1]}$ calculated from  first 20 (red), 50 (green), and 300 (blue) terms, compared with the numerical solution (black). The lower boundary of the domain of convergence is indicated by the vertical dashed line. The horizon is located at ${r_h=-1}$. The insert in the upper right corner shows the numerical value to much lower value of the coordinate $r$, indicating a possible divergence as ${r \rightarrow -\infty}$, that is as ${\bar r \to 0}$.}
	\label{fig:bach}
\end{figure}

There are \emph{three classes} of solutions \emph{containing the Schwarzschild black hole as a special case}, namely the ${[0,0]}$ class with \emph{four} free parameters and the classes ${[0,1]}$ and ${[-1,3]^{\infty}}$, both with \emph{two} free parameters, see Table~\ref{tab:3}. (The class ${[-1,2]}$ contains \emph{only} the Schwarzschild solution.) The solution ${[0,0]}$ describes a generic point of a static, spherically symmetric spacetime in QG, including also black-hole and wormhole solutions. A natural question is whether the solutions ${[0,1]}$ and ${[-1,3]^{\infty}}$ describe \emph{the same black hole at two different regions} (near the horizon and near the singularity, respectively). We have not arrived at a definite answer yet. Nevertheless the Bach invariant \eqref{invB} for the class ${[-1,3]^{\infty}}$ approaches a \emph{finite constant} as ${|r| \rightarrow \infty}$ corresponding to ${\bar r \to 0}$, see expression \eqref{BInv2INFfinal}, while analytical and numerical results describing the behavior of the Bach invariant of the ${[0,1]}$ class of solutions as the value of $r$ decreases below the horizon seems to suggest that in this case the Bach invariant is \emph{unbounded}, see Figure \ref{fig:bach}. If this is indeed the case, then the classes ${[0,1]}$ and ${[-1,3]^{\infty}}$ must describe \emph{distinct} generalizations of the Schwarzschild black hole admitting a nontrivial Bach tensor.

\newpage

\section{Main physical properties of the Schwarzschild--Bach black holes}
\label{physics}

\subsection{Specific observable effects on test particles caused by the Bach tensor}
\label{geodeviation}

In this section we demonstrate that the two  parts $\B_1$, $\B_2$ of the Bach tensor \eqref{B1}, \eqref{B2}, entering the invariant \eqref{invB}, that distinguish the Schwa--Bach and the Schwarz\-schild black holes, can be explicitly observed via a \emph{specific influence on particles}. It is well known that a \emph{relative motion} of freely falling test particles (observers) directly encodes specific components of the spacetime curvature, such as the tidal deformation in the vicinity of a black hole, or a transverse effect of gravitational waves measurable by a laser interferometer detector. This is described by the \emph{equation of geodesic deviation}, see \cite{BicakPodolsky:1999,PodolskySvarc:2012} for a recent review with historical remarks and description of the formalism that we are going to employ here.

\subsubsection{Interpreting solutions to Quadratic Gravity using geodesic deviation}

To obtain physically measurable information about the relative  motion, we have to choose an \emph{orthonormal frame} ${\{\bolde_{(0)}, \bolde_{(1)}, \bolde_{(2)}, \bolde_{(3)}}\}$ such that ${\bolde_{(a)}\cdot\bolde_{(b)}=\eta_{ab}}$, where the time-like vector ${\bolde_{(0)}=\boldu}$ is observer's $4$-velocity. Projecting the equation of geodesic deviation onto this frame we obtain
\be
\ddot Z^{(\rm{i})}= R^{(\rm{i})}_{\quad(0)(0)(\rm{j})}\,Z^{(\rm{j})} \,,\qquad
\rm{i},\,\rm{j}=1,\,2,\,3\,, \label{InvGeoDev}
\ee
where
\begin{equation}
\label{PhysAccel}
\ddot Z^{(\rm{i})} \equiv e^{(\rm{i})}_a\,\frac{\Dif^2 Z^a}{\dd\, \tau^2}
  =e^{(\rm{i})}_a\,{Z^a}_{;cd}\, u^c u^d  \,, \qquad \hbox{and} \qquad
  {R_{(\rm{i})(0)(0)(\rm{j})}\equiv R_{abcd} \,e^a_{(\rm{i})}u^b u^c e^d_{(\rm{j})}} \,.
\end{equation}
Spacetime curvature, characterized by the Riemann tensor, can then be decomposed into the
traceless Weyl tensor, the Ricci tensor, and the scalar curvature $R$.
Its projection (\ref{PhysAccel}) gives
\begin{align}\label{DecompFrame}
R_{(\rm{i})(0)(0)(\rm{j})}=C_{(\rm{i})(0)(0)(\rm{j})}+\tfrac{1}{2}\big(R_{(\rm{i})(\rm{j})}
-\delta_{\rm{i}\rm{j}}\,R_{(0)(0)}\big)-\tfrac{1}{6}\,R\,\delta_{\rm{i}\rm{j}} \,.
\end{align}
Moreover, the vacuum field equations \eqref{fieldeqsEWmod} of Quadratic Gravity (including the Einstein--Weyl theory), ${R_{ab}=4k\, B_{ab}}$ implying ${R=0}$, can be employed.
Substituting these relations into (\ref{DecompFrame}), we finally obtain the \emph{invariant form of the equation of geodesic deviation} (\ref{InvGeoDev}) as
\begin{align}
\ddot{Z}^{(\rm{i})}=  C_{(\rm{i})(0)(0)(\rm{j})}\,Z^{(\rm{j})}
+2k\big(B_{(\rm{i})(\rm{j})}\,Z^{(\rm{j})}-B_{(0)(0)}\,Z^{(\rm{i})}\big)\,. \label{InvGeoDevExpl}
\end{align}
Of course, ${C_{(\rm{i})(0)(0)(\rm{j})}=C^{(\rm{i})}_{\quad(0)(0)(\rm{j})}}$ and ${B_{(\rm{i})(\rm{j})}=B^{(\rm{i})}_{\quad(\rm{j})}}$ since the spatial part of the frame is Cartesian.
The Weyl tensor projections ${C_{(\rm{i})(0)(0)(\rm{j})}}$ can be further decomposed and expressed in terms of the Newman--Penrose scalars $\Psi_A$ with respect to the (real) \emph{null frame}  ${\{\boldk, \boldl, \boldm_{i} \}}$ which is defined by
\begin{align}
 \boldk=\ssqrt(\boldu+\bolde_{(1)})\,, \qquad \boldl=\ssqrt(\boldu-\bolde_{(1)})\,,
 \qquad \boldm_{i}=\bolde_{(i)} \quad \hbox{for} \quad i=2,\,3 \,. \label{NullFrame}
\end{align}
Thus, $\boldk$ and $\boldl$ are future oriented null vectors, and $\boldm_{i}$ are two spatial vectors orthogonal to them, normalized as
${\boldk\cdot\boldl=-1}$ and ${\boldm_{i}\cdot\boldm_{j}=\delta_{ij}}$. Such a generic decomposition was found in~\cite{BicakPodolsky:1999,PodolskySvarc:2012}.

Using these results, we obtain the corresponding \emph{general form of the equation of geodesic deviation} (\ref{InvGeoDevExpl}) \emph{in Quadratic Gravity/the Einstein--Weyl theory}:
\begin{align}
\ddot{Z}^{(1)} = & \quad \Psi_{2S}\,Z^{(1)}+ \tfrac{1}{\sqrt{2}}(\Psi_{1T^j}-\Psi_{3T^j})\,Z^{(j)} \nonumber\\
& \qquad +2k\,\big[(B_{(1)(1)}-B_{(0)(0)})\,Z^{(1)}+B_{(1)(j)} \,Z^{(j)}\,\big]\,,\label{InvGeoDevFinal1}\\
\ddot{Z}^{(i)} = & - \tfrac{1}{2}\Psi_{2S}\,Z^{(i)} + \tfrac{1}{\sqrt{2}}(\Psi_{1T^i}-\Psi_{3T^i})\,Z^{(1)} -\tfrac{1}{2}(\Psi_{0^{ij}}+\Psi_{4^{ij}})\,Z^{(j)} \nonumber\\
& \qquad +2k\,\big[\,B_{(i)(1)} \,Z^{(1)}+B_{(i)(j)} \,Z^{(j)}-B_{(0)(0)}\,Z^{(i)}\,\big]\,, \label{InvGeoDevFinal2}
\end{align}
where we have used the relation ${\Psi_{2T^{(ij)}}=\tfrac{1}{2}\Psi_{2S}\,\delta_{ij}}$ valid in ${D=4}$, see \cite{PodolskySvarc:2012}.
This system of equations admits a \emph{clear physical interpretation}: The Newtonian component $\Psi_{2S}$ of the gravitational field causes classical tidal deformations, $\Psi_{3T^i}, \Psi_{1T^i}$ are responsible for longitudinal motions, while $\Psi_{4^{ij}}, \Psi_{0^{ij}}$ represent the transverse effects of gravitational waves (propagating in the directions $\bolde_{(1)}, -\bolde_{(1)}$, respectively). The additional specific effects caused by the nonvanishing Bach tensor are encoded in the frame components $B_{(a)(b)}$.

\subsubsection{Geodesic deviation in the Schwarzschild--Bach black hole spacetimes}

Let us concentrate on the spherically symmetric black hole metric in the form \eqref{BHmetric}, or
\eqref{confrelation} with \eqref{Kundt seed xy}. In particular, we introduce the ``interpretation'' orthonormal frame associated with a \emph{radially falling observer}, i.e., assuming ${\dot{x}=0=\dot{y}}$. Such a frame reads
\begin{align}
& \bolde_{(0)}\equiv \boldu= \dot{r}\,\partial_r +\dot{u}\,\partial_u \,, \nonumber \\
& \bolde_{(1)}= \tfrac{1}{2}\big[( {\Omega^2\dot{u}} )^{-1}-{\cal H}\dot{u}\big]\partial_r -\dot{u}\,\partial_u \,, \nonumber \\
& \bolde_{(i)}= \Omega^{-1}\big[1+\tfrac{1}{4}(x^2+y^2)\big]\partial_i \,, \label{OrtFrame}
\end{align}
where the normalisation of observer's four-velocity ${\boldu\cdot\boldu=-1}$ implies ${\dot{r}=\tfrac{1}{2}\big[({\Omega^2\dot{u}})^{-1}+{\cal H}\dot{u}\big]}$. Using (\ref{NullFrame}), the associated null interpretation frame thus takes the form
\begin{equation}
 \boldk = \frac{1}{\sqrt{2}\,\dot{u}\,\Omega^2}\,\partial_r \,, \qquad
 \boldl = \frac{\dot{u}\,{\cal H}}{\sqrt{2}}\,\mathbf{\partial}_r+\sqrt{2}\dot{u}\,\partial_u\,, \qquad
 \boldm_i = \Omega^{-1}\big[1+\tfrac{1}{4}(x^2+y^2)\big]\partial_i \,. \label{NullIntFrame}
\end{equation}
A direct calculation shows that \emph{the only nonvanishing Weyl tensor component} with respect to \eqref{NullIntFrame} is
\begin{equation}
\Psi_{2S} \equiv  C_{abcd}\; k^a\, l^b\, l^c\, k^d
=\tfrac{1}{6}\,\Omega^{-2}({\cal H}''+2)\,. \label{Psi2Int}
\end{equation}
This is consistent with the fact that the spherically symmetric black hole metric \eqref{BHmetric} is of algebraic type D.
The explicit Bach tensor projections with respect to the orthonormal frame (\ref{OrtFrame}) are
\begin{align}
B_{(0)(0)}&= \frac{1}{24\,\Omega^6\dot{u}^2}\Big[-(1-\Omega^2{\cal H}\dot{u}^2)^2\,{\cal H}''''+2\Omega^2\dot{u}^2({\cal H}'{\cal H}'''-\tfrac{1}{2}{{\cal H}''}^2+2)\Big]\,, \\
B_{(1)(1)}&= \frac{1}{24\,\Omega^6\dot{u}^2}\Big[-(1+\Omega^2{\cal H}\dot{u}^2)^2\,{\cal H}''''-2\Omega^2\dot{u}^2({\cal H}'{\cal H}'''-\tfrac{1}{2}{{\cal H}''}^2+2)\Big]\,, \\
B_{(0)(1)}&= -\frac{1}{24\,\Omega^6\dot{u}^2}\,(1-\Omega^4{\cal H}^2\dot{u}^4)\,{\cal H}'''' \,,
\qquad B_{(0)(i)}= 0\,, \\
B_{(i)(j)}&= \frac{\delta_{ij}}{12\,\Omega^4}({\cal H}{\cal H}''''+{\cal H}'{\cal H}'''-\tfrac{1}{2}{{\cal H}''}^2+2)\,,
\qquad B_{(1)(i)}= 0 \,.
\end{align}
Therefore, the equation of geodesics deviation \eqref{InvGeoDevFinal1}, \eqref{InvGeoDevFinal2} explicitly becomes
\begin{align}
\ddot{Z}^{(1)} = & \hspace{6mm} \frac{1}{6}\, \Omega^{-2}\big({\cal H}''+2\big)\,Z^{(1)}\,
-\,\frac{1}{3}\,k\,\Omega^{-4}\big({\cal H}{\cal H}''''+{\cal H}'{\cal H}'''-\tfrac{1}{2}{{\cal H}''}^2+2\big)Z^{(1)} \,, \label{InvGeoDevBH1}\\
\ddot{Z}^{(i)} = & - \frac{1}{12}\,\Omega^{-2}\big({\cal H}''+2\big)\,Z^{(i)}
+\frac{1}{12}\,k\,\Omega^{-4}\big((\Omega^2\H\dot{u}^2)^{-1}+\Omega^2{\cal H}\dot{u}^2\big){\cal H}{\cal H}''''\,Z^{(i)} \,. \label{InvGeoDevBHi}
\end{align}
We conclude that there is a classical \emph{tidal deformation} caused by the \emph{Weyl curvature} \eqref{Psi2Int} proportional to ${\Omega^{-2}({\cal H}''+2)}$, i.e., the square root of the invariant \eqref{invC}. Moreover, in
Quadratic Gravity (with ${k\not=0}$) there are \emph{two additional effects} caused by the
presence of a nonvanishing \emph{Bach tensor}. The first can be observed in the longitudinal component
of the acceleration \eqref{InvGeoDevBH1}, while the second can be observed in the transverse components
\eqref{InvGeoDevBHi}. Interestingly, up to a constant they are exactly the square roots of the two parts of the invariant \eqref{invB}, that is the amplitudes $\B_1$, $\B_2$ given by \eqref{B1}, \eqref{B2}.

The influence of these two distinct components $\B_1$ and $\B_2$ of the Bach tensor $B_{ab}$ on test particles is even more explicitly seen in the geodesic deviation of \emph{initially static test particles with} ${\dot{r}=0}$. The 4-velocity normalization then implies ${\Omega^2\H\,\dot{u}^2=-1}$, which simplifies \eqref{InvGeoDevBH1}, \eqref{InvGeoDevBHi} to
\begin{align}
\ddot{Z}^{(1)} = & \hspace{6mm} \frac{1}{6}\, \Omega^{-2}\big({\cal H}''+2\big)\,Z^{(1)}
-\frac{1}{3}\,k\,\Omega^{-4}\big(\B_1+\B_2\big)Z^{(1)} \,, \label{InvGeoDevBH1r0}\\
\ddot{Z}^{(i)} = & - \frac{1}{12}\,\Omega^{-2}\big({\cal H}''+2\big)\,Z^{(i)}
-\frac{1}{6}\,k\,\Omega^{-4}\,\B_1\,Z^{(i)} \,. \label{InvGeoDevBHir0}
\end{align}
From these expressions, it immediately follows that the first component $\B_1$ of the Bach tensor
is directly observed in the \emph{transverse} components of the
acceleration \eqref{InvGeoDevBHir0} along ${\bolde_{(2)}, \bolde_{(3)}}$, that is
${\partial_x, \partial_y}$ (equivalent to ${\partial_\theta, \partial_\phi}$),
while the second component $\B_2$ only occurs in the \emph{radial} component \eqref{InvGeoDevBH1r0} along
$\bolde_{(1)}= -\dot{u}\,(\partial_u+\H\,\partial_r)= -\H\,\Omega'\,\dot{u}\,\partial_{\bar r}$,
proportional to $\partial_{\bar r}$.

Interestingly, \emph{on the horizon there is only the radial effect} given by ${\B_2(r_h)}$ since
${\B_1(r_h)=0}$ due to \eqref{B1} and \eqref{horizon}, see also \eqref{bonhorizon}.

It can also be proven by direct calculation that the specific character of $\B_1, \B_2$
\emph{cannot mimic} the Newtonian tidal effect in the Schwarzschild solution, i.e., cannot
be ``incorporated'' into the first terms ${\Omega^{-2}\big({\cal H}''+2\big)}$ in \eqref{InvGeoDevBH1r0}, \eqref{InvGeoDevBHir0}. Therefore, by measuring the free fall of a set of test particles, it is \emph{possible
to distinguish} the pure Schwarzschild black hole from the Schwarzschild--Bach black hole
geometry which has nonvanishing Bach tensor ${B_{ab}\ne 0}$.

\subsection{Thermodynamic properties: horizon area, temperature, entropy}

It is also important  to determine main geometrical and thermodynamic
properties of the family of Schwarzschild--Bach black holes.
The \emph{horizon} in these spherically symmetric spacetimes is generated by
the rescaled null Killing vector ${\xi\equiv\sigma\partial_u=\sigma\partial_t}$, considering the time-scaling freedom \eqref{scaling-t} represented by a parameter $\sigma$. Thus it appears at \emph{zero of the metric function} ${\H(r)}$, where the norm of $\xi$ vanishes, see \eqref{horizon}. In the explicit form \eqref{Omega_[0,1]}, \eqref{H_[0,1]} this is clearly located at ${r=r_h}$ since ${\H(r_h)=0}$. By simply integrating the angular coordinates of the metric \eqref{BHmetric}, we immediately obtain the \emph{horizon area} as
\be
{\cal A} = 4\pi\,\Omega^2(r_h)= \frac{4\pi}{r_h^2}= 4\pi\,{\bar r}_h^2 \,.
\label{horizon_area}
\ee

The only nonzero derivatives of $\xi$ are
${\xi_{u;r}=-\xi_{r;u}=\frac{1}{2}\sigma(\Omega^2\H)'}$, and thus
${\xi^{\,r;u}=-\xi^{\,u;r}=\Omega^{-4}\xi_{u;r}}$. From the
definition \cite{Wald:1984} of \emph{surface gravity}
${\kappa^2\equiv-\frac{1}{2}\,\xi_{\mu;\nu}\,\xi^{\,\mu;\nu}}$,
we obtain ${\kappa=-\frac{1}{2}\sigma(\H'+2\H\,\Omega'/\Omega)}$. On the
horizon, where ${\H=0}$, using \eqref{H_[0,1]} this simplifies to
\be
\kappa/\sigma = -\frac{1}{2}\,\H'(r_h) = -\frac{r_h}{2} =\frac{1}{2\, {\bar r}_h} \,.
\label{surface_gravity}
\ee
It is \emph{the same expression as for the Schwarzschild solution}
(in which case ${\kappa=1/4m}$). The standard expression for \emph{temperature} of
the black hole horizon ${T\equiv\kappa/(2\pi)}$, which is valid even in higher-derivate
gravity theories \cite{FanLu:2015}, thus yields
\be
T/\sigma = -\frac{r_h}{4\pi}
  = \frac{1}{4\pi\, {\bar r}_h} \,,
\label{temperature}
\ee
\emph{independent of the Bach parameter $b$}.

However, in higher-derivative theories it is \emph{not} possible to use
the usual formula ${S=\frac{1}{4}{\cal A}}$ to determine the \emph{black hole horizon
entropy}. Instead, it is necessary to apply the generalized
formula derived by Wald \cite{Wald:1993,IyerWald:1994}, namely
\be
S=\frac{2\pi}{\kappa}\,\oint \mathbf{Q}\,,
\label{WaldS}
\ee
where the Noether charge 2-form $\mathbf{Q}$ on the horizon~is
\bea
{\bf Q} \rovno \pul \varepsilon_{\mu\nu\alpha\beta}\,Q^{\mu\nu}\,\dd x^{\alpha}\wedge\dd x^{\beta} \,,\nonumber\\
Q^{\mu\nu} \rovno 2X^{\mu\nu\rho\sigma}\,\xi_{\rho;\sigma}
+4{X^{\mu\nu\rho\sigma}}_{;\rho}\,\xi_\sigma   \quad \hbox{and}\quad
X^{\mu\nu\rho\sigma}\equiv\frac{\partial {\cal L}}{\partial R_{\mu\nu\rho\sigma}}
\,, \label{NoetherCharge}
\eea
in which ${\cal L}$ is the Lagrangian of the theory. In the case of Quadratic Gravity \eqref{actionQG}, it can be shown that
\bea
X^{\mu\nu\rho\sigma} \rovno
\frac{1}{16\pi}\bigg[
\Big(\gamma+\frac{2}{3}(2\alpha+3\beta) R \Big)g^{\nu[\sigma}g^{\rho]\mu}
-4\alpha\, g^{\kappa[\nu}g^{\mu][\rho}g^{\sigma]\lambda}R_{\kappa\lambda}
\bigg]\,. \label{Xabcd}
\eea
Subsequent lengthy calculation for the metric \eqref{BHmetric} with ${\Lambda=0}$ then leads to
\be
\mathbf{Q}(r_h) = -\frac{1}{16\pi}\,\Omega^2\,\H'
\bigg[\gamma +\frac{4}{3}k\alpha\,\frac{\B_1+\B_2}{\Omega^4}\bigg]\bigg|_{r=r_h}
\sin\theta\,\dd\theta\wedge\dd\phi\,. \label{NoetherCharge2}
\ee
Evaluating the integral \eqref{WaldS}, and using \eqref{horizon_area}, \eqref{surface_gravity}, \eqref{bonhorizon}, we finally obtain
\be
S = \frac{1}{4G}\,{\cal A}\,\Big(1-4kr_h^2\,b\Big)
  = \frac{1}{4G}\,{\cal A}\,\Big(1-4k\,  \frac{b}{\bar r_h^2}\Big) \,. \label{entropySB}
\ee
This explicit formula for the Schwarzschild--Bach black hole entropy
agrees with the numerical results presented in
\cite{LuPerkinsPopeStelle:2015}, with the identification
${k=\alpha}$ and ${b=\delta^*}$. In fact, it gives a geometrical
interpretation of the ``non-Schwarzschild parameter''
$\delta^*$
 as the dimensionless Bach parameter $b$ that
determines the \emph{value of the Bach tensor on the horizon}
 $r_h$, see relations \eqref{bonhorizon}. Of course,
for the Schwarzschild black hole (${b=0}$) or in Einstein's
General Relativity (${k=0}$) we recover the standard expression
${S=\frac{1}{4G}\,{\cal A}}$. Notice also from \eqref{entropySB} that for a given ${b \ne 0}$, the \emph{deviation} from this standard Schwarzschild entropy \emph{is larger when the Schwarzschild--Bach black holes are smaller} because they have smaller $\bar r_h$.

\section{Conclusions}
\label{conclusions}

The class of spherically symmetric black holes in Quadratic Gravity and the Einstein--Weyl theory was studied in many previous works, in particular \cite{Stelle:1978,Holdom:2002,LuPerkinsPopeStelle:2015, LuPerkinsPopeStelle:2015b, PerkinsPhD}, often by numerical methods applied to complicated field equations corresponding to the standard form of the spherical metric (\ref{Einstein-WeylBH}). In \cite{PodolskySvarcPravdaPravdova:2018, SvarcPodolskyPravdaPravdova:2018}, using a convenient form of the line element (\ref{BHmetric}) conformal to a simple Kundt seed, we obtained a surprisingly simple form of the field equations (\ref{Eq1}), (\ref{Eq2}). This enabled us to find an \emph{explicit form} of their exact solutions. Moreover, we identified the \emph{Bach tensor} as the key ingredient which makes the Schwarzschild solution geometrically distinct from the other branch of ``non-Schwarzschild'' ones. This is a direct consequence of the extension of Einstein's theory to include higher derivative corrections.

The present paper contains a thorough analysis of all such solutions and their derivation, including the details which had to be omitted in our brief letter \cite{PodolskySvarcPravdaPravdova:2018}.

We have started  with the conformal-to-Kundt metric ansatz (\ref{BHmetric}). Together with the Bianchi identities, this leads to a compact form of the Quadratic Gravity field equations (\ref{fieldeqsEWmod}), assuming ${R=0}$, namely the autonomous system of \emph{two ordinary differential equations} (\ref{Eq1}) and (\ref{Eq2}) for two metric functions ${\Omega(r)}$ and ${{\cal H}}(r)$.  They have been solved in terms of \emph{power series} representing these metric functions, expanded around any \emph{fixed point} $r_0$ (\ref{rozvojomeg0}), (\ref{rozvojcalH0}), or using the \emph{asymptotic expansion} (\ref{rozvojomegINF}), (\ref{rozvojcalHINF}), respectively. The field equations have become the algebraic constraints (\ref{KeyEq1}), (\ref{KeyEq2}) in the fixed point case (near ${r_0}$), and (\ref{KeyEq1INF}), (\ref{KeyEq2INF}) in the asymptotic region (as ${r \to \infty}$). Their dominant orders restrict the admitted solutions to (\ref{4classes}) and (\ref{2classes}), respectively. The detailed discussion of all the possible six main classes, together with a suitable fixing of the gauge freedom, can be found in subsequent Sections \ref{description} and \ref{description_INF}. The classes are summarized in Tables~\ref{tbl:01} and \ref{tbl:02} in Section~\ref{summary}.

The most prominent case corresponds to the spherically symmetric \emph{black hole spacetimes with} (in general) \emph{nonvanishing Bach tensor}. This solution has been expanded around the event horizon, see Subsection \ref{SchwaBach_[n,p]=[0,1]}. The metric functions ${\Omega(r)}$ and ${{\cal H}}(r)$ are given by the series (\ref{Omega_[0,1]}), (\ref{H_[0,1]}) with the initial coefficients specified by (\ref{alphasgammaIIbinitial}), and all other coefficients determined by the recurrent relations (\ref{alphasIIbgeneral}). Thus we have obtained the two-parametric family of black holes characterized by the radial position ${r_h}$ of the horizon and by the additional parameter $b$. The new Bach parameter distinguishes this more general \emph{Schwarzschild--Bach} solutions (${b\neq0}$) from the classical Schwarzschild spacetime with vanishing Bach tensor (${b=0}$). The main mathematical properties of the Schwarzschild--Bach metric functions are presented and visualized in Section~\ref{discussion-and-figures}. Subsequent Section~\ref{physics} contains the physical and geometrical analysis. We have discussed specific behavior of freely falling test observers, described by the equation of geodesic deviation, and demonstrated that their \emph{relative motion} encodes the presence of the Bach tensor. The physical investigation is completed by a fully explicit evaluation of the \emph{thermodynamic quantities}. In particular, the expression for entropy (\ref{entropySB}) exhibits the key role of the Bach parameter ${b}$.

Finally, for convenience, in Section~\ref{summary} we have also \emph{summarized all the admitted classes} of solutions, including their physical interpretation, the number of free parameters and, most importantly, relations to previous works. See, in particular, Table~\ref{tab:3}.

We hope that our approach to spherically symmetric vacuum solutions to Quadratic Gravity and the Einstein--Weyl theory may elucidate some of their properties that are not easily accessible by numerical simulations. Of course, we are aware of many remaining open questions. For example, complete analytic identification of the same physical solution in distinct classes and their mutual relations are still missing. It is also of physical interest to understand the effect of nontrivial Bach tensor in the Schwarzschild--Bach spacetimes on perihelion shift and light bending, studied thoroughly during the last century in Einstein's theory using the Schwarzschild solution.

\section*{Acknowledgements}

This work has been supported by the Czech Science Foundation
Grants No. GA\v{C}R 17-01625S (JP, R{\v S}) and 19-09659S (VP, AP), and the Research Plan RVO:
67985840 (VP, AP).

\newpage

\appendix

\section{The Ricci and Bach tensors for the Kundt seed}
\label{derivingRBseed}

We start with the seed Kundt metric (\ref{Kundt seed xy}). Its nontrivial metric components $g_{ab}$ are
\begin{equation}\label{Einstein-WeylBHC}
 g_{xx}^\Kdt = g_{yy}^\Kdt = \textstyle{\left(1+\frac{1}{4}(x^2+y^2)\right)^{-2}}\,,\qquad
 g_{ru}^\Kdt = -1 \,,\qquad
 g_{uu}^\Kdt = {\cal H}\,,
\end{equation}
so that the contravariant components $g^{ab}$ read
\begin{equation}\label{contraEinstein-WeylBHC}
 g^{xx}_\Kdt = g^{yy}_\Kdt = \textstyle{\left(1+\frac{1}{4}(x^2+y^2)\right)^2}\,,\qquad\quad
 g^{ru}_\Kdt = -1 \,,\qquad
 g^{rr}_\Kdt = -{\cal H} \,.
\end{equation}
Recall that the spatial 2-metric ${g_{ij}}$ is a round sphere of unit radius, with the Gaussian curvature ${K=1}$
and thus its Ricci scalar is ${{\cal R}=2K=2}$.
The nontrivial Christoffel symbols for this metric are
\BE
 \Gamma^r_{ru} = {\textstyle -\frac{1}{2}{\cal H}'} \,, \qquad
 \Gamma^r_{uu} = {\textstyle \frac{1}{2}{\cal H}{\cal H}'} \,,  \qquad
 \Gamma^u_{uu} = {\textstyle \frac{1}{2}{\cal H}'} \,, \qquad
 \Gamma^k_{ij} = {\textstyle \!\,^{S}\Gamma^k_{ij}} \,, \label{ChristoffelEnd}
\EE
where ${\,^{S}\Gamma^k_{ij}\equiv\frac{1}{2}g^{kl}(2g_{l(i,j)}-g_{ij,l})}$ are the symbols with respect to the spatial metric $g_{ij}$ of the 2-sphere.
The only nontrivial Riemann curvature tensor components are
\BE
 R_{ruru}^\Kdt = {\textstyle -\frac{1}{2}{\cal H}''} \,, \qquad
 R_{kilj}^\Kdt = g_{kl}g_{ij}-g_{kj}g_{il} \,,
\EE
and the only nontrivial Ricci tensor components of (\ref{Einstein-WeylBHC}) are
\begin{eqnarray}
 R_{ru}^\Kdt \rovno -\pul\,{\cal H}'' \,, \label{Ricci ru5} \\
 R_{uu}^\Kdt \rovno -{\cal H}\,R_{ru}^\Kdt \,,\label{Ricci uu5}\\
 R_{xx}^\Kdt = R_{yy}^\Kdt \rovno g_{xx} \,, \label{Ricci ij5}
\end{eqnarray}
while the Ricci scalar reads
\be
R^\Kdt ={\cal H}''  + 2  \,,
\ee
so that the only nontrivial Weyl tensor components are
\begin{eqnarray}
 C_{ruru}^\Kdt \rovno {\textstyle -\frac{1}{6} R} \,, \label{WeyliK}\\
 C_{riuj}^\Kdt \rovno {\textstyle \frac{1}{12} R \,g_{ij}} \,, \\
 C_{kilj}^\Kdt \rovno {\textstyle \frac{1}{6}R\,(g_{kl}g_{ij}-g_{kj}g_{il}) } \,, \\
 C_{uiuj}^\Kdt \rovno  -{\cal H}\, C_{riuj}  \,. \label{WeylfK}
\end{eqnarray}
The nonzero components of the Bach tensor are
\begin{eqnarray}
B_{rr}^\Kdt \rovno {\textstyle -\frac{1}{6}\,{\cal H}'''' } \,, \label{Bach rr}\\
B_{ru}^\Kdt \rovno {\textstyle \frac{1}{12}\,
\big(2\,{\cal H}{\cal H}''''+{\cal H}'{\cal H}'''-{\textstyle\frac{1}{2}}{{\cal H}''}^2 +2\big) } \,, \label{Bach ru}\\
B_{uu}^\Kdt \rovno  -{\cal H}\,B_{ru}^\Kdt \,, \label{Bach uu}\\
B_{xx}^\Kdt = B_{yy}^\Kdt \rovno {\textstyle \frac{1}{12}}\,g_{xx}\,
\big({\cal H}{\cal H}''''+{\cal H}'{\cal H}'''-{\textstyle\frac{1}{2}}{{\cal H}''}^2 +2\big)   \,,\label{Bach xx}
\end{eqnarray}
involving up to the 4th derivative of the metric function $\H(r)$.

\section{The Ricci and Bach tensors for the conformal metric}
\label{App_derivingFE}

Taking the  class of Kundt geometries (\ref{Kundt seed xy}) as a seed, we can generate the metric of spherically symmetric geometries by the conformal transformation (\ref{confrelation}), that is
\be
\dd   s^2 = \Omega^2(r)\,\bigg[
\frac{\dd x^2+\dd y^2}{\big(1+\frac{1}{4}(x^2+y^2)\big)^2}
-2\,\dd u\,\dd r+{\cal H}(r)\,\dd u^2 \bigg]\,.  \label{BHmetric-xy}
\ee

Now, it is well-known \cite{Wald:1984} that under a conformal transformation of the seed metric
\be
  g_{ab}=\Omega^2 \,g_{ab}^\Kdt \,,
\label{confrel}
\ee
the Ricci scalar and the Ricci and Bach tensors transform as
\begin{eqnarray}
  R      \rovno \Omega^{-2}R^\Kdt-6\Omega^{-3} \square\Omega\,, \label{OmRiccscalar}\\
  R_{ab} \rovno R_{ab}^\Kdt - 2\Omega^{-1}\nabla_a\nabla_b\Omega - \Omega^{-1} g_{ab}^\Kdt\square\Omega+ \Omega^{-2}(4\Omega_{,a}\Omega_{,b}-g_{ab}^\Kdt g^{cd}_\Kdt\Omega_{,c}\Omega_{,d})\,, \label{OmRicc}\\
  B_{ab} \rovno \Omega^{-2}B_{ab}^\Kdt\,. \label{OmBach}
\end{eqnarray}

For the  Kundt seed metric $g_{ab}^\Kdt$ (\ref{Einstein-WeylBHC}), its Ricci and Bach tensors
$R_{ab}^\Kdt$ and $B_{ab}^\Kdt$ are given by (\ref{Ricci ru5})--(\ref{Ricci ij5}) and (\ref{Bach rr})--(\ref{Bach xx}), respectively. The nontrivial derivatives (with respect to the Kundt seed) of the conformal factor $\Omega(r)$ are, in view of (\ref{ChristoffelEnd}),
\begin{eqnarray}
&& \Omega_{,r} \equiv \Omega'\,, \nonumber\\
&& \nabla_r\nabla_r \Omega= \Omega''\,,\quad
\nabla_r\nabla_u \Omega= {\textstyle\frac{1}{2}}{\cal H}'\Omega' = \nabla_u\nabla_r \Omega\,,\quad
\nabla_u\nabla_u \Omega=  {-\textstyle\frac{1}{2}}{\cal H}{\cal H}'\Omega' \,,\\
&& \square\Omega  =  -({\cal H}\Omega''+{\cal H}'\Omega')\,.  \nonumber
\end{eqnarray}
Employing (\ref{OmRicc}), the nonvanishing Ricci tensor components of the metric (\ref{BHmetric-xy}) are thus
\begin{eqnarray}
 R_{rr} \rovno -2\Omega^{-2}\big(\Omega\Omega''-2{\Omega'}^2\big) \,, \label{RT_R rr}\\
 R_{ru} \rovno -\pul \Omega^{-2}\big(\Omega^2 {\cal H}\big)'' \,, \label{RT_R ru}\\
 R_{uu} \rovno  -{\cal H}\, {R}_{ru} \,, \label{RT_R uu}\\
 R_{xx} =  {R}_{yy} \rovno
\Omega^{-2}g_{xx}\,
\big[ \big({\cal H}\Omega\Omega'\big)'+\Omega^2 \big]   \,,\label{RT_R xx}
\end{eqnarray}
and using (\ref{OmRiccscalar}) we obtain
\be
 R = 6\Omega^{-3} \big[{\cal H}\Omega''+{\cal H}'\Omega'
+{\textstyle \frac{1}{6}} ({\cal H}''+2)\Omega \big] \,.
\label{barR}
\ee
The nonvanishing Bach tensor components $B_{ab}$ are obtained by a trivial rescaling (\ref{OmBach}) of (\ref{Bach rr})--(\ref{Bach xx}).

\section{Derivation and simplification of the field equations}
\label{analysingFE}

The vacuum field equations  in the Einstein--Weyl theory and also general Quadratic Gravity for the metric $g_{ab}$ are (\ref{fieldeqsEWmod}), that is
\begin{equation}
  R_{ab} = 4k\,  B_{ab} \,.
 \label{EWfield equations}
\end{equation}
Using the expressions (\ref{RT_R rr})--(\ref{RT_R xx}) and (\ref{OmBach}) with (\ref{Bach rr})--(\ref{Bach xx}), these \emph{field equations explicitly read}
\begin{align}
\Omega\Omega''-2{\Omega'}^2 & = \tfrac{1}{3}k\, {\cal H}'''' \,, \label{Neq_rr} \\
\big(\Omega^2 {\cal H}\big)'' & = -\tfrac{2}{3}k \big(2\,{\cal H}{\cal H}''''+{\cal H}'{\cal H}'''-{\textstyle\frac{1}{2}}{{\cal H}''}^2 +2\big) \,, \label{Neq_ru} \\
\big({\cal H}\Omega\Omega'\big)'+\Omega^2 & = \tfrac{1}{3}k \,\big({\cal H}{\cal H}''''+{\cal H}'{\cal H}'''-{\textstyle\frac{1}{2}}{{\cal H}''}^2 +2 \big) \,. \label{Neq_xx}
\end{align}
The equations (\ref{Neq_rr}), (\ref{Neq_ru}), (\ref{Neq_xx}) represent the nontrivial components $rr$, $ru$, $xx$ (identical to $yy$), respectively.
The $uu$ component of the field equations is just the ${(-{\cal H})}$-multiple of (\ref{Neq_ru}).

Moreover, recall that the trace of the field equations (\ref{EWfield equations}) is ${ {R}=0}$, cf. (\ref{R=0}).
Using (\ref{barR}) we obtain the explicit condition
\begin{equation}
\T\equiv{\cal H}\Omega''+{\cal H}'\Omega'+{\textstyle \frac{1}{6}} ({\cal H}''+2)\Omega = 0 \,.
 \label{traceC}
\end{equation}
It can be checked that this is a direct \emph{consequence} of equations
(\ref{Neq_rr})--(\ref{Neq_xx}). Notice that it is a \emph{linear differential equation for the function} ${\cal H}(r)$,
and also linear differential equation for $\Omega(r)$.

We have thus obtained \emph{three nontrivial field equations} (\ref{Neq_rr})--(\ref{Neq_xx}) for \emph{two unknown functions} $\Omega(r)$ and ${\cal H}(r)$, and also their consequence (\ref{traceC}).
Therefore, this coupled system seems to be overdetermined. However, now we prove that the key metric functions  $\Omega(r)$ and ${\cal H}(r)$ are, in fact, \emph{solutions of just two coupled equations}.

To this end, let us introduce the \emph{auxiliary symmetric tensor} $J_{ab}$ defined as
\BE
J_{ab}\equiv   R_{ab}-\pul R\,g_{ab} - 4k\,B_{ab} \,.
\label{defJab}
\EE
Using $J_{ab}$, the vacuum field equations (\ref{GenQGFieldEq}) of Quadratic Gravity (assuming a constant $R$ and ${\Lambda=0}$) or Einstein--Weyl gravity (with ${\beta=0=\Lambda}$) are simply
\BE
J_{ab}=0 \,.
\label{Jab=0}
\EE

Now, by employing the contracted \emph{Bianchi identities} ${\nabla^b R_{ab}=\frac{1}{2} R_{,a}}$
and the conservation property of the Bach tensor ${\nabla^b B_{ab}=0}$, see (\ref{Bachproperties}), we obtain
\BE
\nabla^b J_{ab}\equiv 0 \,.
\label{BianchiIdEW}
\EE
Interestingly, this is actually a \emph{geometrical identity} which is valid
without employing any field equations, namely (\ref{Jab=0}), or (\ref{EWfield equations}) in particular.

An explicit evaluation of the identity (\ref{BianchiIdEW}) for the metric $ {g}_{ab}$ of the form (\ref{BHmetric-xy})
leads to the following equations, which are \emph{always satisfied}:
\begin{align}
& \nabla^b J_{rb} = - \Omega^{-3}\Omega'\big(J_{ij}\,{g}^{ij}
  +{\cal H} J_{rr}\big)-\Omega^{-2}\big({\cal H} J_{rr,r}+ J_{ru,r}+\tfrac{3}{2}{\cal H}' J_{rr}\big)
  \hspace{-25mm}&\equiv 0 \,, \label{BI_r} \\
& \nabla^b J_{ub} = -2\Omega^{-3}\Omega'\big(J_{uu}+{\cal H} J_{ru}\big)-\Omega^{-2}\big(J_{uu}+
 {\cal H} J_{ru}\big)_{,r} &\equiv 0 \,, \label{BI_u} \\
& \nabla^b J_{ib} = \Omega^{-2} J_{ik||l}\,{g}^{kl} &\equiv 0 \,. \label{BI_i}
\end{align}
Here the spatial covariant derivative $_{||}$ is calculated with respect to the spatial part $g_{ij}$ of the Kundt
seed metric (\ref{Einstein-WeylBHC}). Moreover, a direct calculation of $J_{ab}$ defined by (\ref{defJab}) gives
\begin{equation}
J_{uu}=-{\cal H}\, J_{ru}\,, \qquad  J_{xx}=\mathcal{J}(r)\, g_{xx}
 = J_{yy}  \,, \label{EqDep}
\end{equation}
where the function $\mathcal{J}(r)$ is defined as
\begin{equation}
\mathcal{J} \equiv \Omega^{-2}\,\big[\big({\cal H}\Omega\Omega'\big)'+\Omega^2
-3\T\Omega
-\tfrac{1}{3}k \,\big({\cal H}{\cal H}''''+{\cal H}'{\cal H}'''-{\textstyle\frac{1}{2}}{{\cal H}''}^2 +2 \big)\big] \,, \label{calJ}
\end{equation}
and
\begin{align}
J_{rr} =  2\Omega^{-2}&\,\big[-\Omega\Omega''+2{\Omega'}^2+\tfrac{1}{3}k\,{\cal H}''''\big] \,, \label{bJrr}\\
J_{ru} =  \Omega^{-2}&\,\big[-\tfrac{1}{2}\big(\Omega^2 {\cal H}\big)''
+3\T\Omega
 -\tfrac{1}{3}k \big(2\,{\cal H}{\cal H}''''+{\cal H}'{\cal H}'''-{\textstyle\frac{1}{2}}{{\cal H}''}^2 +2\big)\big] \label{bJru}\,.
\end{align}

By substituting the relations (\ref{EqDep}) into (\ref{BI_u}) and (\ref{BI_i}), it can  be seen immediately
that these two conditions are automatically satisfied.
Interestingly, the remaining Bianchi identity (\ref{BI_r}) gives a \emph{nontrivial} result. If the metric functions
$\Omega(r)$ and ${\cal H}(r)$ satisfy the two field equations ${J_{rr}=0}$ and ${J_{ru}=0}$
then necessarily ${J_{ij}\,g^{ij}\equiv0}$, that is
${J_{xx}\,g^{xx}+J_{yy}\,g^{yy}=2\mathcal{J}(r)=0}$
and thus ${J_{xx}=0=J_{yy}}$.

Therefore, we conclude that \emph{all field equations} for the metric (\ref{BHmetric-xy}) \emph{reduce
just to two key equations}, namely ${J_{rr}=0}$ and ${J_{ru}=0}$. Since ${g^{ab}J_{ab}=0}$, it also implies ${R=0}$ and thus ${\T=0}$, cf.~\eqref{traceC}.
This coupled system of two equations completely determines all possible exact vacuum solutions  of the type
(\ref{BHmetric-xy}) in Einstein--Weyl gravity, and since ${  R=0}$, also in a general Quadratic Gravity.
The key point is that, due to the Bianchi identities, the two key equations \emph{imply} the nontrivial field equations ${J_{xx}=0= J_{yy}}$ since necessarily ${\mathcal{J}=0}$, that is using (\ref{calJ})
\begin{equation}
\big({\cal H}\Omega\Omega'\big)'+\Omega^2-3\T\Omega
= \tfrac{1}{3}k \big({\cal H}{\cal H}''''+{\cal H}'{\cal H}'''-{\textstyle\frac{1}{2}}{{\cal H}''}^2 +2 \big) \,. \label{J=0}
\end{equation}
The equation ${J_{rr}=0}$ is exactly equation \eqref{Neq_rr}, and equation \eqref{Neq_ru} is simply ${J_{ru}=0}$ with ${\T=0}$.
Finally, substituting ${\T=0}$ into \eqref{J=0}, we immediately obtain \eqref{Neq_xx}. This completes the proof of the equivalence.

To integrate the field equations, it is necessary to solve the equation \eqref{Neq_rr}. Simultaneously, we must solve the equation
\be
\big(\Omega^2 {\cal H}\big)'' -6\T\Omega =
 -\tfrac{2}{3}k \big(2\,{\cal H}{\cal H}''''+{\cal H}'{\cal H}'''-{\textstyle\frac{1}{2}}{{\cal H}''}^2 +2\big) \label{bJru0}\,.
\ee
Remarkably, this equation \emph{can further be simplified} by expressing the term ${{\cal H}''''}$ from \eqref{Neq_rr}. We thus finally obtain \emph{two very simple field equations}
\begin{align}
\Omega\Omega''-2{\Omega'}^2 = &\ \tfrac{1}{3}k\,{\cal H}'''' \,, \label{Eq1C}\\
\Omega\Omega'{\cal H}'+3\Omega'^2{\cal H}+\Omega^2
 = &\ \tfrac{1}{3}k \big({\cal H}'{\cal H}'''-{\textstyle\frac{1}{2}}{{\cal H}''}^2 +2 \big)\,, \label{Eq2C}
\end{align}
for the two metric functions $\Omega(r)$ and ${\cal H}(r)$. Alternatively, instead of solving the single equation (\ref{Eq2C}), it is also possible to solve \emph{any two of the three equations} (\ref{Neq_ru}), (\ref{Neq_xx}), (\ref{traceC}).

\end{document}